\documentclass[11pt]{article}
\pdfoutput=1
\usepackage{amsmath,amssymb,latexsym,graphicx}
\usepackage{color}
\definecolor{MyDarkBlue}{rgb}{0,0.08,0.45}
\usepackage{hyperref}
\hypersetup{
colorlinks=true,
citecolor=MyDarkBlue,
linkcolor=MyDarkBlue,
urlcolor=MyDarkBlue,
pdfauthor={Bernard de Wit and Stefanos Katmadas },
pdftitle={Near-horizon analysis of D=5 BPS black holes and
    rings},
pdfsubject={hep-th}
}
\newcommand{\ft}[2]{{\textstyle\frac{#1}{#2}}}

\def\slash#1{\rlap{\hbox{$\mskip 1 mu /$}}#1}      % good slash for
\def\Slash#1{\rlap{\hbox{$\mskip 3 mu /$}}#1}      % " upper
\newsavebox{\uuunit}
\sbox{\uuunit}
    {\setlength{\unitlength}{0.825em}
     \begin{picture}(0.6,0.7)
        \thinlines
        \put(0,0){\line(1,0){0.5}}
        \put(0.15,0){\line(0,1){0.7}}
        \put(0.35,0){\line(0,1){0.8}}
       \multiput(0.3,0.8)(-0.04,-0.02){10}{\rule{0.5pt}{0.5pt}}
     \end {picture}}

\numberwithin{equation}{section}
\begin{document}
\begin{titlepage}
\begin{center}
\hfill ITP-UU-09/78 \\
\hfill SPIN-09/36  \\
%\hfill {\tt hep-th/yymmnnn}\\
\vskip 15mm

%%%%%%%%%%%%%%%%%%%%%%%%%%%%%%%%%%%%%%%%%%
{\Large \textbf{Near-horizon analysis of D=5 BPS black holes and
    rings}}
%%%%%%%%%%%%%%%%%%%%%%%%%%%%%%%%%%%%%%%%%%%
\vskip 12mm

\textbf{Bernard de Wit and Stefanos Katmadas}

\vskip 4mm
$^b${\em Institute for Theoretical Physics} and {\em Spinoza
  Institute,\\ Utrecht University, Utrecht, The Netherlands}\\
{\tt  B.deWit@uu.nl\;, \; S.Katmadas@uu.nl} \\[1mm]
\end{center}

\vskip .3in
%%%%%%%%%%%%%%%%%%%%%%%%%%%%%%%%%%%%%%%%%%%%%%%%%%%%%
\begin{center} {\bf ABSTRACT } \end{center}
\begin{quotation}\noindent
  A comprehensive analysis is presented based exclusively on
  near-horizon data to determine the attractor equations and the
  entropy of BPS black holes and rings in five space-time dimensions,
  for a Lagrangian invariant under eight supersymmetries with
  higher-derivative couplings. For spinning black holes the results
  only partially agree with the results of previous work, where often
  additional input was used beyond the near-horizon behaviour. A
  number of discrepancies remains, for example, pertaining to small
  black holes and to large spinning black holes, which are related to
  the presence of the higher-derivative couplings. Arguments are
  presented to explain some of them. For the black rings, the analysis
  is intricate due to the presence of Chern-Simons terms and due to
  the fact that the gauge fields are not globally defined. The
  contributions from the higher-derivative couplings take a systematic
  form in line with expectations based on a variety of arguments.
\end{quotation}

\vfill
%%%%%%%%%%%%%%%%%%
%\flushleft{\today}
%%%%%%%%%%%%%%%%%%
\end{titlepage}
%%%%%%%%%%%%%%%%
\eject
%%%%%%%%%%%%%%%%
\tableofcontents
\pagebreak
%%%%%%%%%%%%%%%%%%%%%%%%%%%%%%%%%%%%%%%%%%%%%%%%%%%%%%%%%%%%%%%%%%%%%%
%%%%%%%%%%%%%%%%%%%%%%%%%%%%%%%%%%%%%%%%%%%%%%%%%%%%%%%%%%%%%%%%%%%%%%
\section{Introduction}
\label{sec:introduction}
\setcounter{equation}{0}
%%%%%%%%%%%%%%%%%%%%%%%%%%%%%%%%%%%%%%%%%%%%%%%%%%%%%%%%%%%%%%%%%%%%%%
The attractor phenomenon for BPS black branes
\cite{Ferrara:1995ih,Strominger:1996kf,Ferrara:1996dd} is caused by
full supersymmetry enhancement at the horizon, which induces stringent
restrictions on the values of the fields and the space-time geometry.
When supersymmetry is realized off shell, the resulting attractor
equations can be analyzed in a way that is independent of the action.
In this way universal results can be obtained even when the action
contains higher-derivative couplings, as was first demonstrated for
$N\!=\!2$ supergravity in four dimensions \cite{Lopes Cardoso:1998wt}.

In five space-time dimensions, supersymmetric attractors come in two
varieties, associated with the near-horizon geometry of the rotating
black hole \cite{Breckenridge:1996is, Gauntlett:1998fz}, and of the
black ring \cite{Elvang:2004rt}. In the context of the two-derivative
effective action, these attractors have been studied in
\cite{Gauntlett:1998fz, Gauntlett:2002nw,Gauntlett:2004qy,
  Kraus:2005gh,Goldstein:2007km,Hanaki:2007mb}, using
mostly on-shell methods. It is possible to include higher-derivative
couplings into the conventional two-derivative supergravity action,
but those require the use of off-shell methods. Such a
(four-derivative) supersymmetric action has been constructed in
\cite{Hanaki:2006pj}.  Both the two- and the four-derivative couplings
involve a Chern-Simons term, which is a characteristic feature of
five-dimensional supergravity. In the two-derivative case the
Chern-Simons term is cubic in the gauge fields, whereas the
higher-derivative mixed Chern-Simons term involves also the
spin connection field. As a result the Lagrangian is only gauge
invariant up to a total derivative, a feature that causes certain
technical complications.

A study of BPS black holes and black rings that includes these
higher-derivative interactions was initiated some time ago in
\cite{Kraus:2005vz,Castro:2007hc,Castro:2007ci,Castro:2008ne}. In
these works, a number of black hole solutions was constructed, and the
corresponding attractors were studied by taking the near-horizon
limit. In addition, the entropy function formalism \cite{Sen:2005wa}
was used to determine the macroscopic entropy of these black holes,
after reducing to four dimensions to restore gauge invariance of the
action. A corresponding analysis for black rings was hampered by the
difficulty in obtaining full asymptotically flat solutions.

In this work, we present a comprehensive treatment of five-dimensional
$N\!=\!2$ attractors in the presence of the same four-derivative
couplings, using the tools provided by the off-shell calculus. This
analysis of the near-horizon behaviour thus relies only on the full
supersymmetry enhancement and does not take into account the more
global aspects of possible solutions. In particular, no assumptions
are made concerning the existence of interpolating solutions towards
asymptotic infinity, and no use is made of any information from
outside the near-horizon region. This is in line with the idea that
the entropy of black branes should be determined fully by the horizon
properties, in the spirit of the Bekenstein-Hawking area
law.\footnote{%%%%%%%%%%%%%%%%%%%%%%%%%%%%%%%%%%%%%%%%%%%%
  See, however, \cite{Banerjee:2009uk}, for a possibly different
  perspective. } %%%%%%%%%%%%%%%%%%%%%%%%%%%%%%%%%%%%%%%%%
As in the four-dimensional analysis, we find that the allowed
space-time geometry is the same as for the two-derivative theory,
which in the case at hand is described by the $AdS_2\times S^2\times
S^1$ geometry of \cite{LozanoTellechea:2002pn}.  Because this geometry
interpolates between the black hole and the black ring attractors, we
can treat both types of five-dimensional attractors in a unified way
for a large part of the analysis.

The higher-derivative corrections in the action enter into the
expressions for both the entropy and the attractor equations
pertaining to electric charges and angular momenta. For the Wald
entropy \cite{Wald:1993nt,Jacobson:1993vj,Iyer:1994ys} we obtain a
universal formula expressed in terms of the horizon fields, which
applies to both black holes and rings.  This is an intriguing result,
because the derivation in these two cases proceeds rather differently
due to a number of subtleties associated with the mixed
gauge-gravitational Chern-Simons term. Our treatment of this mixed
Chern-Simons term is inspired by, but not completely identical to, the
approach followed in \cite{Tachikawa:2006sz}. The existence of a
universal entropy formula is in line with previous results based on
the entropy function upon reduction to four dimensions, and we confirm
this by confronting the results with the four-dimensional near-horizon
analysis.

The charges and the angular momenta can also be defined at the
horizon. In view of the first law of black hole mechanics, this
requires the use of the same Noether potential that enters into the
determination of the Wald entropy. The evaluation of the full Noether
potential is rather involved, and, as alluded to above, the relevant
potentials do not take the same form for black rings and for black
holes. The electric charges defined at the horizon are conserved by
construction (although they are not invariant under large gauge
transformations in the case of black rings).  Subtleties arise with
the proper definition of the gauge fields in the presence of the
Chern-Simons terms, and those have important implications on the
attractor equations for black hole and black ring charges.

There exists an extended literature on how to define the electric
charges (for a general discussion, see for instance
\cite{Marolf:2000cb}). In the case at hand, it is worth mentioning
that the results of this paper differ from those of
\cite{Castro:2007hc,Castro:2007ci,Castro:2008ne}. Many of these
differences reside in the definition of the electric charges that was
adopted in these references, which was based on the asymptotic
fall-off of the electric fields at spatial infinity. As a result the
charges for asymptotically Taub-NUT solutions are different from the
charges for asymptotically flat solutions, and furthermore these
charges depend on the distance from the horizon.  In contrast, the
charges employed in this paper are insensitive to the asymptotic
structure of the space-time and do not depend on the distance from the
horizon.

As mentioned above, the BPS near-horizon geometries come in two
varieties. In the case of a spacelike horizon cross section with
spherical topology, we recover the $AdS_2\times S^3$ near-horizon
geometry of the rotating black hole
\cite{Breckenridge:1996is,Gauntlett:1998fz}. In the other case we find
the $AdS_3\times S^2$ near-horizon geometry of the supersymmetric
black ring \cite{Elvang:2004rt}. The latter constitutes a special
limit of the generic BPS near-horizon geometry for which the spacelike
cross section of the horizon has the topology of $S^2\times S^1$, as
is appropriate for a ring. Unlike the black hole, the black ring
carries two independent angular momenta associated with rotations in
two orthogonal planes. There are some other new features
related to the non-contractible $S^1$. The first one
concerns the fact that this background allows for non-trivial magnetic
charges on the circle, because magnetic charges are not pointlike in
five dimensions, but are stringlike objects. Hence the ring carries
magnetic dipole charges. The second one concerns the non-trivial
moduli associated with Wilson lines along the circle. We present a
careful treatment of the gauge fields in this topology, which enables
us to recover the correct electric charges and their associated
attractor equations, following the strategy of \cite{Hanaki:2007mb}.
Using this same strategy we also establish the modified electric
charges that are additive.

Our results for the entropy agree with the results of microscopic
counting for large black holes
\cite{Breckenridge:1996is,Vafa:1997gr,Huang:2007sb} and for black
rings \cite{Bena:2004tk,Cyrier:2004hj}. So far this agreement holds
for static black holes, because at present there exists no analytic
expression for the microscopic entropy of rotating black
holes.\footnote{%%%%%%%%%%%%%%%%%%%%%%%%%%%%%%%%%%%%%%%%%%%%%%%%%%%
  In theories with 16 supersymmetries explicit expressions are
  available \cite{Castro:2008ys,Banerjee:2008ag}. It is an interesting
  question as to whether there exist asymptotic limits thereof which
  will agree with the results of this paper. \label{footnote:N=4}
} %%%%%%%%%%%%%%%%%%%%%%%%%%%%%%%%%%%%%%%%%%%%%%%%%%%%%%%%%%%%%%%%%
In contrast, we disagree with the microscopic counting
\cite{Huang:2007sb} for small black holes, whose macrocsopic entropy
depends sensitively on the higher-derivative couplings. This puzzle
may be characteristic for small black holes; also in four dimensions,
the supergravity description of small black holes was often
problematic, although there the problems did not pertain to the
leading contribution. For black rings the effect of the
higher-derivative couplings was not included in
\cite{Bena:2004tk,Cyrier:2004hj}, but we will exactly reproduce the
result known from the $(4,0)$ conformal field theory, which leads to
an expression proportional to $\sqrt{c_L\,\hat q_0}$ (c.f.
\cite{Maldacena:1997de}).  Here, $c_L$ is the central charge, which
can be expressed in the dipole charges and which includes the effect
of the higher-derivative correction. Its form is in agreement with
arguments based on the $AdS_3$ near-horizon geometry
\cite{Kraus:2005vz,Kraus:2005zm, Kraus:2006wn}.  Furthermore,
$\hat{q}_0$ is an appropriately defined quantity expressed in the
angular momenta and the charges.  As noted in \cite{Bena:2005ae}, this
quantity is naturally written in terms of the aforementioned modified
electric charges that are additive.

On the other hand, our results for rotating black holes only partially
agree with the macroscopic results of
\cite{Castro:2007hc,Castro:2007ci,Castro:2008ne}, as was already
alluded to above in connection with the definition used for the
electric charges. Our results also disagree with the prediction of
\cite{Guica:2005ig} for the first-order contribution from the angular
momentum to the black hole entropy, based on the addition of the Euler
density to the supergravity action. In this work, also a correction to
the black ring entropy was determined based on the Euler density.  In
hindsight, it is difficult to see how the Euler density could possibly
capture, at the same time, all contributions to the entropy, the
angular momenta and the electric charges for both black holes and
black rings, as it does not include the contributions from the mixed
Chern-Simons term, which, especially for black rings, is responsible
for subtle effects.

Another issue concerns the connection between corresponding black hole
solutions and their associated entropy in four and in five dimensions.
This connection is motivated by the fact that the four-dimensional
theory can be obtained by dimensional reduction on a circle from the
five-dimensional one
\cite{Gaiotto:2005gf,Behrndt:2005he,Hanaki:2007mb}, although there may
be subtleties. One such subtlety, related to the supersymmetry
preserved by five-dimensional attractors upon reducing to four
dimensions, was already discussed in
\cite{Goldstein:2008fq}. Following this reasoning, our
five-dimensional attractor equations should be related to the
four-dimensional attractors with a specific $R^2$-coupling. Indeed, we
find agreement with four dimensions in the case of the black ring,
except that the quantity $\hat q_0$ in four dimensions will only
depend on the unmodified electric charges. We explain the reason for
this fact, which is of topological origin and not related to the
presence of higher-derivative couplings. For the case of the rotating
black hole, we find a clear discrepancy in the contributions from the
higher-derivative couplings to the electric charges. A similar, though
somewhat different, deviation from the four-dimensional situation was
observed in \cite{Castro:2007hc,Castro:2007ci,Castro:2008ne}. One way
to understand the discrepancy in the electric charges follows from the
observation that the reduction to four dimensions of the
higher-derivative term will involve an extra vector multiplet
associated with the Kaluza-Klein photon, which will also be subject to
higher-derivative interactions.  So far, such interactions have never
been considered directly in four dimensions, and in fact the precise
form of these four-dimensional couplings is not fully known although
there are indications that they should not affect the formula for the
Wald entropy for BPS black holes. But most likely they will have an
effect on the electric charges, and this may resolve the present
discrepancy between the electric charges in four and five dimensions.

This paper is organized as follows.  Section \ref{sec:superconf-mult}
contains a concise but comprehensive review of the superconformal
transformation rules for the supermultiplets of interest. In section
\ref{sec:tensor-calculus} the product rules and supersymmetric density
formulae are presented, using the notation of this paper. Section
\ref{sec:bps-attor-eqs} is devoted to the derivation of the attractor
equations and their consequences. The attractor equations are derived
in subsection \ref{sec:supersymmetry}, and the resulting geometry is
discussed in subsection \ref{sec:space-time-geometry}. Subsequently
the horizon values of the gauge fields and the linear supermultiplets
are discussed in subsections \ref{sec:gauge-fields} and
\ref{sec:linear-multiplets}, respectively. The invariant action is
given in section \ref{sec:lagr-electr-charg} from which the attractor
equations for the electric charges can be determined. In section
\ref{sec:entropy-angular-momentum} we discuss the entropy and angular
momentum for black holes and rings. As it turns out the mixed
Chern-Simons term requires a different approach for black holes and
black rings. Therefore, after a review of the more generic situation
for black holes, a subsection \ref{sec:an-alternative-form} is devoted
to the alternative treatment of the mixed Chern-Simons term that is
required for the black rings. The final results of this paper for
spinning black holes and black rings, together with a discussion and a
comparison to results in the literature, are presented in section
\ref{sec:entropy-black-holes} and section
\ref{sec:entropy-black-rings}, respectively. Readers who are not
primarily interested in the technical details, may proceed directly to
these two sections. There are two appendices: appendix
\ref{App:5D-conv} introduces the spinor and space-time notation used
in the first part of this paper, and appendix \ref{App:conf-sg}
contains a brief review of extended conformal supergravities in five
space-time dimensions.

%%%%%%%%%%%%%%%%%%%%%%%%%%%%%%%%%%%%%%%%%%%%%%%%%%%%%%%%%%%%%%%%%%%%%%
\section{Superconformal multiplets}
\label{sec:superconf-mult}
\setcounter{equation}{0}
%%%%%%%%%%%%%%%%%%%%%%%%%%%%%%%%%%%%%%%%%%%%%%%%%%%%%%%%%%%%%%%%%%%%%%
A convenient method for dealing with off-shell formulations of
supergravity theories is provided by the superconformal multiplet
calculus. This calculus was originally set up for $N\!=\!2$
supergravity in $d\!=\!4$ dimensions \cite{de Wit:1979ug,de
  Wit:1980tn,de Wit:1984pk,de Wit:1984px}, following early work for
$N\!=\!1, d\!=\!4$ supergravity \cite{Kaku:1978nz,Kaku:1978ea}. The
$N\!=\!1$ case was worked out more fully in \cite{Kugo:1982cu}, and
shortly thereafter the formalism was also applied to $N\!=\!1,d\!=\!6$
supergravity in \cite{Bergshoeff:1985mz}.  For $d\!=\!5$ dimensions
superconformal methods were developed relatively recently by several
groups
\cite{Bergshoeff:2001hc,Fujita:2001kv,Bergshoeff:2004kh,Hanaki:2006pj},
and these results were exploited in the work of
\cite{Kraus:2005vz,Castro:2007hc,Castro:2008ne}. However, these groups
use different field and symmetry definitions, which have features that
are qualitatively different from the conventions used in $d=4$
dimensions.  Obviously this poses no problem of principle, but in
order to make the connection with the four-dimensional theory as
direct as possible, we have chosen to adopt slightly different
conventions.

In this section we give a self-contained summary of the transformation
rules of superconformal multiplets in five space-time dimensions,
namely the Weyl multiplet, the vector multiplet, the linear multiplet
and the hypermultiplet for supergravity in five space-time dimensions
with eight supercharges. With the exception of the hypermultiplet,
these multiplets define off-shell representations of the algebra of
superconformal transformations. We refer to appendix \ref{App:5D-conv}
for spinor and space-time conventions. Some additional material about
the Weyl multiplets in four and five dimensions with eight and sixteen
supercharges is presented in appendix \ref{App:conf-sg}. The fields of
conformal supergravity are dual to the components of the
supermultiplet of currents, and are subject to a number of gauge
transformations directly related to the conservation laws of these
currents. The bosonic gauge transformations are those of the conformal
group, diffeomorphisms, local Lorentz transformations with generators
$M_{ab}$, scale transformations with generator $D$ and special
conformal transformations (also called conformal boosts) with
generators $K_a$. Furthermore there are local R-symmetry
transformations. In five space-time dimensions, the R-symmetry group
equals $\mathrm{USp}(2N)$ so that for simple supergravity we have
$\mathrm{USp}(2)\cong\mathrm{SU}(2)$. The fermionic gauge
transformations are the conventional Q- and the special conformal
S-supersymmetry transformations.

%%%%%%%%%%%%%%%%%%%%%%%%%%%%%%%%%%%%%%%%%%%%%%%%%%%%%%%%%%%%%%%%%%%%%%
\subsection{The Weyl multiplet}
\label{sec:Weyl-multiplet}
%%%%%%%%%%%%%%%%%%%%%%%%%%%%%%%%%%%%%%%%%%%%%%%%%%%%%%%%%%%%%%%%%%%%%%
The Weyl multiplet of five-dimensional simple conformal supergravity
is shown in table~\ref{tab:countWeyl}. The independent fields consist
of the f\"unfbein $e_\mu{}^a$, the gravitino field $\psi_\mu{}^i$, the
dilatational gauge field $b_\mu$, the R-symmetry gauge fields $V_{\mu
  i}{}^j$ (which is an anti-hermitean, traceless matrix in the
$\mathrm{SU}(2)$ indices $i,j$) and a tensor field $T_{ab}$, a scalar
field $D$ and a spinor field $\chi^i$. The three gauge fields
$\omega_\mu{}^{ab}$, $f_\mu{}^a$ and $\phi_\mu$, associated with local
Lorentz transformations, conformal boosts and S-supersymmetry,
respectively, are not independent and will be discussed later. The
infinitesimal Q, S and K transformations of the independent
fields, parametrized by spinors $\epsilon^i$ and $\eta^i$ and a vector
$\Lambda_\mathrm{K}{}^a$, respectively, are as follows, 
\begin{eqnarray}
  \label{eq:Weyl-susy-var}
  \delta e_\mu{}^a &=&  \ft 12\bar\epsilon_i \gamma^a \psi_\mu{}^i\,,
  \nonumber\\ 
  \delta \psi_{\mu}^i  &=& 
  {\cal  D}_\mu \epsilon^i + \tfrac1{4}\mathrm{i}\,
  T_{ab}( 3\,\gamma^{ab}\gamma_\mu-\gamma_\mu\gamma^{ab}) \epsilon^i
  -\mathrm{i}  \gamma_\mu\eta^i \,, \nonumber\\ 
  \delta V_{\mu i}{}^j &=& 3 \mathrm{i}
  \bar\epsilon_{i} \phi_{\mu}{}^{j}
  -8\bar\epsilon_{i}\gamma_\mu\chi^{j} -3 \mathrm{i}
  \bar\eta_{i}\psi_\mu{}^{j} + 
  \delta^i{}_j\,[-\ft{3}{2}\mathrm{i}\bar\epsilon_{k}\phi_{\mu}{}^{k}
  +4\bar\epsilon_{k}\gamma_\mu\chi^{k}+\ft{3}{2}\mathrm{i}
  \bar\eta_{k}\psi_\mu^{k}] \,, \nonumber \\
  \delta b_\mu &=&{}
  \ft12\mathrm{i} \bar\epsilon_i\phi^i_\mu -2 \bar\epsilon_i\gamma_\mu
  \chi^i + \ft12\mathrm{i} \bar\eta_i\psi^i_\mu +2\Lambda _{K}{}^a
  e_{\mu a} \,, \nonumber\\
  \delta T_{ab} &=&  \ft23 \mathrm{i} \bar\epsilon_i \gamma_{ab}
  \chi^i -\ft{1}{8} \mathrm{i} \bar\epsilon_i R_{ab}{}^i(Q)\,,
  \nonumber\\  
  \delta \chi^i &=&  
  \ft 14 \epsilon^i D +\ft{1}{128} 
  R_{\mu\nu j}{}^{i}(V) \gamma^{\mu\nu} \epsilon^j 
  + \ft3{128}\mathrm{i}(3\, \gamma^{ab} \Slash{D}
  +\Slash{D}\gamma^{ab})T_{ab} \, \epsilon^i \nonumber\\
  &&{}
  -\ft 3{32} T_{ab}T_{cd}\gamma^{abcd}\epsilon^i 
                                %+ \ft16 T^2 \epsilon^i   
  +\ft3{16} T_{ab}\gamma^{ab} \eta^i  \,, \nonumber\\
  \delta D &=&{}
  \bar\epsilon_i \Slash{D} \chi^i - \mathrm{i}
  \bar\epsilon_i  T_{ab}\gamma^{ab} \chi^i - \mathrm{i}
  \bar\eta_i\chi^i \,. 
\end{eqnarray}
Under local scale transformations the various fields and
transformation parameters transform as indicated in table
\ref{tab:weyl-multiplet}. The derivatives $\mathcal{D}_\mu$ are
covariant with respect to all the bosonic gauge symmetries with the
exception of the conformal boosts. In particular we note
\begin{equation}
  \label{eq:D-epsilon}
\mathcal{D}_{\mu} \epsilon^i = \big( \partial_\mu - \tfrac{1}{4}
\omega_\mu{}^{cd} \, \gamma_{cd} + \tfrac1{2} \, b_\mu\big)
\epsilon^i + \tfrac1{2} \,{V}_{\mu j}{}^i \, \epsilon^j  \,, 
\end{equation}
where the gauge fields transform under their respective gauge
transformations according to
$\delta\omega_\mu{}^{ab}=\mathcal{D}_\mu\lambda^{ab}$, $\delta b_\mu=
\mathcal{D}_\mu\Lambda_D$ and $\delta V_{\mu i}{}^j= \mathcal{D}_\mu
\Lambda_i{}^j$, with $(\Lambda_i{}^j)^\ast\equiv \Lambda^i{}_j=
- \Lambda_j{}^i$. The derivatives $D_\mu$ are covariant with
respect to all the superconformal symmetries. 
  
%%%%%%%%%%%%%%%%%%%%%%%%%%%%%%%%%%%%%%%%%%%%%%%%%%%%%%%%%%%%%%%%
%
\begin{table}[t]
\centering
\begin{tabular}{|c|ccccccc|ccc|ccc|} 
\hline 
 & &\multicolumn{8}{c}{Weyl multiplet} & &
 \multicolumn{2}{c}{parameters} & \\  \hline 
 field & $e_\mu{}^{a}$ & $\psi_\mu{}^i$ & $b_\mu$ &
 $\mathcal{V}_{\mu\,i}{}^j$ & $T_{ab} $ & 
 $ \chi^i $ & $D$ & $\omega_\mu{}^{ab}$ & $f_\mu{}^a$ &$\phi_\mu{}^i$&
 $\epsilon^i$ & $\eta^i$  
 & \\ \hline
$w$  & $-1$ & $-\tfrac12 $ & 0 &  0 & 1 & $\tfrac{3}{2}$ & 2 & 0 &
1 & $\tfrac12 $ & $-\tfrac12$  & $\tfrac12$ & \\ \hline 
\end{tabular}
\caption{\small Weyl weights $w$ of the Weyl multiplet component
    fields and the supersymmetry transformation parameters.}
\label{tab:weyl-multiplet}
\end{table}
%%%%%%%%%%%%%%%%%%%%%%%%%%%%%%%%%%%%%%%%%%%%%%%%%%%%%%%%%%%%%%%

In order to discuss the dependent gauge fields, we first introduce the
following curvature tensors,
\begin{eqnarray}
  \label{eq:curvatures}
  R_{\mu\nu}{}^a(P) &=& 2\,\mathcal{D}_{[\mu}e_{\nu]}{}^a
  -\tfrac12 \bar\psi_{[\mu i}\gamma^a \psi_{\nu]}{}^i
  \,,\nonumber\\[.4ex]   
  R_{\mu\nu}{}^{ab}(M) &=& 2\,\partial_{[\mu}\omega_{\nu]}{}^{ab}
  -2\, \omega_{[\mu}{}^{ac} \omega_{\nu]c}{}^{b} -8\, e_{[\mu}{}^{[a}
  f_{\nu]}{}^{b]} + \mathrm{i} \bar\psi_{[\mu
  i}\gamma^{ab}\phi_{\nu]}{}^i \nonumber\\
  && - \tfrac14 \mathrm{i} T^{cd}
  \,\bar\psi_{[\mu i}(6\gamma^{[a}\gamma_{cd}\gamma^{b]}
  -\gamma^{ab}\gamma_{cd}-\gamma_{cd}\gamma^{ab})\psi_{\nu]}{}^i
  \nonumber\\ 
  &&
  -\tfrac12\bar\psi_{[\mu i}(\gamma_{\nu]} R^{abi}(Q) + 2\,\gamma^{[a}
  R_{\nu]}{}^{b]i}(Q)) + 8\,
  e_{[\mu}{}^{[a}\,\bar\psi_{\nu]i}\gamma^{b]}\chi^i \,,\nonumber\\[.4ex]  
  R_{\mu \nu}(D) &=&{} 2\,\partial_{[\mu} b_{\nu]}-4\,f_{[\mu}{}^a
  e_{\nu]a}   
  - \mathrm{i} \bar{\psi}_{[\mu i} \phi_{\nu]}{}^i + 4\,\bar\psi_{[\mu
  i}\gamma_{\nu]}\chi^i \,. \nonumber\\[.4ex]  
  R_{\mu\nu i}{}^j (V) &=& 2\, \mathcal{\partial}_{[\mu}V_{\nu]i}{}^j
  - V_{[\mu i}{}^k V_{\nu] k}{}^j   \nonumber \\
  &&
  - 6 \mathrm{i}\,\bar\psi_{[\mu i} \phi_{\nu]}{}^j +16 
  \bar\psi_{[\mu i}\gamma_{\nu]} \chi^j
  +\delta_i{}^j\big[3 \mathrm{i}\,\bar\psi_{[\mu k} \phi_{\nu]}{}^k -8 
  \bar\psi_{[\mu k}\gamma_{\nu]} \chi^k \big] \,,\nonumber\\[.4ex]  
  R_{\mu\nu}{}^i(Q) &=& 2\, \mathcal{D}_{[\mu}\psi_{\nu]}{}^i -2 \mathrm{i}\,
  \gamma_{[\mu} \phi_{\nu]}{}^i  +\ft12\mathrm{i}\, T_{ab}(
  3\,\gamma^{ab}\gamma_{[\mu} -\gamma_{[\mu}\gamma^{ab})
  \psi_{\nu]}{}^i  \,. 
\end{eqnarray}
The conventional constraints (which are not invariant under Q- and
S-supersymmetry) are as follows, 
\begin{eqnarray}
  \label{eq:conv-constraints}
  R_{\mu\nu}{}^a(P) &=& 0\,,\nonumber \\
  \gamma^\mu R_{\mu\nu}{}^i(Q) &=& 0\,,\nonumber\\
  e_a{}^\mu\, R_{\mu\nu}^{ab}(M) &=& 0 \,. 
\end{eqnarray}
These conditions determine the gauge fields $\omega_\mu{}^{ab}$,
$f_\mu{}^a$ and $\phi_\mu{}^i$. The conventional constraints lead to
additional constraints on the curvatures when combined with the
Bianchi identities. In this way one derives $R_{[abc]d}(M) =0=
R_{ab}(D)$ and the pair-exchange property $R_{abcd}=R_{cdab}$ from the
first and the third constraint.  The second constraint, which implies
also that $\gamma_{[\mu\nu} R_{\rho\sigma]}{}^i(Q) =0$, determines the
curvature $R_{\mu\nu}{}^i(S)$, which we refrained from defining
previously. It turns out to be proportional to $R_{\mu\nu}{}^i(Q)$ and
derivatives thereof,
\begin{eqnarray}
  \label{eq:R(S)} 
  R_{\mu\nu}{}^i(S) &=& -\mathrm{i} \Slash{D} 
  R_{\mu\nu}{}^i(Q) - {\rm i} \gamma_{[\mu} 
  D^{\rho }  R_{\nu]\rho}{}^i(Q)
  - 4\,\gamma_{\mu\nu} T^{\rho\sigma} R_{\rho\sigma}{}^i(Q)  \nonumber\\
  &&  +18\, \gamma_\sigma T^{\rho\sigma} \gamma_{[\mu} 
  R_{\nu]\rho}{}^i(Q) -5 \, T^{\rho\sigma}\gamma_{\rho\sigma}  
  R_{\mu\nu}{}^i(Q)  - 12\, T^\rho{}_{[\mu} R_{\nu]\rho}{}^i(Q) \,.
\end{eqnarray}
The remaining curvature $R_{\mu\nu}{}^a(K)$ does not play a role in
what follows.

Whereas the first constraint is invariant under S- but not under
Q-supersymmetry, the other two constraints are invariant under neither
supersymmetry. This implies that the dependent gauge fields will
acquire terms in their transformation rules proportional to the
constrained curvature tensors,
\begin{eqnarray}
  \label{eq:variations-PSK-fields}
  \delta\omega_\mu{}^{ab}&=& \mathcal{D}_\mu\lambda^{ab}+ 4
  \Lambda_K{}^{[a} e_\mu{}^{b]} - \ft12 \mathrm{i}
  \epsilon_i\gamma^{ab}\phi_\mu{}^i + \ft12 \mathrm{i}
  \eta_i\gamma^{ab}\psi_\mu{}^i \nonumber\\
  &&{}
  + \tfrac18 \mathrm{i} T^{cd}
  \,\bar\epsilon_i(6\gamma^{[a}\gamma_{cd}\gamma^{b]}
  -\gamma^{ab}\gamma_{cd}-\gamma_{cd}\gamma^{ab})\psi_\mu{}^i 
  \nonumber\\[.4ex]
  &&{}
  +\tfrac14\bar\epsilon_i(\gamma_\mu R^{abi}(Q) + 2\,\gamma^{[a}
  R_\mu{}^{b]i}(Q)) + 4\,
  e_\mu{}^{[a}\,\bar\epsilon_i\gamma^{b]}\chi^i 
  \,,\nonumber\\[.4ex]
  \delta\phi_\mu{}^i&=& \mathcal{D}_\mu\eta^i
  +\tfrac14\mathrm{i}\,T_{ab}(\gamma_\mu\gamma^{ab}-
  \gamma^{ab}\gamma_\mu)\eta^i  
  +\mathrm{i}f_\mu{}^a\gamma_a \epsilon^i -\mathrm{i}\Lambda_K{}^a
  \gamma_a\psi_\mu{}^i \nonumber\\
  &&-\ft1{48}\mathrm{i}
  (2\,\gamma^{ab}\gamma_\mu-\gamma_\mu\gamma^{ab}) R_{abj}{}^i(V)\epsilon^j
  + \ft {1}{4}\big(\Slash{D}T^{ab}\gamma_{ab}\gamma_\mu  +
     D_a T^{ab}\gamma_\mu\gamma_b\big)\epsilon^i  \nonumber\\ 
  &&  +\mathrm{i}\big(-\tfrac34 T^{ab}T^{cd}\gamma_{\mu abcd} +
  T_{\mu a}T_{bc}\gamma^{abc}  -4\, T_{\mu a}T^{ab}\gamma_b 
    - \tfrac34 \gamma_\mu T^2\big) \epsilon^i \nonumber\\
  &&-\ft94 \mathrm{i}\,\bar\epsilon_j\psi_\mu{}^j\, \chi^i +
  \tfrac74\mathrm{i}\, \bar\epsilon_j\gamma_a\psi_\mu{}^j\, 
  \gamma^a\chi^i - \ft 18 \mathrm{i}
  \bar\epsilon_k\gamma^{ab}\psi_\mu{}^{k}\big(\gamma_{ab}\chi^i
  +\ft 14  R_{ab}{}^i(Q) \big)  \nonumber \\
  &&
  +\ft 14 \mathrm{i}
  \bar\epsilon_k\gamma^{ab}\psi_\mu{}^{i}\big(\gamma_{ab}\chi^k
    + \ft 14  R_{ab}{}^k(Q) \big) \,,\nonumber\\[.4ex]  
\delta f_\mu{}^{a}&=& \mathcal{D}_\mu\Lambda_K{}^a +\ft12
  \eta_i\gamma^a\phi_\mu{}^i  + \cdots\,, 
\end{eqnarray}
where here and henceforth $T^2\equiv (T_{ab})^2$.  With these results
we obtain the following Q- and S-variations that will be needed in due
course,
\begin{eqnarray}
  \label{eq:delta-R-Q+V}
  \delta R_{ab}{}^i(Q) &=&{}  -\ft1{24}(\gamma^{cd}\gamma_{ab} -
    4\,\delta_a{}^{[c} \delta_b{}^{d]})\, R_{cd i}{}^i(V) \epsilon^j 
                     - \ft 14 R_{ab}{}^{cd}(M) \gamma_{cd}\epsilon^i
    \nonumber\\  
    &&{} 
    + \tfrac12 \mathrm{i}\big(3\,D_{[a} T^{cd}
    \gamma_{cd}\gamma_{b]} -  D_{[a} T^{cd} \gamma_{b]}\gamma_{cd}-
    \gamma_{[a}\Slash{D}T^{cd} \gamma_{cd}\gamma_{b]} - 
    D_c T^{cd}\gamma_{ab}\gamma_d \big) \epsilon^i \nonumber\\ 
    && {}
    -2\big(T_{ab} T_{cd}\gamma^{cd} +
    T_{ac}T_{bd}\gamma^{cd}+2\,T_{c[a}T^{cd} \gamma_{b]d}+
    \tfrac14 T^2\gamma_{ab}\big)\epsilon^i\nonumber\\ 
    &&{}
      + \big(\gamma^{cd}\gamma_{ab} -4\,\delta_a{}^{[c}
      \delta_b{}^{d]}\big)\eta^i \, T_{cd}\,, \nonumber \\[.4ex]
    \delta R_{ab i}{}^j(V) &=& 3\mathrm{i}\,\bar\epsilon_{i}
       R_{ab}{}^{j}(S) 
       +16\,\bar\epsilon_{i}\gamma_{[a}D_{b]}\chi^{j}
       -4\mathrm{i}\,\bar\epsilon_{i}\big(3\,\gamma_{[a}\gamma^{cd}
       \gamma_{b]}-\gamma^{cd} \gamma_{ab}\big)\chi^{j} \,T_{cd}
       \nonumber\\  
       &&{}  
       - 3\mathrm{i}\,\bar\eta_{i}R_{ab}{}^{j}(Q)
    -16\mathrm{i}\,\bar\eta _{i} \gamma _{ab }\chi^{j} - 
    \mathrm{trace}\;. 
\end{eqnarray}

The above transformations coincide with those of
\cite{Bergshoeff:2001hc,Bergshoeff:2004kh}, upon including a
$T$-dependent S-supersymmetry transformation into the Q-supersymmetry
variations and rescaling the tensor field by a factor $4/3$. The
difference with the conventions of \cite{Fujita:2001kv,Hanaki:2006pj}
are a bit more involved. The commutator of two Q-supersymmetry
transformations closes into the superconformal transformations as
follows,
\begin{equation}
  \label{eq:QQ-comm}
  \left[\delta_Q(\epsilon_1), \delta_Q(\epsilon_2) \right] =
  \xi^\mu\hat D_\mu + \delta_M(\lambda)+ \delta_S(\eta)+
  \delta_K(\Lambda_K)\,,  
\end{equation}
where $\xi^\mu\hat D_\mu$ denotes the effect of a supercovariant
general coordinate transformation. The parameters appearing on the
right-hand side associated with the general coordinate transformation
and the Lorentz transformation are given by
\begin{eqnarray}
  \label{eq:QQ-parameters}
  \xi^\mu &=&{} \ft 12\,\bar\epsilon_{2i}\gamma^\mu\epsilon_1{}^i \,,
  \nonumber \\ 
  \lambda^{ab} &=&{} \ft18\mathrm{i}T^{cd}\,\bar\epsilon_{2i}
  (6\gamma^{[a}\gamma_{cd}\gamma^{b]} 
  -\gamma^{ab}\gamma_{cd}-\gamma_{cd}\gamma^{ab})\epsilon_1{}^i  \,,
\end{eqnarray}
whereas the parameters associated with S-supersymmetry and conformal
boosts are not given as they are not needed below.  The commutator of
two S-supersymmetry transformations and the commutator of a Q- and an
S-supersymmetry transformation close as follows,
\begin{eqnarray}
  \label{eq:SS+SQ}
  \left[\delta_S(\eta_1),\delta_S(\eta_2) \right]&=&
  \delta_K(\Lambda_K)\,, \nonumber\\ 
  \left[\delta_S(\eta), \delta_Q(\epsilon) \right] &=&
  \delta_D(\Lambda_D)+ \delta_M(\lambda)+\delta_R(\Lambda) +
  \delta_K(\tilde\Lambda_K) \,, 
\end{eqnarray}
where
\begin{eqnarray}
  \label{eq:SS+SQ-parameters}
  \Lambda_K{}^a&=&\ft{1}{2}\bar\eta_{2i}\gamma^a\eta_1{}^i  \,,
  \nonumber \\
  \Lambda_D&=& \ft{1}{2}\mathrm{i}\,\bar{\epsilon}_i\eta^i
  \,, \nonumber\\
  \lambda^{ab}&=& -\ft{1}{2}\mathrm{i}\,\bar\epsilon_i\gamma^{ab}\eta^i
  \,, \nonumber\\ 
   \Lambda_i{}^j &=& 3 \mathrm{i}\,(\bar\epsilon_i\eta^j -
  \ft12\delta_i{}^j \bar\epsilon_k\eta^k)\,, \nonumber\\
  \tilde\Lambda_K{}^a&=& -
  \ft1{16}\,\bar\epsilon_i[5\gamma^a\gamma^{bc}-4\gamma^{bc}\gamma^a]\eta^i
  \,T_{bc} \,. 
\end{eqnarray}
Furthermore, we note the following commutation relation, 
\begin{equation}
  \label{eq:KQ=S}
  \left[\delta_K(\Lambda_K),\delta_Q(\epsilon) \right]=
  \delta_S(\mathrm{i}\,\Slash{\Lambda}_K\epsilon^i) \,.  
\end{equation}

%%%%%%%%%%%%%%%%%%%%%%%%%%%%%%%%%%%%%%%%%%%%%%%%%%%%%%%%%%%%%%%%%%%%%%
\subsection{The vector supermultiplet}
\label{sec:vector-multiplet}
%%%%%%%%%%%%%%%%%%%%%%%%%%%%%%%%%%%%%%%%%%%%%%%%%%%%%%%%%%%%%%%%%%%%%%
The vector supermultiplet consists of a real scalar $\sigma$, a gauge
field $W_\mu$, a triplet of (auxiliary) fields $Y^{ij}$, and a fermion
field $\Omega^i$. Under superconformal transformations these fields
transform as follows, 
\begin{eqnarray}
  \label{eq:sc-vector-multiplet}
  \delta \sigma &=&{}
  \ft1{2}\mathrm{i}\bar{\epsilon}_i\Omega^i \,,
  \nonumber \\ 
  \delta\Omega^i &=&{}
  - \ft14 (\hat{F}_{ab}- 4\,\sigma T_{ab}) \gamma^{ab} \epsilon^i
  -\ft1{2}\mathrm{i} \Slash{D} \sigma\epsilon^i -\varepsilon_{jk}\,
  Y^{ij} \epsilon^k 
  %+ \sigma \,T_{ab}\gamma^{ab}\epsilon^i+
  + \sigma\,\eta^i \,,   \nonumber\\ 
  \delta W_\mu&=&{}
  \ft12 \bar{\epsilon}_i\gamma_\mu\Omega^i -\ft12 \mathrm{i}
  \sigma \,\bar\epsilon_i \psi^i_\mu \,, \nonumber\\ 
  \delta Y^{ij}&=&{}  
  \ft12 \varepsilon^{k(i}\, \bar{\epsilon}_k \Slash{D} \Omega^{j)} 
  + {\mathrm{i}}\varepsilon^{k(i}\, \bar\epsilon_k (-\ft1{4}
  T_{ab}\gamma^{ab}\Omega^{j)}+ 4 \sigma \chi^{j)})
  -\ft1{2}{\mathrm{i}}  \varepsilon^{k(i}\, \bar{\eta}_k
  \Omega^{j)} \,.  
\end{eqnarray}
where $(Y^{ij})^\ast\equiv Y_{ij}= \varepsilon_{ik}\varepsilon_{jl}
Y^{kl}$, and the supercovariant field strength is defined as, 
\begin{equation}
  \label{eq:W-field-strength}
 \hat F_{\mu\nu}= \partial_\mu W_\nu - \partial_\nu W_\mu -
\bar\Omega_i\gamma_{[\mu} \psi_{\nu]}{}^i +\ft12
\mathrm{i}\sigma\,\bar\psi_{[\mu i} \psi_{\nu]}{}^i \,.
\end{equation}
The commutator of two Q-supersymmetry transformations closes as in
\eqref{eq:QQ-comm} modulo an extra gauge transformation, $\delta W_\mu
=\partial_\mu(\frac12\mathrm{i}
\sigma\,\bar\epsilon_{2i}\epsilon_1{}^i)$. We also note the
transformation rule,
\begin{eqnarray}
  \label{eq:delta-F+sigma-T}
  \delta(\hat F_{ab} -4\,\sigma T_{ab})&=&{}- 
  \bar\epsilon_i\gamma_{[a}D_{b]}\Omega^i 
  -\tfrac83\,\sigma\,\bar\epsilon_i\gamma_{ab}\chi^i \nonumber\\
  &&{} 
  +\ft14\mathrm{i}\bar\epsilon_i (3\,\gamma_{[a}\gamma^{cd}\gamma_{b]}
  -\gamma^{cd}\gamma_{ab} - 8\,\delta_a^c\delta_b^d)\Omega^i\,T_{cd}
  +\mathrm{i}\bar\eta_i\gamma_{ab}\Omega^i\,. 
\end{eqnarray}
The fields behave under local scale transformations according to the
weights shown in table~\ref{tab:w-weights-matter}.

%%%%%%%%%%%%%%%%%%%%%%%%%%%%%%%%%%%%%%%%%%%%%%%%%%%%%%%%%%%%%%%%%%%%
%
\begin{table}[t]
\centering
\begin{tabular}{|c|cccc| } 
\hline 
 &  \multicolumn{4}{c|}{vector multiplet}  \\  \hline 
 field & $\sigma$ & $W_\mu$ & $\Omega_i$  & $Y_{ij}$  \\ \hline
$w$  & $1$ & $0$ &$\tfrac{3}{2}$ & 2  \\ \hline
\hline 
 & \multicolumn{4}{c|}{linear multiplet}  \\  \hline
 field & $L^{ij}$& $E_a$ & $\varphi_i$ & $N$  \\ \hline
$w$  & $3$ & $4$ & $\tfrac{7}{2}$ &  4 \\ \hline\hline
 & \multicolumn{4}{c|}{hypermultiplet}  \\  \hline
 field & $A_i{}^\alpha$ &  & $\zeta^\alpha$ &   \\ \hline
$w$  & $\tfrac{3}{2}$ & ~ & $2$ &  ~   \\ \hline
\end{tabular}
\caption{\small Weyl weights $w$ of the vector multiplet, the tensor
  (linear) multiplet, and the hypermultiplet component fields.} 
\label{tab:w-weights-matter} 
\end{table}
%%%%%%%%%%%%%%%%%%%%%%%%%%%%%%%%%%%%%%%%%%%%%%%%%%%%%%%%%%%%%%%%%%%%%%

%%%%%%%%%%%%%%%%%%%%%%%%%%%%%%%%%%%%%%%%%%%%%%%%%%%%%%%%%%%%%%%%%%%%%%
\subsection{The linear supermultiplet}
\label{sec:linear-multiplet}
%%%%%%%%%%%%%%%%%%%%%%%%%%%%%%%%%%%%%%%%%%%%%%%%%%%%%%%%%%%%%%%%%%%%%%
The linear multiplet consists of a triplet of scalars $L^{ij}$, a
divergence-free vector $\hat E_a$, an (auxiliary) scalar $N$, and a fermion
field $\varphi^i$. The superconformal transformation rules for these
fields are as follows,
\begin{eqnarray}
  \label{eq:linear-multiplet}
  \delta L^{ij}&=&{}
  -\mathrm{i} \,\varepsilon^{k(i}\,\bar\epsilon_k \varphi^{j)}
  \,,  \nonumber\\
  \delta \varphi^{i}&=&{}
  -\ft12 \mathrm{i}\, \varepsilon_{jk}\Slash{D} L^{ij}\epsilon^k+ \ft12
  (N-\mathrm{i}\hat{\Slash{E}})\epsilon^i 
  %-\varepsilon_{jk}  T^{ab}\gamma_{ab} L^{ij}\epsilon^k 
  + 3\varepsilon_{jk} L^{ij} \eta^k  \, ,\nonumber\\ 
  \delta \hat E_a &=&{}
  -\ft{1}{2}\mathrm{i}\, \bar\epsilon_{i}\gamma_{ab} D^b \varphi^{i}
  +\tfrac{1}{8} 
  \bar\epsilon_{i}(3\gamma_a\gamma^{bc} + \gamma^{bc}\gamma_a)
  \varphi^{i}T_{bc}  -2 
  \bar\eta_{i}\gamma_a\varphi^{i} \, ,\nonumber\\ 
  \delta N &=&{}
  \ft12\bar\epsilon_i \Slash{D} \varphi^i + \ft{3}{4}\mathrm{i}
  \bar\epsilon_i \gamma^{ab}\varphi^iT_{ab} -4 \mathrm{i} \,
  \varepsilon_{jk}\,\bar\epsilon_i\chi^k L^{ij}
  +\tfrac32 \mathrm{i}\bar\eta_i\varphi^i  \,. 
\end{eqnarray}
The constraint on $\hat E^a$,
\begin{equation}
\label{eq:Con.E}
  D_a\hat E^a =0\,, 
\end{equation}
can be solved in terms of a three-rank anti-symmetric tensor gauge
field $E_{\mu\nu\rho}$, which transforms as follows under the
superconformal transformations,
\begin{equation}
  \label{eq:susy-E}
  \delta E_{\mu\nu\rho}=
  \tfrac12\bar\epsilon_i\gamma_{\mu\nu\rho}\varphi^i -
  \tfrac32\mathrm{i}\,\bar\epsilon_i\gamma_{[\mu\nu}
  \psi_{\rho]}{}^k\, \varepsilon_{jk} L^{ij}\,. 
\end{equation}
These transformations close according to the commutation relations
\eqref{eq:QQ-comm}, \eqref{eq:SS+SQ} and \eqref{eq:KQ=S}, up to a
tensor gauge transformation, $\delta E_{\mu\nu\rho} = \partial_{[\mu}
(-\tfrac23 \mathrm{i}
\,\bar\epsilon_{2i}\gamma_{\nu\rho]}\epsilon_1{}^k
\,\varepsilon_{jk}L^{ij})$.  The supercovariant field strength
associated with $E_{\mu\nu\rho}$ equals
\begin{equation}
  \label{eq:E-field-strength}
  \hat E^\mu= \tfrac16 \mathrm{i}\,e^{-1}
  \varepsilon^{\mu\nu\rho\sigma\lambda} 
  \Big[\partial_\nu E_{\rho\sigma\lambda} -\tfrac12 \bar\psi_{\nu
  i}\gamma_{\rho\sigma\lambda} \varphi^i +
  \tfrac34\mathrm{i}\,\bar\psi_{\nu i}\gamma_{\rho\sigma}
  \psi_\lambda{}^k\,\varepsilon_{jk}L^{ij} \Big] \,. 
\end{equation}
The behaviour under local scale transformations follow from the
weights shown in table~\ref{tab:w-weights-matter}. The tensor field
$E_{\mu\nu\rho}$ is inert under scale transformations and thus carries
zero weight.  

%%%%%%%%%%%%%%%%%%%%%%%%%%%%%%%%%%%%%%%%%%%%%%%%%%%%%%%%%%%%%%%%%%%%%%
\subsection{Hypermultiplets}
\label{sec:hypermultiplet}
%%%%%%%%%%%%%%%%%%%%%%%%%%%%%%%%%%%%%%%%%%%%%%%%%%%%%%%%%%%%%%%%%%%%%%
Hypermultiplets are necessarily associated with target spaces of
dimension $4r$ that are hyperk\"ahler cones
\cite{deWit:1999fp,deWit:2001dj}. The supersymmetry transformations
are most conveniently written in terms of the sections
$A_i{}^\alpha(\phi)$, where $\alpha= 1,2,\ldots,2r$,
\begin{eqnarray} 
  \label{eq:hypertransf}
  \delta A_i{}^\alpha&=& \mathrm{i}\,\bar\epsilon_i\zeta^\alpha\,,
  \nonumber\\ 
  \delta\zeta^\alpha &=& -\ft12 \mathrm{i}\Slash{D}
  A_i{}^\alpha\epsilon^i 
          %- \ft13 T^{ab}\gamma_{ab}  A^{i\alpha}\epsilon^j\varepsilon_{ij}
  + \tfrac3{2} A_i{}^\alpha\eta^i \,.
\end{eqnarray}
The $A_i{}^\alpha$ are the local sections of an
$\mathrm{Sp}(r)\times\mathrm{Sp}(1)$ bundle. The existence of such an
associated bundle is known from general arguments \cite{Swann}.  We
also note the existence of a covariantly constant skew-symmetric
tensor $\Omega_{\alpha\beta}$ (and its complex conjugate
$\Omega^{\alpha\beta}$ satisfying
$\Omega_{\alpha\gamma}\Omega^{\beta\gamma}= -\delta_\alpha{}^\beta$),
and the symplectic Majorana condition for the spinors reads as
$C^{-1}\bar\zeta_\alpha{}^\mathrm{T} = \Omega_{\alpha\beta}
\,\zeta^\beta$. Covariant derivatives contain the $\mathrm{Sp}(r)$
connection $\Gamma_A{}^\alpha{}_\beta$, associated with rotations of
the fermions. The sections $A_i{}^\alpha$ are pseudo-real, i.e. they
are subject to the constraint, $A_i{}^\alpha \varepsilon^{ij}
\Omega_{\alpha\beta} = A^j{}_\beta\equiv (A_j{}^\beta)^\ast$. The
information on the target-space metric is contained in the so-called
hyperk\"ahler potential,
\begin{equation}
  \label{eq:hyperkahler-pot}
  \varepsilon_{ij} \,\chi  = \Omega_{\alpha\beta} \,A_i{}^\alpha
  A_j{}^\beta \,. 
\end{equation}
For the local scale transformations we refer again to the weights
shown in table~\ref{tab:w-weights-matter}. The hypermultiplet does not
exist as an off-shell supermultiplet. Closure of the superconformal
transformations is only realized upon using fermionic field equations,
but this fact does not represent a serious problem in what follows.

%%%%%%%%%%%%%%%%%%%%%%%%%%%%%%%%%%%%%%%%%%%%%%%%%%%%%%%%%%%%%%%%%%%%%
\section{Tensor calculus}
\label{sec:tensor-calculus}
\setcounter{equation}{0}
%%%%%%%%%%%%%%%%%%%%%%%%%%%%%%%%%%%%%%%%%%%%%%%%%%%%%%%%%%%%%%%%%%%%%

In the previous section we introduced various superconformal
multiplets. With the exception of the hypermultiplets, these
multiplets are truly off-shell, so that the superconformal symmetries
close without the need for imposing field equations. The tensor
calculus for these multiplets consists of various multiplication rules
and decompositions, as well as invariant density formulas. With these
results one can construct a rather general class of invariant actions.

In five space-time dimensions the linear multiplets play an important
role. At the linearized level in flat space, linear and vector
supermultiplets are related. For instance, starting with the field
$Y^{ij}$ belonging to a vector supermultiplet, one can generate a
linear multiplet upon the following identification,
\begin{eqnarray}
  \label{eq:vector-to-linear}
  L^{ij} &\to& 2\,Y^{ij}\,, \nonumber\\
  \varphi^i &\to& \mathrm{i} \slash{\partial}\Omega^i\,, \nonumber\\
  E_\mu &\to&  \partial^\nu F_{\nu\mu} \,, \nonumber\\
  N &\to&  \Box \sigma \,, 
\end{eqnarray}
as the reader can easily verify by explicit calculation. At this point
one can generate a new vector multiplet, by starting with the field
$N$ and identifying it with a new field $\sigma$, etcetera, at the
price of including higher and higher powers of derivatives. It is easy
to see that the linear multiplet precisely corresponds to the field
equations of the vector multiplet. Conversely, the vector multiplet
will arise as the field equations of the linearized tensor multiplet
action in flat space. This relationship is clearly embodied in the
density formula that we will define at the end of this section (c.f.
\eqref{eq:vector-tensor-density}) and this feature has been exploited
extensively in four space-time dimensions, both for $N=1$ and in $N=2$
supersymmetry, but in five dimensions the restrictions are much
stronger.

In the superconformal setting the relationship between vector and
linear supermultiplets must, however, be modified in view of the
additional restrictions posed by superconformal symmetries. For
instance, the fields $L^{ij}$ and $Y^{ij}$ behave differently under
scale transformations, and, moreover, $L^{ij}$ is invariant under
S-supersymmetry, whereas $Y^{ij}$ is not. Nevertheless the
relationship can still be established provided one gives up linearity.
In the absence of the superconformal background fields in flat
space-time, the first component of the correspondence
\eqref{eq:vector-to-linear} is then replaced by (for a single
multiplet),
\begin{equation}
  \label{eq:nonlinear-L-to-Y}
  L^{ij} \to 2\,\sigma\,Y^{ij}
  +\tfrac12\mathrm{i}\,\varepsilon^{k(i}\bar\Omega_k \Omega^{j)} \,,
\end{equation}
To establish the existence of this composite linear multiplet one
verifies that the lowest component has the correct Weyl weight and is
S-supersymmetric, and furthermore, that its supersymmetry variation is
expressed in terms of a simple doublet spinor which can then act as
the representative of the linear multiplet spinor $\varphi^i$. If
these criteria are not met, then one will not be dealing with a linear
multiplet consisting of $8+8$ degrees of freedom, but with a much
larger multiplet. When dealing with several vector multiplets, labeled
by indices $I,J,\dots= 1,2,\ldots,n$, the expression
\eqref{eq:nonlinear-L-to-Y} generalizes only slightly. It remains
quadratic on the vector multiplets and depends on it in a symmetric
fashion. Hence we start with
\begin{equation}
  \label{eq:full-L}
  L^{ij (IJ)} = 2\, \sigma^{(I}\, Y^{ijJ)} +\tfrac12\mathrm{i}
  \,\varepsilon^{k(i} \,\bar\Omega_k{}^{(I} \Omega^{j)J)} \,, 
\end{equation}
For clarity of the notation, we will henceforth suppress the explicit
indices $(I,J)$ on the right-hand side. In the presence of several
vector multiplets, $\sigma^2$ generalizes to $\sigma^{(I}\sigma^{J)}$,
$\sigma\Omega^i$ to $\sigma^{(I}\Omega^{i J)}$, etcetera.

The other components of the corresponding linear multiplet follow by
applying successive supersymmetry variations and one finds the
following expressions, all manifestly quadratic in the vector
multiplet components, 
\begin{eqnarray}
  \label{eq:3-linear-vector-components}
  \varphi^{i(IJ)} &=& \mathrm{i} \,\sigma\Slash{D}\Omega^i + \tfrac12
  \mathrm{i}\, \Slash{D}\sigma\,\Omega^i - 8\,\sigma^2
  \,\chi^i + Y^{ij}\varepsilon_{jk} \Omega^k 
  - \tfrac14(\hat F_{ab}- 6\,\sigma\,T_{ab})\gamma^{ab} \Omega^i
  \,, \nonumber \\[.2ex]
  E^{a(IJ)} &=& \tfrac18\mathrm{i} \varepsilon^{abcde} \hat F_{bc}
  \hat F_{de} +D_b(\sigma\, \hat F^{ba} -6\,\sigma^2\, T^{ba})+ \cdots
  \,,   \nonumber \\[.2ex] 
  N^{(IJ)} &=& \tfrac12 D^aD_a \sigma^2 -\tfrac12 (D_a\sigma)^2
  + \vert Y^{ij}\vert^2  \nonumber \\
  &&{}
  -\tfrac14 \hat F_{ab}\hat F^{ab} 
  +6\,\sigma\,\hat F_{ab} T^{ab} -\sigma^2\big(4\,D + \tfrac{39}2
  T^2\big)+ \cdots \,, 
\end{eqnarray}
where supercovariant terms of higher-order in the fermion fields have
been suppressed. It is also possible to derive the expression for the
three-rank tensor gauge field associated with this multiplet, by
requiring \eqref{eq:susy-E}, 
\begin{eqnarray}
  \label{eq:tensor-from.vector2}
  E_{\mu\nu\rho}^{(IJ)} &=&{} \tfrac12\mathrm{i}\,e\,
   \varepsilon_{\mu\nu\rho\sigma\lambda} 
   \,(\sigma\,\hat F^{\sigma\lambda} -6 \,\sigma^2\,
   T^{\sigma\lambda}) + \tfrac32 W_{[\mu} F_{\nu\rho]}
   +\ft14\bar\Omega_i\gamma_{\mu\nu\rho}\Omega^i \nonumber \\
   &&{}
   -\tfrac32\mathrm{i}\sigma\,
   \bar\Omega_i\gamma_{[\mu\nu}\psi_{\rho]}{}^i 
   -\tfrac34 \sigma^2\,\bar\psi_{[\mu i} \gamma_\nu\psi_{\rho]}{}^i \,. 
\end{eqnarray}
The above construction can be generalized to non-abelian vector multiplets
as well. In principle, a linear multiplet can also be constructed from
hypermultiplets, but the resulting linear multiplet will not be fully
realized off-shell.  

The same strategy can also be used for constructing a linear multiplet
from the square of the Weyl multiplet. In view of the fact that the
transformations for the Weyl multiplet fields are not linear, this
construction is considerably more complicated than the one above. The
starting point, as before, is to define a composite field
$L^{\mathrm{W}ij}$ in terms of the Weyl multiplet fields, which
satisfies all the requirements for the lowest-dimensional component of
a superconformal linear multiplet. This linear multiplet has
originally been determined in \cite{Hanaki:2006pj} (with different
conventions). In the conventions of this paper we found the following
result,
\begin{equation}
  \label{eq:tensor-from.weyl2}
  L^{ij \mathrm{W}}= - \,\varepsilon^{k(i}\left[\tfrac1{32} \mathrm{i}  \bar
  R_{abk}(Q)\, R^{j)\,ab}(Q) + \ft{32}{3}\mathrm{i} \bar\chi_k\chi^{j)} -
  \tfrac14 T^{ab}\,R_{ab k}{}^{j)} (V) \right]\,.
\end{equation}
which, indeed, is S-invariant and transforms under
Q-supersymmetry into a spinor doublet. Furthermore it scales with Weyl
weight 3, as is appropriate for a linear multiplet. By applying
successive supersymmetry transformations, we identify the other
components of this linear multiplet, 
\begin{eqnarray}
  \label{eq:varphi-E-N-Weyl}
  \varphi^{i\mathrm{W}}&=&{}  
  \tfrac1{64} R_{ab}{}^{cd}(M) \,\gamma_{cd} R^{abi}(Q) 
  +\tfrac1{32} R_{abj}{}^i(V) R^{abj}(Q) - \tfrac34 T^{ab}
  R_{ab}{}^i(S) \nonumber\\  
  &&{} 
  -\tfrac16 R_{abj}{}^i(V) \gamma^{ab} \chi^j 
  -\tfrac{3}{8}\mathrm{i}\, D_a T_{bc}\,\gamma^c R^{abi}(Q) 
  + \tfrac3{16} T^{ab}T_{cd}\,\gamma^{cd} R_{ab}{}^i(Q) 
  \nonumber\\ 
  &&{}
    +4\mathrm{i}\,T^{ab}\gamma_a D_b\chi^i 
    -\tfrac12\mathrm{i} \big(\gamma^{ab}\Slash{D}T_{ab}
  +3\,\Slash{D}\gamma^{ab}T_{ab}\big)\chi^i +
  \tfrac83\big(2\,D+3\,T^2\big)\chi^i \,,
  \nonumber\\[.5ex] 
  \hat E^{a\mathrm{W}}&=& -\tfrac1{128}\mathrm{i} \,\varepsilon^{abcde}\left[
  R_{bc}{}^{fg}(M)\,R_{defg}(M)+ \tfrac13
  R_{bc j}{}^i(V) \,R_{de i}{}^j(V)\right]  \nonumber\\
  &&{} 
  + \tfrac3{2}\mathrm{i} \,\varepsilon^{abcde}\,
  D_b\big[T_{cf}D^fT_{de} +\tfrac32T_{cf} D_dT_e{}^f\big] 
  \nonumber\\ 
  &&{}
  - D_b\Big[\tfrac38  R(M)_{cd}{}^{ab}\,T^{cd} +
  2\,T^{ab}\,D +\tfrac{3}{4}T^{ab} \,T^2 - 9\, 
  T^{ac}T_{cd}T^{db} \Big]
%\nonumber\\  &&{} 
  + \cdots\,, \nonumber\\[.5ex]
  N^\mathrm{W}&=&{} 
     \tfrac1{64} R_{ab}{}^{cd}(M)\,R_{cd}{}^{ab}(M) +\tfrac1{96}
  R_{ab j}{}^i(V) \,R^{ab}{}_i{}^j(V) + 
  \tfrac{15}{8}T^{ab}T_{cd}\,R_{ab}{}^{cd}(M)   \nonumber\\
  &&{} 
  + 3\,T^{ab}D^cD_aT_{bc} -\tfrac32 \big(D_aT_{bc}\big)^2 + \tfrac32 
   D_cT_{ab}\,D^aT^{cb} 
   \nonumber\\
   &&{} 
   -\tfrac{9}{4}\mathrm{i} \varepsilon_{abcde} T^{ab} T^{cd}D_fT^{fe} 
  +\tfrac83 D^2 + 8\, T^2\,D - \tfrac{33}8 (T^2)^2 + \tfrac{81}2
  (T^{ac}T_{bc})^2 
   \nonumber\\
  &&{}
  + \cdots\,, 
\end{eqnarray}
where the dots refer to fermionic terms, which we will not need for
what follows.\footnote{ %%%%%%%%%%%%%%%%%%%%%%%%%%%%%%%%%%%%%%%%%
  Note that \eqref{eq:3-linear-vector-components} and
  \eqref{eq:varphi-E-N-Weyl} contain second-order superconformally
  covariant derivatives. For convenience we exhibit the bosonic
  structure of three such expressions,
  \begin{eqnarray}
  \label{eq:DD(sigma+T)}
  D_\mu D_a \sigma &=& \mathcal{D}_\mu \mathcal{D}_a \sigma + 2\,
  f_{\mu a}\, \sigma\,, \nonumber\\
  D_\mu D_a A_i{}^\alpha &=& \mathcal{D}_\mu \mathcal{D}_a A_i{}^\alpha + 3\,
  f_{\mu a}\, A_i{}^\alpha \,, \nonumber\\  
  D_\mu D_a T_{cd} &=& \mathcal{D}_\mu \mathcal{D}_a T_{cd} -
  4\,f_{\mu[c} T_{d]a} + 4\, f_\mu{}^e\,\eta_{a[c} T_{d]e} + 2\,
  f_{\mu a} T_{cd} \,. 
  \end{eqnarray}
} %%%%%%%%%%%%%%%%%%%%%%%%%%%%%%%%%%%%%%%%%%%%%%%%%%%%%%%%%%%%%%%

In order to represent a linear multiplet, the vector
$\hat{E}^{a\mathrm{W}}$ should satisfy the constraint $D_a\hat
E^{a\mathrm{W}}=0$, as a consequence of which this vector can be
expressed in terms of a three-rank tensor field
$E_{\mu\nu\rho}^\mathrm{W}$. In principle, we can determine the full
expression of this composite tensor by verifying its supersymmetry
transformation \eqref{eq:susy-E}. This is how we originally obtained
\eqref{eq:tensor-from.vector2}. For the Weyl multiplet, however, this
calculation is considerably more involved, so that we restrict
ourselves to the expression for the purely bosonic terms. The result
reads as follows,
\begin{eqnarray}
  \label{eq:tensor-potential-from-weyl2}
  E_{\mu\nu\rho}^\mathrm{W}  &=& - \tfrac3{16} \omega_{[\mu}{}^{ab} 
  \left(\partial_\nu\omega_{\rho]\,ab} - \tfrac23
  \omega_{\nu\,ac}\,\omega_{\rho]}{}^c{}_b \right) - \tfrac1{16}
  V_{[\mu i}{}^j \left(\partial_\nu V_{\rho j}{}^i - \tfrac13 V_{\nu j}{}^k\,
  V_{\rho] k}{}^i \right)  \nonumber\\
  &&{} -  9\, \left(T_{\sigma[\mu}\,\mathcal{D}^\sigma T_{\nu\rho]} +
  \tfrac32 T_{\sigma[\mu}\,\mathcal{D}_\nu T_{\rho]}{}^\sigma
  \right) 
  \nonumber\\
  &&{} + \mathrm{i} \,e\,\varepsilon_{\mu\nu\rho\sigma\lambda}\left(
  \tfrac3{16} R(M)_{\kappa\tau}{}^{\sigma\lambda} T^{\kappa\tau}  +
  T^{\sigma\lambda} \,D + 
  \tfrac3{8} T^{\sigma\lambda} T^2 -
  \tfrac9{2}T^{\sigma\kappa}T_{\kappa\tau}T^{\tau\lambda} \right)  
    \nonumber\\
  &&{}
  + \cdots\,, 
\end{eqnarray}
where the dots represent the fermionic contributions. It is not
difficult to verify that this expression is invariant under scale
transformations and conformal boosts, up to tensor gauge
transformations and up to terms proportional to fermions (we recall
that the spin connection depends both on $b_\mu$ and $\psi_\mu{}^i$),
and that the tensor field strength corresponding to it reproduces the
bosonic terms in $\hat E^{a \mathrm{W}}$ shown in
\eqref{eq:varphi-E-N-Weyl}.

Finally, we note the existence of a superconformally invariant density
for a product of a vector with a tensor supermultiplet. The
corresponding expression takes the following form,
\begin{eqnarray} 
  \label{eq:vector-tensor-density}
  e^{-1}\mathcal{L}_\mathrm{vt} 
  &=&\big(Y_{ij} -\tfrac12
  \bar\Omega_i\gamma^\mu\psi_{\mu}{}^k\,\varepsilon_{kj}\big) L^{ij}
  +\sigma \big(N -\tfrac{1}{2}\bar\varphi_i\gamma^\mu\psi_\mu{}^i\big)
  + \mathrm{i}\,\bar\Omega_i\varphi^i 
  \nonumber\\[.2ex] 
  &&{}
  +\tfrac16 \mathrm{i}e^{-1} \varepsilon^{\mu\nu\rho\sigma\lambda} \,W_\mu
  \, \partial_\nu E_{\rho\sigma\lambda} 
   +\tfrac{1}{4}\mathrm{i}\,\sigma\,L^{ij}\,
  \bar\psi_{\mu i}\gamma^{\mu\nu}\psi_\nu{}^{k}\varepsilon_{kj}  \,. 
\end{eqnarray} 
By using the composite linear multiplets defined previously, this
density formula thus enables the construction of superconformally
invariant actions. This represents a standard way of constructing
actions that is also well-known in the context of four space-time
dimensions. We will make use of this result in section
\ref{sec:lagr-electr-charg}.

%%%%%%%%%%%%%%%%%%%%%%%%%%%%%%%%%%%%%%%%%%%%%%%%%%%%%%%%%%%%%%%%%%%%%%
\section{BPS attractor equations}
\label{sec:bps-attor-eqs}
\setcounter{equation}{0}
%%%%%%%%%%%%%%%%%%%%%%%%%%%%%%%%%%%%%%%%%%%%%%%%%%%%%%%%%%%%%%%%%%%%%%
In this section we derive the conditions for full supersymmetry of the
field configuration. Here we follow the systematic approach outlined
for four space-time dimensions in \cite{LopesCardoso:2000qm}.  In this
section the analysis is done entirely at the off-shell level and we
obtain the full space-time geometry. Our analysis differs from the one
of \cite{Castro:2008ne}, where on-shell information was already
introduced at an earlier stage of the calculation. Only in the next
section \ref{sec:lagr-electr-charg} we will make use of the
supersymmetric action. Although our analysis is different in spirit
and covers a much larger class of supergravity theories, the results
turn out to overlap substantially with those of
\cite{Gauntlett:2002nw}.

%%%%%%%%%%%%%%%%%%%%%%%%%%%%%%%%%%%%%%%%%%%%%%%%%%%%%%%%%%%%%%%%%%
\subsection{Supersymmetry}
\label{sec:supersymmetry}
%%%%%%%%%%%%%%%%%%%%%%%%%%%%%%%%%%%%%%%%%%%%%%%%%%%%%%%%%%%%%%%%%%
To analyze supersymmetry one chooses a purely bosonic field
configuration and requires that the supersymmetry variation of all 
fermion fields vanish up to a uniform S-supersymmetry
transformation. In this context it is convenient to define two
`compensating' spinor fields, $\zeta^i_\mathrm{V}$ and
$\zeta^i_\mathrm{H}$, belonging to the vector multiplet sector and the
hypermultiplet sector, respectively, which transform linearly under
S-supersymmetry, 
\begin{equation}
  \label{eq:zeta-v+h}
  \zeta^i_\mathrm{V}= \frac{1}{C(\sigma)} \, C_{IJK} \sigma^I\sigma^J
  \,\Omega^{iK} \,, \qquad \zeta^i_\mathrm{H} = - \frac{2}{3\,\chi}
  \varepsilon^{ij}\,
  \Omega_{\alpha\beta} A_j{}^\alpha \zeta^\beta   \,. 
\end{equation}
Here we have introduced a symmetric three-rank tensor $C_{IJK}$ and a
corresponding function $C(\sigma)=C_{IJK}\sigma^I\sigma^J\sigma^K$.
The tensor $C_{IJK}$ must be non-vanishing, but other than that there
are no immediate restrictions.

It is straightforward to write down the supersymmetry variations of
these two spinor fields (which both carry scaling weights equal to
$\tfrac12$), 
\begin{eqnarray}
  \label{eq:delta-zeta-v+h}
    \delta\zeta^i_\mathrm{V}&=&  \big(T_{ab}  -\tfrac1{12}F_{ab}{}^I 
    \partial_I\ln C(\sigma) \big) \gamma^{ab}\epsilon^i
    -\tfrac1{6} \mathrm{i} \Slash{\mathcal{D}}\ln
    C(\sigma)\,\epsilon^i - \tfrac13  
    \varepsilon_{jk} Y^{ijI}\partial_I\ln C(\sigma) \, \epsilon^k 
    +\eta^i\,,  \nonumber \\ 
    \delta\zeta^i_\mathrm{H}&=& -\tfrac16\mathrm{i}
    \Slash{\mathcal{D}} \ln\chi\, \epsilon^i +
    \tfrac13\mathrm{i}\slash{k}_j{}^i\,\epsilon^j  +\eta^i\,,
\end{eqnarray}
where here and henceforth we suppress terms proportional to the
fermion fields. Furthermore we made use of the identity
\cite{deWit:1999fp},
\begin{equation}
  \label{eq:ADA}
  \chi^{-1}\Omega_{\alpha\beta}\,A_i{}^\alpha\mathcal{D}_\mu
  A_j{}^\beta = 
  \tfrac12 \varepsilon_{ij} \,\mathcal{D}_\mu\ln\chi + 
  k_{\mu i}{}^k\varepsilon_{kj}\,, 
\end{equation}
where $k_{\mu j}{}^i$ is proportional to the $\mathrm{SU}(2)$ Killing
vectors of the underlying hyperk\"ahler cone.

We now require that the S-supersymmetric linear combinations,
$\zeta^i_\mathrm{V}-\zeta^i_\mathrm{H}$, $\zeta^\alpha - \tfrac32
A_i{}^\alpha\,\zeta^i_\mathrm{H}$, $\Omega^{iI} -
\sigma^I\zeta^i_\mathrm{V}$, $\varphi^i -
3\,\varepsilon_{jk}L^{ij}\zeta_\mathrm{V}^k$, and $\chi^i - \tfrac3{16}
T_{ab} \gamma^{ab}\zeta^i_\mathrm{V}$, do not transform under
Q-supersymmetry. This leads to the following conditions,
\begin{equation}
  \label{eq:bps-1}
  \begin{array}{rcl} 
  \mathcal{D}_\mu(\chi^{-1/2}A_i{}^\alpha) &=& 0\,,\\
  \partial_\mu(C^{-1/3}(\sigma) \,\sigma^I) &=& 0\,,  \\
  \mathcal{D}_\mu \big(C^{-1}(\sigma) \,L^{ij} \big) &=& 0\,, \\
  C(\sigma)\,\chi^{-1} &=& \mathrm{constant}\,, \\
    F_{ab}{}^I &=& 4\,\sigma^I T_{ab}\,,  \\
  \mathcal{D}_{[a}\big(C^{1/3}(\sigma)\,T_{bc]}\big) &=&0\,,\\
  \mathcal{D}_b \big(C^{2/3}(\sigma) T^{ba}\big) &=& \mathrm{i}
  \varepsilon^{abcde} T_{bc}T_{de} \,C^{2/3}(\sigma) \,, 
  \end{array}
  \qquad
  \begin{array}{rcl}
    k_{\mu j}{}^i&=&0\,,\\
    R_{\mu\nu i}{}^j(V)&=&0\,, \\
    Y^{ijI}&=& 0 \,,\\
    N&=&0\,,\\
    \hat E^a &=& 0\,,\\
    D&=&0\,, \\
    ~&~& 
  \end{array}
\end{equation}
which were also given in \cite{Castro:2008ne} in the conventions of
\cite{Fujita:2001kv,Hanaki:2006pj}. However, there are further
constraints in view of the fact that {\it all} fermionic quantities
must vanish under supersymmetry. Experience from the corresponding
analysis in four space-time dimensions \cite{LopesCardoso:2000qm}
indicates that one must also consider the variations of
$R_{ab}{}^i(Q)- (T_{cd}\gamma_{cd}\gamma_{ab}-
4\,T_{ab})\zeta_\mathrm{V}^i$, and of $D_\mu\zeta_\mathrm{H}$.
Combining the result of the first variation with the previous results,
one finds,
\begin{eqnarray}
  \label{eq:bps-2}
  \mathcal{D}_cT_{ab} &=& \tfrac12\mathrm{i}\,\eta_{c[a}\varepsilon_{b]defg}\,
  T^{de}T^{fg} \nonumber\\
  &&{}
  - \tfrac13\left[2\,\mathcal{D}_{[a} \ln C(\sigma) \, T_{b]c} -
    \mathcal{D}_c \ln C(\sigma) \, T_{ab} 
    - 2\,\mathcal{D}^d \ln C(\sigma) \, T_{d[a}\,\eta_{b]c} \right]
  \,,\nonumber \\
  R_{ab}{}^{cd}(M) &=& -2\left[T^2\, \delta_{ab}{}^{cd} +
  4\,T_{ab}T^{cd} +4\, T_{[a}{}^c T_{b]}{}^d - 8\, T_{e[a}
  T^{e[c}\delta_{b]}{}^{d]} \right] \,.
\end{eqnarray}
In addition one considers the variation of the S-invariant
combination, $D_\mu\zeta_\mathrm{H}^i -
\tfrac16[\delta_j^i\Slash{D}\ln \chi\gamma_\mu
-2\slash{\hat{k}}_j{}^i\gamma_\mu -6\mathrm{i} \,T_{\mu a}\gamma^a]
\zeta_\mathrm{H}^j$, subject to the conditions \eqref{eq:bps-1}. This
confirms the consistency of the previous results and, in addition,
gives rise to one more condition,
\begin{equation}
  \label{eq:bps-3}
  f_\mu{}^a = -\tfrac16\mathcal{D}_\mu \mathcal{D}^a \ln \chi
  +\tfrac1{18} \mathcal{D}_\mu\ln\chi \mathcal{D}^a\ln\chi  
  - 4\, T_{\mu b}T^{ab} +
  \tfrac14\left[3\,T_{bc}T^{bc} - \tfrac19 (\mathcal{D}_b\ln\chi)^2
  \right] \,e_\mu{}^a \,. 
\end{equation}
Using the arguments presented in \cite{LopesCardoso:2000qm}, we
conclude that the above equations \eqref{eq:bps-1}, \eqref{eq:bps-2}
and \eqref{eq:bps-3} comprise all the conditions for a supersymmetric
field configuration consisting of the Weyl multiplet, vector
multiplets, linear multiplets and hypermultiplets, without imposing
equations of motion. Because the fermionic equations of motion must be
satisfied, simply because of supersymmetry, most of the bosonic
equations of motion must be satisfied as well. There are, however,
exceptions, such as the equation of motion associated with the scalar
field $D$ belonging to the Weyl multiplet, which does not appear as
the supersymmetry variation of a fermionic expression.

Combining the second equation of \eqref{eq:bps-2} with
\eqref{eq:bps-3}, we derive the following equation,
\begin{eqnarray}
  \label{eq:R-omega}
  \mathcal{R}_{ab}{}^{cd}(\omega,e) &=& 2\,e_a{}^\mu e_b{}^\nu
  \left(\partial_{[\mu}\omega_{\nu]}{}^{cd} 
  - \omega_{[\mu}{}^{ce} \omega_{\nu]e}{}^{d}\right)  \nonumber\\
  &=&{}
  -8\big(T_{ab}T^{cd} + T_a{}^{[c} T_b{}^{d]}\big) +
  \delta_{[a}^c\delta_{b]}^d \big(4 \,T_{ef}T^{ef} - \tfrac29
  (\mathcal{D}_e\ln \chi)^2 \big)   \nonumber\\
  &&{}
  -\delta_{[a}{}^{[c} \left(16 \,
  T_{b]e}   T^{d]e} +\tfrac43 \mathcal{D}_{b]} \mathcal{D}^{d]}
  \ln \chi   -\tfrac49 \mathcal{D}_{b]}\ln \chi \mathcal{D}^{d]} \ln
  \chi  \right) \,. 
\end{eqnarray}

%%%%%%%%%%%%%%%%%%%%%%%%%%%%%%%%%%%%%%%%%%%%%%%%%%%%%%%%%%%%%%%%%%%%
\subsection{Space-time geometry}
\label{sec:space-time-geometry}
%%%%%%%%%%%%%%%%%%%%%%%%%%%%%%%%%%%%%%%%%%%%%%%%%%%%%%%%%%%%%%%%%%%%
Before discussing the resulting space-time geometry we have to impose
a number of gauge choices. We set the dilatational gauge field $b_\mu=
0$ (in fact, K-invariance implies that the equations found above are
already independent of $b_\mu$) and furthermore we set the function
$C(\sigma)$ equal to a constant $C$. This implies that also $\chi$
becomes a constant. The ratio of the two constants $C$ and $\chi$ will
eventually be defined by the equation of motion for the field $D$, but
at the moment we proceed without making reference to any particular
Lagrangian. Note that the various fields will still be subject to
constant scale transformations which are a remnant of the full
space-time dependent dilatations. Physical results should, of course,
be insensitive to these scale transformations. In addition we set the
$\mathrm{SU}(2)$ gauge connections to zero, in view of the fact that
their field strength is vanishing (c.f.  \eqref{eq:bps-1}). In this
situation the various scalar fields $\sigma^I$ and $L^{ij}$ are all
constant.

The resulting geometry is now of a special type, as the tensor
$T_{\mu\nu}$ is an example of a conformal Killing-Yano tensor
\cite{Krtous:2006qy}. Locally, in five space-time dimensions, this
tensor generically induces a family of pairs of two-surfaces which together
with the fifth orthogonal dimension foliate the space-time. It also
leads to a Killing vector associated with this fifth dimension and a
symmetric Killing tensor,
\begin{equation}
  \label{eq:Killing-tensor-vector}
  \xi^\mu =
  \mathrm{i} e^{-1}\,\varepsilon^{\mu\nu\rho\sigma\tau} \,
  T_{\nu\rho}\, T_{\sigma\tau}\,, \qquad K_{\mu\nu}= T_{\mu\rho}\,
  T_\nu{}^\rho\,,
\end{equation}
where $e=\det(e_\mu{}^a)$.
Using the properties of the tensor $T_{\mu\nu}$ (in the gauge
indicated above), we obtain the following results for the Riemann tensor
and for the derivative of $T_{\mu\nu}$,
\begin{eqnarray}
  \label{eq:R-tensor-DT}
  \mathcal{R}_{ab}{}^{cd} &=&
  -8\big(T_{ab}T^{cd} + T_a{}^{[c} T_b{}^{d]}\big)  -16\,\delta_{[a}{}^{[c}
  \, T_{b]e}   T^{d]e} + 4\,
  \delta_{[a}^c\delta_{b]}^d \,T_{ef}T^{ef}\,,   \nonumber\\
  \mathcal{D}_\rho T_{\mu\nu} &=& \tfrac12 g_{\rho[\mu}\,\xi_{\nu]}\,. 
\end{eqnarray}
Furthermore we note the results, 
\begin{eqnarray}
  \label{eq:implications}
\begin{array}{rcl}
    \mathcal{D}_\mu \xi_\nu &=& -\mathrm{i}
  e \,\varepsilon_{\mu\nu\rho\sigma\tau} \xi^\rho
  T^{\sigma\tau}\,,  \\
  \mathcal{D}_\rho K_{\mu\nu} &=& -\tfrac12 \xi_{(\mu}T_{\nu)\rho} \,, 
\end{array}
\qquad
\begin{array}{rcl}
  \xi^\mu T_{\mu\nu}&=& 0\,, \\
   T^2 &\equiv& (T_{ab})^2 ~=~ \mathrm{constant} \,. \\
\end{array}
\end{eqnarray}
From these equations it is clear that $\xi^\mu$ is indeed a Killing
vector. Furthermore one may easily verify that the Riemann tensor
satisfies the Bianchi identity. 

If $\xi^\mu$ vanishes then the tensors $T_{\mu\nu}$ and $K_{\mu\nu}$
are covariantly constant and so is the Riemann tensor so that we are
dealing with a locally symmetric space. In this particular case the
space is in fact the product of a two- and a three-dimensional
maximally symmetric space, as the Riemann tensor decomposes into two
Riemann tensors corresponding to these subspaces satisfying $R_{\hat
  a\hat b}{}^{\hat c\hat d}\propto c\, T^2\, \delta_{[\hat a}{}^{\hat c}
\delta_{\hat b]}{}^{\hat d}$, with proportionality $c=-16$ and
$c=4$ for the two- and the three-dimensional subspace, respectively.
Here the indices $\hat a,\hat b,\hat c,\hat d$ refer to the tangent-space
projected onto the two- or three-dimensional subspaces. 

Rather than considering this case any further, we concentrate on the
more general case where $\xi^\mu\not=0$ and return to the limit of
vanishing $\xi^\mu$ at the end. Obviously the line element must
reflect the isometry associated with the Killing vector $\xi^\mu$.
Choosing a coordinate $\psi$ by
$\xi^\mu\partial_\mu=\partial/\partial\psi$, we decompose the
coordinates into $\psi$ and four-dimensional coordinates $x^m$, where
$m=1,2,3,4$, without committing ourselves to a certain signature
yet.\footnote{ %%%%%%%%%%%%%%%%%%%%%%%%%%%%%%%%%%%%%%%%%%%%%%%%
  At this point we are using Pauli-K\"all\'en metric conventions,
  where the signature is determined by making one of the coordinates
  purely imaginary. This enables us to consider all possible
  signatures at once, so that this analysis encompasses the solutions
  for minimal supergravity found in \cite{Gauntlett:2002nw}.
  Momentarily we will assume that the Killing vector $\xi^\mu$ is
  spacelike.
}  %%%%%%%%%%%%%%%%%%%%%%%%%%%%%%%%%%%%%%%%%%%%%%%%%%%%%%%%%%%%
Correspondingly, the tangent-space indices $a= 1,2,\ldots,5$ are
decomposed into $a=5$ and indices $p, q,\ldots=1,2,3,4$. Upon a
suitable local Lorentz transformation, the f\"unfbein is brought into
the form,
\begin{equation}
  \label{eq:5-beine}
  e_\mu{}^5\,\mathrm{d}x^\mu  = \mathrm{e}^g\big[\mathrm{d}\psi +
  \sigma_m\,\mathrm{d}x^m\big] \,, \qquad e_\mu{}^p \,\mathrm{d}x^\mu =
  \mathrm{e}^{-g/2} \,\hat e_m{}^p\,\mathrm{d}x^m \,.
\end{equation}
In view of the isometry corresponding to shifts of the coordinate
$\psi$ we may assume that $g$, $\sigma_m$ and the vierbein field $\hat
e_m{}^p$ do not depend on $\psi$. The corresponding inverse f\"unfbein
components are given by,
\begin{equation}
  \label{eq:inverse-bein}
  e_5{}^\psi = \mathrm{e}^{-g} \,, \qquad e_5{}^m= 0\,,\qquad 
  e_p{}^\psi = -\sigma_p \,\mathrm{e}^{g/2} \,, \qquad 
  e_p{}^m = \mathrm{e}^{g/2}\,\hat e_p{}^m\,, 
\end{equation}
where, on the right-hand side, four-dimensional tangent-space and
world indices are converted by the vierbein $\hat e_m{}^p$ and its
inverse (so that, e.g. $\sigma_p= \hat e_p{}^m\,\sigma_m$, and the
covariant derivative $\nabla_p$ contains the spin connection
$\hat\omega_m{}^{pq}$, associated with the vierbein $\hat e_m{}^p$).
This leads to the following expressions for the spin connection,
$\omega_{abc} \equiv e_a{}^\mu\,\omega_{\mu cd}$,
\begin{eqnarray}
  \label{eq:spin-conn-decomp}
  \omega_{pqr} &=&\mathrm{e}^{g/2}\,\big[\hat\omega_{pqr} +
  \delta_{p[q}\nabla_{r]} g\big]   \,,\nonumber\\ 
  \omega_{5pq} &=& \omega_{qp5}~=~ \tfrac12 
  \mathrm{e}^{2g}\,\mathcal{Q}_{pq}   \,,\nonumber\\ 
  \omega_{55p} &=& -\mathrm{e}^{g/2}\,\nabla_p g \,,
\end{eqnarray}
where $\mathcal{Q}_{pg}$ equals,  
\begin{equation}
  \label{eq:Q-sigma}
  \mathcal{Q}_{pq} =  \hat e_{p}{}^m \hat
  e_{q}{}^n\,\mathcal{Q}_{mn}\,,\quad
  \mathcal{Q}_{mn}=\partial_m\sigma_n - \partial_n\sigma_m\,.  
\end{equation}

Let us now return to \eqref{eq:R-tensor-DT} and consider the second
equation, which we write in tangent-space indices as,
\begin{equation}
  \label{eq:DT-tangent}
  e_c{}^\mu\, \partial_\mu T_{ab} + 2\,\omega_{c[a}{}^d\,T_{b]d} =
  \tfrac12 \delta_{c[a}\delta_{b]5}\,\xi\,,
\end{equation}
where we made use of the fact that $T_{5a}=0$ and defined $\xi =
\mathrm{i}\varepsilon^{pqrs}T_{pq}T_{rs}$, where
$\varepsilon^{pqrs}=\varepsilon^{5pqrs}$ so that $\xi^\psi=
\mathrm{e}^{-g}\,\xi$. Changing the overall sign of the epsilon tensor
is irrelevant as it only corresponds to a sign change of the
coordinate $\psi$.\footnote{%%%%%%%%%%%%%%%%%%%%%%%%%%%%%%%%%
  In Pauli-K\"all\'en notation we now fix convention such that
  $\varepsilon_{\mu\nu\rho\sigma\tau} x^\mu x^\nu x^\rho x^\sigma
  x^\tau = \mathrm{i} \,5! \, x^{0}x^1 x^2 x^3 x^5$. %%%%%%%%%%
} %%%%%%%%%%%%%%%%%%%%%%%%%%%%%%%%%%%%%%%%%%%%%%%%%%%%%%%%%%
Imposing the equations contained in \eqref{eq:DT-tangent} leads to the
following results,
\begin{equation}
  \label{eq:omega-T-constraints}
  \partial_\psi T_{ab}=0\,,\
  \qquad \mathcal{Q}_{pq}= - 2\mathrm{i}
  \mathrm{e}^{-2g} 
  \,\varepsilon_{pqrs} T^{rs} \,,\qquad \nabla_p T_{qr} = 0\,,\qquad g =
  \mathrm{constant}  \,.
\end{equation}

These results are consistent with what is found when considering the
Riemann tensor from the connections \eqref{eq:spin-conn-decomp} upon
comparison with the first equation
\eqref{eq:R-tensor-DT}.\footnote{%%%%%%%%%%%%%%%%%%%%%%%%%%%%%%%
  We note that \eqref{eq:omega-T-constraints} has been derived from
  \eqref{eq:DT-tangent} assuming $\det[T]\not=0$. For $\det[T]=0$ one
  can arrive at the same result by also making use of 
  \eqref{eq:R-tensor-DT} and \eqref{eq:curvature-decom}.
} %%%%%%%%%%%%%%%%%%%%%%%%%%%%%%%%%%%%%%%%%%%%%%%%%%%%%%%%%%%%%%
Here and henceforth we will be assuming that the four-dimensional
subspace has signature $(-,+,+,+)$, so that the Killing vector
$\xi^\mu$ is spacelike and $\xi$ is real. The various curvature
components read,
\begin{eqnarray}
  \label{eq:curvature-decom}
  \mathcal{R}_{pq5r}&=& - \tfrac12
  \mathrm{e}^{5g/2}\,\left[\nabla_r\mathcal{Q}_{pq}
  +\nabla_rg\,\mathcal{Q}_{pq} + \nabla_{[p} g\,\mathcal{Q}_{q]r} -
  \delta_{r[p}\,\mathcal{Q}_{q]s} \nabla^sg \right] \,,\nonumber\\[.5ex]
  \mathcal{R}_{5p5q}&=&  \mathrm{e}^{g} \left[\nabla_p\nabla_q g-
  \tfrac12 \delta_{pq} \,(\nabla_rg)^2  +
  2\,\nabla_pg\,\nabla_qg\right] - \tfrac14 \mathrm{e}^{4g} \,
  \mathcal{Q}_{pr} 
  \mathcal{Q}_q{}^r \,,\nonumber\\[.5ex]
  \mathcal{R}_{pqrs}&=& \mathrm{e}^g \,\mathcal{R}_{pqrs}(\hat\omega)
  - 2\,\mathrm{e}^g \,\delta_{[p[r}\left[\nabla_{s]}\nabla_{q]}g +
  \tfrac12   \nabla_{s]}g \,\nabla_{q]}g  
  -\tfrac14 \delta_{s]q]}\,(\nabla_ug)^2 \right]  \nonumber\\
  &&{}
  + \tfrac12\,\mathrm{e}^{4g}\left[\mathcal{Q}_{pq} \mathcal{Q}_{rs}
  - \mathcal{Q}_{p[r} \mathcal{Q}_{s]q} \right] \,,
\end{eqnarray}
where the right-hand side is consistently written in four-dimensional
notation. Obviously $\mathcal{R}_{pq5r}$ must vanish in order to be
consistent with the first equation \eqref{eq:R-tensor-DT}, and this is
indeed what is implied by the earlier results
\eqref{eq:omega-T-constraints}. Likewise the expression for
$\mathcal{R}_{5p5q}$ is consistent with the corresponding equation
\eqref{eq:R-tensor-DT}. Hence we are left to analyse the last equation
of \eqref{eq:curvature-decom}, which determines the four-dimensional
Riemann tensor $\mathcal{R}(\hat\omega)$ according to
\begin{equation}
  \label{eq:4d-Riemann}
  \mathcal{R}_{pqrs}(\hat\omega)= - 16\,\mathrm{e}^{-g} \Big[
  4\,\delta_{[p[r} \,T_{s]t}T_{q]}{}^t - \delta_{p[r}\delta_{s]q}\,T^2
  \Big] \,. 
\end{equation}
The Ricci scalar, $\mathcal{R}_{ab}{}^{ab}(\hat\omega)=0$. Further
inspection shows that this Riemann tensor corresponds to a product of
two two-dimensional spaces with equal radii, namely $AdS_2$ and $S^2$.
The geometry thus takes the form of a circle (parametrized by the
coordinate $\psi$) non-trivially fibered over an $AdS_2\times S^2$
base space.  We now adopt four-dimensional coordinates by writing the
respective metrics in the standard form of a Bertotti-Robinson and a
two-sphere metric, with coordinates $t,r$, and $\theta,\varphi$,
respectively, so that the five-dimensional line element takes the
following form ($r$ is non-negative and $\theta$ and $\varphi$ have
periodicity $\pi$ and $2\pi$, respectively),
\begin{eqnarray}
  \label{eq:line-element-sigma}
  ds^2&=&\frac1{16\,v^2} \Big(-r^2 \mathrm{d}t^2
  +\frac{\mathrm{d}r^2}{r^2} 
  +\mathrm{d}\theta^2  +\sin^2\theta \,\mathrm{d}\varphi^2\Big) +
  \mathrm{e}^{2g} \big(\mathrm{d}\psi +\sigma \big)^2 \,,
  \nonumber\\[.5ex] 
  \sigma&=&{}  -\frac1{4\, v^2}\,\mathrm{e}^{-g}\,
  \big(T_{23}\,r\,\mathrm{d}t - T_{01}\,
  \cos\theta\,\mathrm{d}\varphi \big) \;, 
\end{eqnarray}
corresponding to,
\begin{equation}
  \label{eq:Q}
  Q_{tr}= \frac{1}{4\,v^2}\,\mathrm{e}^{-g}\,T_{23}\,, \qquad 
    Q_{\theta\varphi}= - 
    \frac{1}{4\,v^2}\,\mathrm{e}^{-g}\,T_{01}\,\sin\theta\,. 
\end{equation}
Here and henceforth we use the definition, 
\begin{equation}
  \label{eq:v+sigma}
  v = \sqrt{(T_{01})^2 + (T_{23})^2} \;,
\end{equation}
where $T_{01}$ and $T_{23}$ are the nonvanishing components of the
tensor field $T_{ab}$, where the local Lorentz indices are
$(0,1,2,3)$. Note that the vierbein fields can be chosen diagonally;
their values can be read off from \eqref{eq:line-element-sigma},
\begin{equation}
  \label{eq:diag-vierbein}
  e_m{}^p \,\mathrm{d}x^m = \frac{1}{4\,v} \Big(r\,\mathrm{d}t,\,
  \frac{\mathrm{d}r}{r},\, \mathrm{d}\theta  ,\, 
  \sin\theta\,\mathrm{d}\varphi \Big) \,, \qquad (p=0,1,2,3)\,. 
\end{equation}
In this Lorentz frame, the fields $T_{ab}$ are constant. For future
use we also list the nonvanishing spin-connection fields,
\begin{eqnarray}
  \label{eq:omega-mu-full}
  \omega_m{}^{pq} &=& \stackrel{\circ}{\omega}_m{}^{pq} +\tfrac12
  \sigma_m\,\mathrm{e}^{3g}\,\mathcal{Q}^{pq}\,,  \nonumber\\
  \omega_m{}^{p5} &=&  \tfrac12 e_{mq}  \,
  \mathrm{e}^{2g}\,\mathcal{Q}^{pq} \,,  \nonumber\\
  \omega_\psi{}^{pq} &=&{}  \tfrac12
  \mathrm{e}^{3g}\,\mathcal{Q}^{pq}\,,
\end{eqnarray}
where $\stackrel{\circ}{\omega}_t{}^{01}=- r$ and
$\stackrel{\circ}{\omega}_\varphi{}^{23}= \cos\theta$. 

Observe that $\sigma^I$, $T_{ab}$, $v$ and $\mathrm{e}^{-g}$ transform
with weight $+1$ under the (constant) scale transformations inherited
from the five-dimensional dilatations. As a result, the metric
\eqref{eq:line-element-sigma} scales uniformly with weight $-2$ and
the one-form $\sigma$ is inert under scale transformations. Note that
$\sigma$ is determined up to a four-dimensional gauge transformation
associated with shifts of the coordinate $\psi$ with a function
depending on the four-dimensional coordinates. Such diffeomorphisms
leave the form of the line element invariant.

Let us now further discuss the line element
\eqref{eq:line-element-sigma}. Assuming that $T_{01}\not=0$, we can
rewrite the line element in the form,
\begin{eqnarray}
  \label{eq:GH-BMPV}
  ds^2&=&- \frac{\rho^4}{16\,v^2}\left(\frac{T_{01}}{v}\,\mathrm{d}t +
  \frac{T_{23}}{v\,\rho^2}\Big(\cos\theta\,\mathrm{d}\varphi +
  \frac{1}{p^0}\,\mathrm{d}\psi \Big)\right)^2 \nonumber\\
  &&{}
  + \frac1{4\,v^2\rho^2}\left(\mathrm{d}\rho^2 +
  \frac{\rho^2}{4}\Big(\mathrm{d}\theta^2  + \mathrm{d}\varphi^2
  +\frac1{(p^0)^2} \,\mathrm{d}\psi^2  + \frac{2}{p^0} 
  \,\cos\theta\,\mathrm{d}\varphi\,\mathrm{d}\psi\Big) \right) \;, 
\end{eqnarray}
where we used the definitions
\begin{equation}
  \label{eq:rho-p0}
  \rho= \sqrt{r} \,,\qquad 
  p^0= \frac{\mathrm{e}^{-g}}{4\,v^2}\,T_{01}\,. 
\end{equation}
To make $p^0$ unambiguous we fix the periodicity interval for $\psi$
to $4\pi$. The second term of the line element then corresponds to a
flat metric, up to an overall warp factor $(2v\rho)^{-2}$. To see this
we combine the four Cartesian coordinates into two complex ones, which
we parametrize as,
\begin{equation}
  \label{eq:flat-R4}
  z_1= \rho\,\cos{\theta/2}\, \exp \tfrac12\mathrm{i} [\psi/p^0
  +\varphi] \,,\qquad 
    z_2= \rho\,\sin{\theta/2}\, \exp \tfrac12\mathrm{i} [\psi/p^0
  -\varphi] \,.
\end{equation}
Clearly for $\vert p^0\vert=1$ we cover the whole four-dimensional
space $\mathbb{R}^4$. For $\vert p^0\vert \not=1$ we have a conical
singularity at the origin. In all cases the three-dimensional horizon
is located at $r=0$ and its cross-sectional area is equal to
\begin{equation}
  \label{eq:area}
  A_3= \int_{\Sigma_\mathrm{hor}} = \pi^2 v^{-2}\, \mathrm{e}^g\,.  
\end{equation}
Observe that this result is not invariant under the scale
transformations introduced earlier, which simply reflects the fact
that the line element is not invariant either. Furthermore the
bi-normal tensor at the horizon is the same in all cases when given
with tangent space indices. Its only non-vanishing components are,
\begin{equation}
  \label{eq:binormal}
  \varepsilon_{01}= \pm 1\,,
\end{equation}
so that $\varepsilon_{\mu\nu}\varepsilon^{\mu\nu}= -2$. Both
\eqref{eq:area} and \eqref{eq:binormal} can be derived by first
determining the bi-normal tensor and the cross-sectional area in a
coordinate frame that is non-singular at the horizon, and subsequently
converting the results to the singular frame used in the text.

The line element \eqref{eq:GH-BMPV} describes the near-horizon
geometry of the spinning charged black hole \cite{Breckenridge:1996is}
(see also, \cite{Gauntlett:1998fz}), and we observe that the rotation
is associated with a globally defined one-form on $S^3$, in view of
$\mathrm{Im}\,[z_1\,\mathrm{d}z_1{}^\ast+z_2\,\mathrm{d}z_2{}^\ast] =
\rho^2[(p^0)^{-1} \mathrm{d}\psi +\cos\theta\,\mathrm{d}\varphi]$. 
Clearly the angular momentum of the black hole is proportional to
$T_{23}$. When $T_{23}=0$ we are dealing with a static black hole and
the near-horizon geometry is given by,
\begin{equation}
  \label{eq:staticBH}
  ds^2=\frac1{16\,v^2}\left(-r^2 \mathrm{d}t^2
  +\frac{\mathrm{d}r^2}{r^2}\right) + \frac1{4\,v^2} \,
  ds^2(S^3/\mathbb{Z}_{p^0})\,. 
\end{equation}

Finally we turn to the case $T_{01}=0$ where we find,
\begin{equation}
  \label{eq:line-element-beta=0}
  ds^2=\frac1{16\,T_{23}{}^2}\frac{\mathrm{d}r^2}{r^2} 
  +\mathrm{e}^{2g}\,\mathrm{d}\psi^2  - \frac{\mathrm{e}^g}{2\,T_{23}}
  \,r\, \mathrm{d}\psi\, \mathrm{d}t + \frac1{16\,T_{23}{}^2} \,
  \mathrm{d}s^2(S^2) \;,
\end{equation}
where $\mathrm{d}s^2(S^2)$ is the line element belonging to the unit
two-sphere. The first three terms constitute a metric which is locally
$AdS_3$ so that the near-horizon geometry is that of $AdS_3\times
S^2$. This is the near-horizon geometry of a supersymmetric black
ring, or, when we drop the identification $\psi\cong\psi+4\,\pi$, of
an infinitely long black string.

%%%%%%%%%%%%%%%%%%%%%%%%%%%%%%%%%%%%%%%%%%%%%%%%%%%%%
\subsection{Gauge fields}
\label{sec:gauge-fields}
%%%%%%%%%%%%%%%%%%%%%%%%%%%%%%%%%%%%%%%%%%%%%%%%%%%%%
According to \eqref{eq:bps-1}, the field strengths $F_{\mu\nu}{}^I$
are determined in terms of the tensor field $T_{ab}$,
\begin{equation}
  \label{eq:F_I}
  F_{tr}{}^I = \frac{\sigma^I}{4\,v^2} \,T_{01}\,,\qquad 
  F_{\theta\varphi}{}^I = \frac{\sigma^I}{4\,v^2} \,T_{23}\,\sin\theta \,.
\end{equation}
At this point we can define magnetic charges associated with
$Q_{\theta\varphi}$ and $F_{\theta\varphi}{}^I$. Employing the same
conventions for these field strengths (apart from a relative sign
between $p^0$ and $p^I$), we define
\begin{equation}
  \label{eq:magnetic-P}
    p^0= \frac{\mathrm{e}^{-g}}{4\,v^2}\,T_{01}\,, \qquad
  p^I= \frac{\sigma^I}{4\,v^2} \,T_{23}\,,
\end{equation}
with the same expression for $p^0$ as given in \eqref{eq:rho-p0}. In
the five-dimensional context, the $p^I$ will play the role of dipole
magnetic charges. They are proportional to $T_{23}$, so they will
vanish for a static black hole. The definition of the electric
charges, which involves the equations of motion, will be discussed in
later sections. From \eqref{eq:F_I} we can determine the vector
potentials,
\begin{equation}
  \label{eq:A-I}
  W_\mu{}^I(x)\, \mathrm{d}x^\mu = -\,\frac{\sigma^I}{4\,
  v^2}\left(T_{01}\,r\,\mathrm{d}t 
  +T_{23}\,\cos\theta\, \mathrm{d}\varphi\right) + 
  \,\mathrm{d}\Lambda^I(x)  \,,
\end{equation} 
up to an abelian gauge transformation, parametrized by
$\Lambda^I(x)$. 

For the spinning black hole, where $T_{01}\not=0$, the gauge
transformation can be chosen such that the gauge potentials are
globally defined on $S^3$. To see this one makes use of the
observation preceding \eqref{eq:staticBH} in the previous subsection,
which leads to,
\begin{equation}
  \label{eq:A-I-S3}
  W_\mu{}^I \mathrm{d}x^\mu = - \frac{\sigma^I}{4\,
  v^2}\left(T_{01}\,r\,\mathrm{d}t 
  +T_{23}\,\Big(\frac{\mathrm{d}\psi}{p^0} +\cos\theta\,
  \mathrm{d}\varphi\Big)\right) \,.
\end{equation} 

In the case of the black ring, where $T_{01}=0$, the gauge
transformations in \eqref{eq:A-I} introduce an uncontractible
component corresponding to Wilson lines around the circle parametrized
by $\psi$. The proper definition of the Wilson line moduli is subtle
due to the presence of the charges $p^I$ and the $S^1\times S^2$
topology, as we shall discuss below. In any case, due to the presence
of large gauge transformations (i.e. gauge transformations that cannot be
connected continuously to the identity), these moduli $a^I$ should be
periodically identified and furthermore they should be defined such
that they are not subject to small gauge transformations. At any rate
the gauge fields are expected to contain the following terms,
\begin{equation}
  \label{eq:A-I-S1}
  W_\mu{}^I \mathrm{d}x^\mu = - p^I\cos\theta\,
  \mathrm{d}\varphi + a^I \mathrm{d}\psi\,.
\end{equation} 
However, unlike in the case of the spinning black holes, the gauge
fields are not globally defined, as is obvious from the fact that the
monopole fields are sourced by Dirac strings. This phenomenon
implies that the gauge fields should be defined in patches, connected
by suitable gauge transformations.  In the context of five space-time
dimensions the Dirac strings are degenerate and one is actually
dealing with Dirac membranes. Just as in the case of Dirac strings,
the Dirac membranes are subject to constraints, some of them related
to charge quantization (to appreciate this, the reader may consult
\cite{Henneaux:1986tt,Bekaert:2002eq}, where some of this is explained
in the context of $2+1$ dimensions). 

For a single black ring and for multiple concentric black rings, the
appropriate sections have been considered in \cite{Hanaki:2007mb},
guided by the explicit ring solutions \cite{Elvang:2004rt} and
\cite{Gauntlett:2004qy}. Although these results were obtained without
taking into account possible higher-derivative interactions, they 
should still apply to the general case, as the choice of the sections and
the corresponding Dirac membranes is entirely based on the topology of
the underlying charge configuration. With this in mind we replace
\eqref{eq:A-I-S1} by the following sections (for a single ring),
\begin{equation}
  \label{eq:A-I-ringsections}
  W_\mu{}^I \mathrm{d}x^\mu =
  - p^I\left[\cos\theta \, \mathrm{d}\varphi
  \pm \mathrm{d}(\varphi+ \tfrac12\psi) \right] +
  a^I \mathrm{d}\psi \,, 
\end{equation} 
where we note that $\cos\theta$ can be extended globally into the ring
coordinate conventionally denoted by $x$
\cite{Elvang:2004rt,Emparan:2006mm}. For $x=1$ and $x=-1$ one is
dealing with the inner and the outer part, respectively, of the
two-dimensional plane that contains the ring. Hence the plus sign in
\eqref{eq:A-I-ringsections} refers to the section that is singularity
free in the outer part of plane, and the minus sign to the section
that is singularity free in the inner part.

The nontrivial, and somewhat unexpected, feature of
\eqref{eq:A-I-ringsections}, is that the gauge transformation between
the two patches involves a $\mathrm{d}\psi$ component, contrary to
what one would expect based on intuition from four dimensions. Indeed,
in the case of an infinite black string, this gauge transformation is
just $\propto p^I\mathrm{d}\varphi$. However, the ring topology
requires a more extended gauge transformation. 

\begin{figure}[ht]
  \centering
  \includegraphics[scale=0.125]{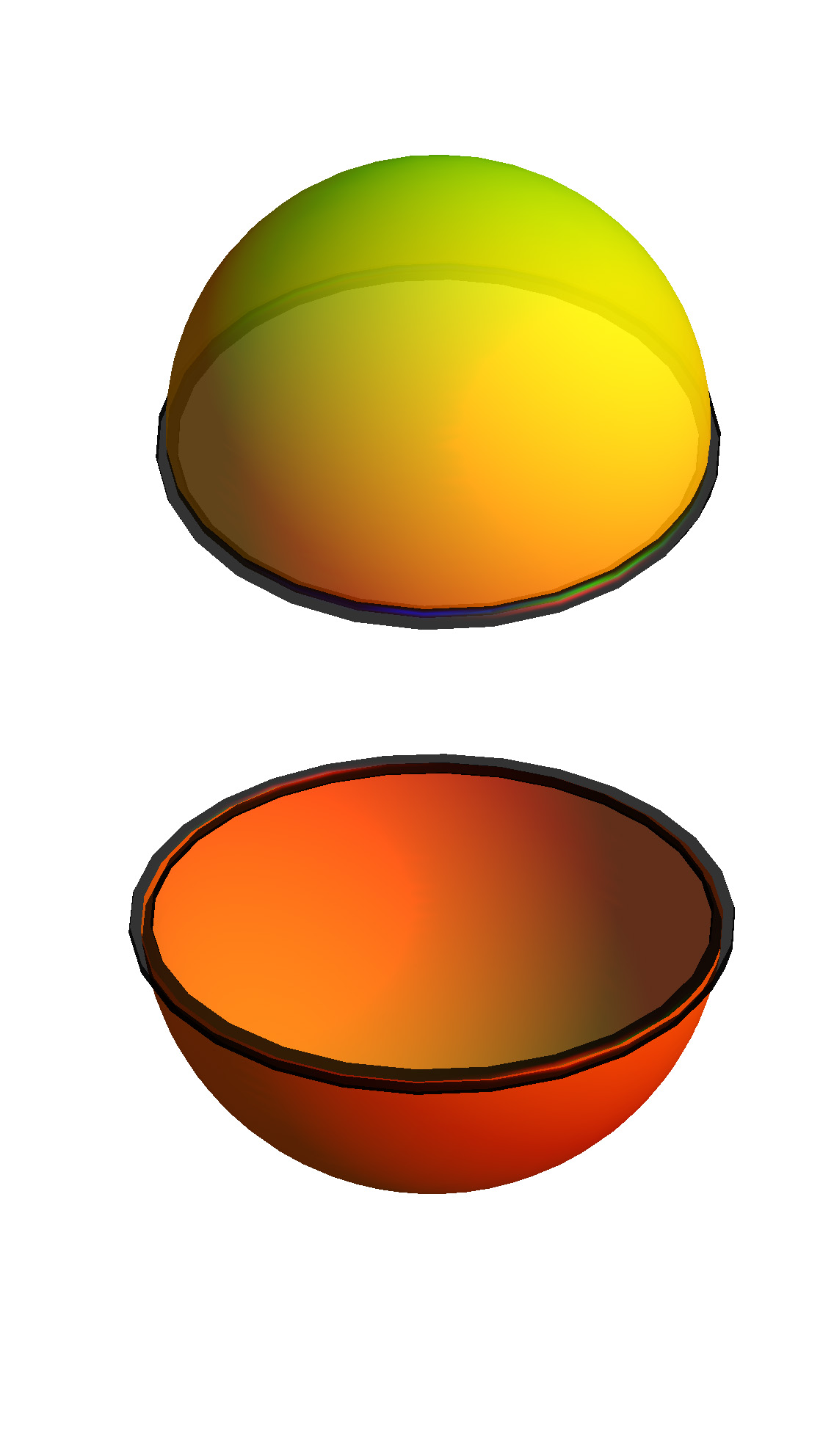}\hspace{3cm} 
 \includegraphics[scale=0.105]{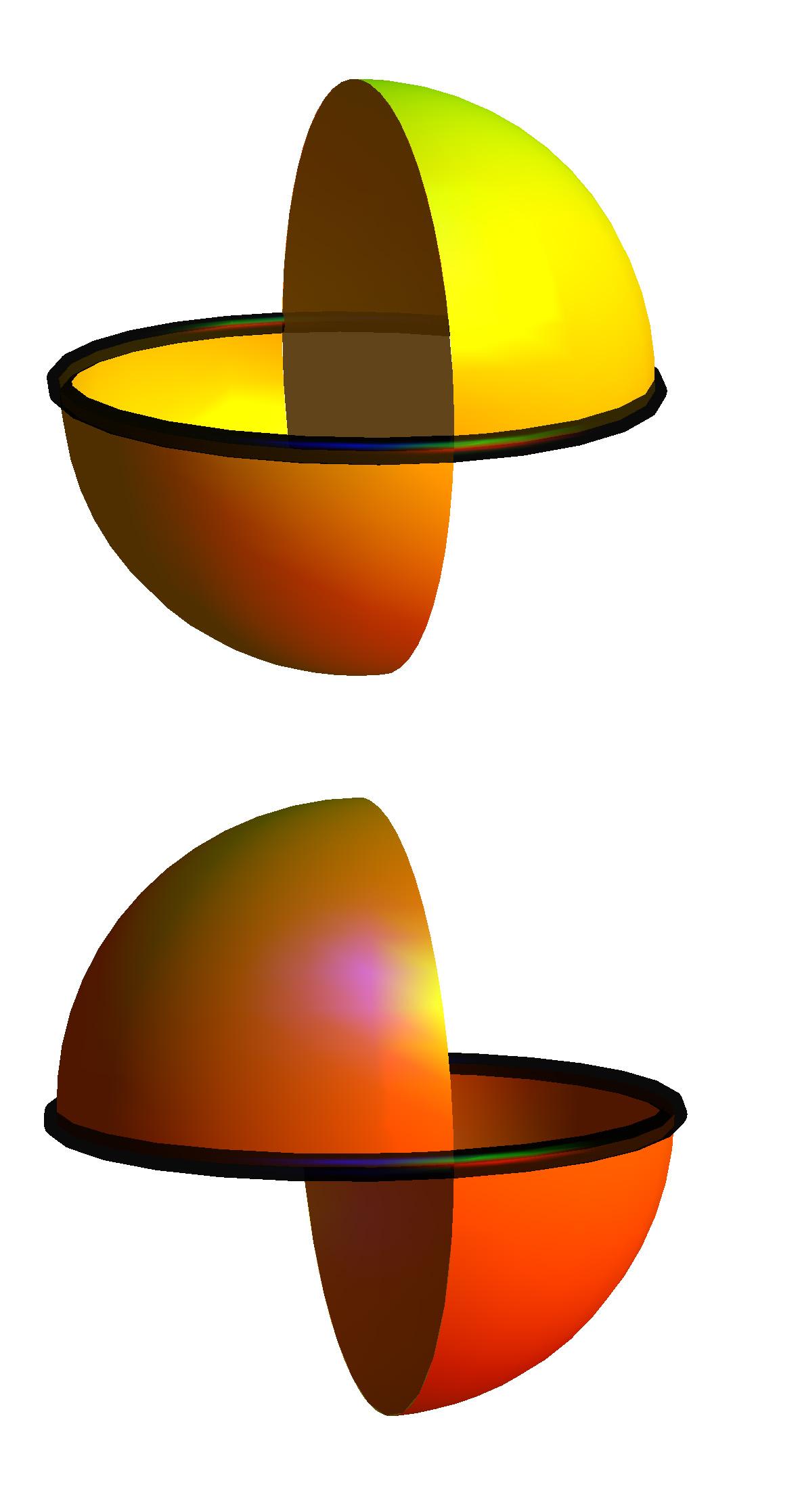}\hspace{1cm} 
  \caption{\sf\small The two figures on the left-hand side correspond
    to the two different gauge field patches based on a topologically
    trivial choice for the two-dimensional Dirac brane.  The
    three-dimensional hypersurface bounded by the two branes is a ball
    $B^3$. The gauge transformation associated with the transition
    between the two patches has only components along the angle
    $\varphi$ not shown in the picture. On the right-hand side the two
    figures show a singular limit of the relevant but non-trivial
    choice for the Dirac brane. The three-dimensional hypersurface
    connecting the two branes is the sum of the $B^3$ above and the
    generator of $H_3(\mathcal{M}^4)$. The corresponding gauge
    transformation has an additional component along the ring circle,
    normal to the plane.  } %%
  \label{fig:Diracbrane}
\end{figure}

One way to understand this difference is to appreciate the fact that,
in order that the Dirac membrane be unobservable, the gauge
transformation between the patches must allow for general deformations
of its worldvolume irrespective of its topology. Choosing a
topologically trivial brane on each patch, say along the north and
south pole of each sphere on the ring (see the two figures on the
left-hand side of Fig.~\ref{fig:Diracbrane}), leads to the gauge
transformation $-2 p^I\mathrm{d}\varphi$ between the patches. This is
also the only possible choice for an infinite string. But in the case
of a proper ring embedded in a four dimensional
space,\footnote{%%%%%%%%%%%%%%%%%%%%%%%%%%%%%%%%%%%%%%%%
  We assume a topologically trivial embedding space, like
  $\mathbb{R}^4$ or Taub-NUT, in the following discussion.}
%%%%%%%%%%%%%%%%%%%%% 
the topology of the spatial manifold $\mathcal{M}^4$ corresponding to
the embedding space minus the ring is nontrivial. Possible Dirac
branes are classified as the boundaries of three-dimensional spatial
hypersurfaces. Thus it is important to know the third homology group
$H_3(\mathcal{M}^4)$, since the Dirac brane can also be the boundary
of a non-trivial hypersurface, as opposed to the trivial one discussed
above.

In the case at hand it can be shown that
$H_3(\mathcal{M}^4)=\mathbb{Z}$, so that the generator of the group is
a hypersurface with no boundary that wraps the ring once. A
corresponding Dirac brane is described as the boundary of the sum of
the topologically trivial hypersurface and this generator. Such a
brane starts at the north pole of the sphere at some point along the
ring. When moving along the $S^1$ of the ring, this brane rotates to
the south pole and subsequently it returns to the north pole when
reaching the point of departure. A singular limit of this surface is
shown on the right-hand side of Fig.~\ref{fig:Diracbrane}.  Using the
construction based on de Rham currents in
\cite{Henneaux:1986tt,Bekaert:2002eq,Bekaert:2002cz}, this leads to a
gauge transformation between the gauge field patches that is
proportional to the Poincar\'e dual of the generator described above.
A component along $\psi$ is obviously necessary due to the plane in
the centre.  The relative coefficient in the gauge transformation
$\mathrm{d}(\varphi+\tfrac12\psi)$ has been fixed by demanding
periodicity of this generator. Finally, note that higher wrappings
would introduce integral multiples of the same one-form, and are
therefore irrelevant in view of the integral shift symmetry of $a^I$.

%%%%%%%%%%%%%%%%%%%%%%%%%%%%%%%%%%%%%%%%%%%%%%%
%%%%%%%%%%%%%%%%%%%%%%%%%%%%%%%%%%%%%%%%%%%%%%%%%

The way to measure the Wilson line moduli $a^I$ now proceeds through
the Chern-Simons charges of the ring, defined by the integral over the
3-cycle associated with the horizon $\Sigma$,
\begin{equation}
  \label{eq:CS-charge}
  Q_I^\mathrm{CS} \propto  \oint_\Sigma \; C_{IJK}\,W^J\wedge F^K\,. 
\end{equation}
It was demonstrated in \cite{Hanaki:2007mb}, by using the sections
\eqref{eq:A-I-ringsections} and carefully evaluating the integral,
that the Chern-Simons charges are linearly related to the moduli
$a^I$, i.e., $Q_I^\mathrm{CS}\propto C_{IJK}\,a^J p^K$. The use of the
sections \eqref{eq:A-I-ringsections} is essential for obtaining this
relationship, so that the $a^I$, which are identified in this way, are
no longer subject to small gauge transformations. Here it is relevant
that the Chern-Simons charges are also invariant under small gauge
transformations. This result is also consistent with large gauge
transformations as both the $a^I$ and the $Q_I^\mathrm{CS}$ change
under a large gauge transformations by an integer (in proper units).

Although it is not the primary purpose of this paper to consider
multi-ring solutions, it is illuminating to briefly consider the
situation of concentric rings \cite{Gauntlett:2004qy}. Labeling the
rings by an index $i$, one introduces the moduli $a^I{}_i$ and the
charges $p^I{}_i$ of the $i$-th ring. Following the same logic as
above, an extended set of sections generalizing
\eqref{eq:A-I-ringsections} can be found that matches the one used in
\cite{Hanaki:2007mb}. One can then derive the following relation,
\begin{equation}
  \label{eq:CS-charge-multi}
  Q_I^\mathrm{CS} \propto   C_{IJK}\bigg[ \sum_i (2\,a^J +p^J)_i \, p^K{}_i
  - \Big(\sum_i p^J{}_i\Big) \Big(\sum_j p^K{}_j\Big)\bigg] \;, 
\end{equation}
which, for a single ring, reduces to the previous result. The above
relation indicates that the Chern-Simons charges are not additive,
unlike the moduli $(a^I)_i$ and the charges $(p^I)_i$ associated with
the various rings. In fact, as we will establish later in section
\ref{sec:entropy-black-rings}, the best way to write this result is as
follows,
\begin{equation}
  \label{eq:CS-charge-multi-rewrite}
  Q_I^\mathrm{CS}-6\,C_{IJK} P^JP^K = -12\, C_{IJK} \sum_i (a^J
  + \tfrac12p^J)_i \, p^K{}_i  \,, 
\end{equation}
where $P^I{}_i= \sum_i p^J{}_i$. This indicates that the expression on
the left-hand side is in fact additive. We will return to this topic
in section \ref{sec:entropy-black-rings}.

%%%%%%%%%%%%%%%%%%%%%%%%%%%%%%%%%%%%%%%%%%%%%%%%%%%%%%%%%%%%%%%%%%%%
\subsection{Linear multiplets}
\label{sec:linear-multiplets}
%%%%%%%%%%%%%%%%%%%%%%%%%%%%%%%%%%%%%%%%%%%%%%%%%%%%%%%%%%%%%%%%%%%%
As a last topic of this section we reconsider the two linear
multiplets constructed in section \ref{sec:tensor-calculus} from the
product of two vector multiplets and from the square of the Weyl
multiplet both vanish for BPS configurations, as the reader can easily
verify. However, the corresponding three-forms, denoted by
$E_{\mu\nu\rho}$, do not necessarily vanish. Since these quantities
will play a role in what follows, we will evaluate some of the
corresponding expressions here.

First of all, we present some components of the tensor field
$E^{(IJ)}_{\mu\nu\rho}$, defined in \eqref{eq:tensor-from.vector2}.
Subject to the BPS conditions, one obtains the following results,
\begin{eqnarray}
  \label{eq:E-attractor-IJ}
  E_{\psi\theta\varphi}^{(IJ)} &=& \frac{\sin\theta}{8\, v^2}
  \Big[-\mathrm{e}^g\, \sigma^I\sigma^J \,T_{01} + \sigma^{(I} W_\psi{}^{J)}
  \,T_{23} \Big]\;, \nonumber \\[.2ex] 
  E_{\psi rt}^{(IJ)} &=&{} - \frac{1}{8\, v^2}
  \Big[\mathrm{e}^g\, \sigma^I\sigma^J\, T_{23} + \sigma^{(I} W_\psi{}^{J)}
  \,T_{01} \Big]\;, \nonumber \\[.2ex] 
  E_{rt\varphi}^{(IJ)} &=&{} - \cos \theta \;
  \frac{T_{01}T_{23}\,\sigma^I\sigma^J}{32\,v^4}\;, \nonumber \\[.2ex] 
  E_{\theta t\varphi}^{(IJ)} &=&{} r \sin \theta \;
  \frac{T_{01}T_{23}\,\sigma^I\sigma^J}{32\,v^4} \;. 
\end{eqnarray}
For $E^\mathrm{W}_{\mu\nu\rho}$, defined in
\eqref{eq:tensor-potential-from-weyl2}, one derives the following
equation upon using the BPS conditions, 
\begin{eqnarray}
  \label{eq:tensor-Weyl-attractor}
  E_{\mu\nu\rho}^\mathrm{W}  &=& - \tfrac3{16} \omega_{[\mu}{}^{ab} 
  \left(\partial_\nu\omega_{\rho]\,ab} - \tfrac23
    \omega_{\nu\,ac}\,\omega_{\rho]}{}^c{}_b \right) \nonumber\\[.2ex] 
  &&{}
  - \tfrac34\mathrm{i} e\,\varepsilon_{\mu\nu\rho\sigma\lambda}
  \left(T^2 T^{\sigma\lambda} + 6\,
  T^{\sigma\kappa}T_{\kappa\tau}T^{\tau\lambda} \right)\,.
\end{eqnarray}
This result leads to the following components, 
\begin{eqnarray}
  \label{eq:W-tensor-potential-attractor}
  E^\mathrm{W}_{\psi\theta\varphi} &=&{} - \tfrac{3}{8} 
  \sin\theta\, \mathrm{e}^g \,T_{01}
  \,. \nonumber \\[.2ex] 
  E^\mathrm{W}_{\psi rt} &=&{} \tfrac{3}{8} \,\mathrm{e}^g \,T_{23}
  \,,\nonumber \\[.2ex]  
  E^\mathrm{W}_{rt \varphi} &=&{} \cos\theta \; 
  \frac{T_{01}T_{23}}{16\, v^2}\,, \nonumber \\[.2ex] 
   E^\mathrm{W}_{\theta t\varphi} &=&{} -r \sin\theta \;
  \frac{T_{01}T_{23}}{16\, v^2}\,. 
\end{eqnarray}
We note that the components listed in \eqref{eq:E-attractor-IJ} and
\eqref{eq:W-tensor-potential-attractor} are consistent with the fact
that these three-forms are closed. Furthermore they are invariant
under the scale transformations mentioned previously.

%%%%%%%%%%%%%%%%%%%%%%%%%%%%%%%%%%%%%%%%%%%%%%%%%%%%%%%%%%%%%%%%%%%%
\section{The Lagrangian and the electric charges}
\label{sec:lagr-electr-charg}
\setcounter{equation}{0}
%%%%%%%%%%%%%%%%%%%%%%%%%%%%%%%%%%%%%%%%%%%%%%%%%%%%%%%%%%%%%%%%%%%%
The construction of the relevant Lagrangian follows from the results
presented in section \ref{sec:tensor-calculus}. Linear multiplets can
be constructed from the products of two multiplets, which can then be
coupled to a vector multiplet by using the superconformal density
formula \eqref{eq:vector-tensor-density}. The linear multiplet
constructed from two vector multiplets will be written by means of a
symmetric three-rank constant tensor $C_{IJK}$, which can be
identified with the tensor that we introduced earlier in
\eqref{eq:zeta-v+h}, although this is by no means essential. Below we
will also use the notation $C(\sigma)=
C_{IJK}\,\sigma^I\sigma^J\sigma^K$.  The lowest component of the
linear multiplet associated with the symmetrized product of two vector
multiplets will thus be identified with $-C_{IJK}\,L^{ij(JK)}$, where
we make use of \eqref{eq:full-L}. Higher components are defined
accordingly. The vector multiplet that couples to the linear multiplet
quadratic in the Weyl multiplet is characterized by constants $c_I$,
so that its scalar field equals $c_I\sigma^I$. Finally we also include
the Lagrangian for hypermultiplets.

After these definitions we introduce the expression for the bosonic
terms in the Lagrangian, with convenient normalizations, decomposed
according to,
\begin{equation}
  \label{eq:L-tot}
  \mathcal{L} = \mathcal{L}_\mathrm{vvv}+\mathcal{L}_\mathrm{hyper} +
  \mathcal{L}_\mathrm{vww} \,.
\end{equation}
Here the Lagrangian cubic in vector multiplet fields equals, 
\begin{eqnarray}
  \label{eq:vvv}
  8\pi^2\mathcal{L}_\mathrm{vvv} &=&{}
  3\,C_{IJK} \sigma^I\Big[\tfrac12 \mathcal{D}_\mu\sigma^J
  \,\mathcal{D}^\mu\sigma^K 
  + \tfrac14 F_{\mu\nu}{}^J F^{\mu\nu K} - Y_{ij}{}^J  Y^{ijK }
  -3\,\sigma^J F_{\mu\nu}{}^K  T^{\mu\nu} \Big] \nonumber \\
%%%%%%%%%%%%
  &&{}
   - \tfrac18\mathrm{i}C_{IJK}\,e^{-1}
  \varepsilon^{\mu\nu\rho\sigma\tau} W_\mu{}^I 
  F_{\nu\rho}{}^J F_{\sigma\tau}{}^K 
  - C(\sigma) \Big[\tfrac18 \mathcal{R} - 4\,D - \tfrac{39}2
  T^2\Big]\,, 
\end{eqnarray}
the Lagrangian for hypermultiplets (one of which plays the role of a
compensating supermultiplet) reads,
\begin{equation}
  \label{eq:lagr-hypers}
  8\pi^2\mathcal{L}_\mathrm{hyper} =  -\tfrac12
  \Omega_{\alpha\beta}\, \varepsilon^{ij} 
  {\cal D}_\mu A_i{}^\alpha\, {\cal D}^\mu
  A_j{}^{\beta}+ \chi\Big[\tfrac{3}{16}\mathcal{R} 
  + 2\, D  + \tfrac3{4} T^2 \Big]\,,
\end{equation}
and the Lagrangian containing the higher-derivative couplings is given
by, 
\begin{eqnarray}
  \label{eq:vww}
  8\pi^2\mathcal{L}_\mathrm{vww} &=&{}
  \tfrac14 c_I Y_{ij}{}^I \, T^{ab} R_{abk}{}^j(V) \,\varepsilon^{ki}
  \nonumber \\[.5ex] 
%%%%%
  &&{}
  + c_I\sigma^I\Big[ 
     \tfrac1{64} R_{ab}{}^{cd}(M)\,R_{cd}{}^{ab}(M) +\tfrac1{96}
  R_{ab j}{}^i(V) \,R^{ab}{}_i{}^j(V) \Big]  \nonumber\\[.5ex]
%%%%
  &&{}
   -\tfrac1{128}\mathrm{i}e^{-1}
  \,\varepsilon^{\mu\nu\rho\sigma\tau}\,c_IW_\mu{}^I\left[  
  R_{\nu\rho}{}^{ab}(M)\,R_{\sigma\tau ab}(M)+ \tfrac13
  R_{\nu\rho j}{}^i(V) \,R_{\sigma\tau i}{}^j(V)\right] \Big]
  \nonumber\\[.5ex] 
%%%%%%%%
  &&{}  + \tfrac3{16} c_I\big(10\,\sigma^I \,T_{ab}- F_{ab}{}^I\big)\,
  R(M)_{cd}{}^{ab}\,T^{cd} \nonumber\\[.5ex]
%%%%%%%%%%
  &&{} 
  + c_I\sigma^I\Big[ 3\,T^{ab}\mathcal{D}^c\mathcal{D}_aT_{bc} -\tfrac32
  \big(\mathcal{D}_aT_{bc}\big)^2 
  + \tfrac32 \mathcal{D}_cT_{ab}\,\mathcal{D}^aT^{cb}
  + \mathcal{R}_{ab}(T^{ac} T^b{}_c - \tfrac12 \eta^{ab} T^2) \Big]  
   \nonumber\\[.5ex]
%%%
   &&{} 
   + c_I\sigma^I\Big[
   \tfrac83 D^2 + 8\, T^2\,D - \tfrac{33}8 (T^2)^2 + \tfrac{81}2
   (T^{ac}T_{bc})^2 \Big]    \nonumber\\[.5ex]
%%%%%%%
  &&{}
   - c_IF_{ab}{}^I\Big[T^{ab}\,D +\tfrac{3}{8}T^{ab} \,T^2 - \tfrac9{2} \, 
  T^{ac}T_{cd}T^{db} \Big] \nonumber\\[.5ex]
%%%%%%%%
  &&{}
  +  \tfrac3{4}\mathrm{i} \,\varepsilon^{abcde}\Big[c_I F_{ab}{}^I
  \big(T_{cf}\mathcal{D}^fT_{de} +\tfrac32 T_{cf}
  \mathcal{D}_dT_e{}^f\big)
   -  3\,c_I\sigma^I T_{ab} T_{cd}\,\mathcal{D}^fT_{fe}\Big]\,.  
\end{eqnarray}
We remind the reader that $\mathcal{R}$ and $\mathcal{R}_{ab}$ refer
to the Ricci scalar and tensor. The factor $8\pi^2$, which equals four
times the volume of the unit sphere $S^3$, has been included to avoid
explicit factors of $\pi$ when defining electric
charges.\footnote{%%%%%%%%%%%%%%%%%%%%%%%%%%% 
  In four space-time dimensions one extracts a factor equal to two
  times the volume of the unit sphere $S^2$. In this way the Coulomb
  potential has the same normalization in four and in five dimensions, 
  without factors of $\pi$. } %%%%%%%%%%%%%%%%%%%%%%%%%%%%%%%%%%%%%%
In the above result there are two terms which cannot be written in a
manifestly gauge invariant form, related to the appearence of
gravitational and $\mathrm{SU}(2)$ Chern-Simons terms. To avoid these
Chern-Simons terms we have chosen to write their contribution in a
form that is explicitly proportional to the gauge fields $W_\mu{}^I$.
This representation may lead to difficulty in case that the gauge
fields are not globally defined, as we shall discuss in due course.

For future use we present the equation of motion for the auxiliary
field $D$ that follows from the above results,
\begin{equation}
  \label{eq:D-field-eq}
  \tfrac{16}3 c_I\sigma^I D + c_I(8\,\sigma^I T_{ab}- F_{ab}{}^I)
  \,T^{ab} +4\,C(\sigma) + 2\, \chi =0\,. 
\end{equation}
On the horizon, this relation yields
\begin{equation}
  \label{eq:attractor-D}
   \chi= - 2\,C(\sigma) - 2\, c_I\sigma^I\,T^2 \,.
\end{equation}

To appreciate the implications of the above results, let us first
consider \eqref{eq:vvv} for a single vector supermultiplet, so that
$C(\sigma)=\sigma^3$. When suppressing the coupling to the fields
$T_{ab}$, $D$ and to the Ricci scalar $R$, we are dealing with a
Lagrangian based on a scalar field $\sigma$, a gauge field $W_\mu$ and
an auxiliary field $Y^{ij}$.  Upon re-introducing the fermion fields,
this Lagrangian is invariant under rigid superconformal
transformations. Note that the overall sign of the Lagrangian is
irrelevant, as it can be absorbed into an overall sign of the vector
supermultiplet fields. To identify the kinetic terms one may expand about
some constant value of the field $\sigma$. The value of this constant
is arbitrary and in fact it can be changed by a scale transformation
that acts on all the fields and belongs to the rigid superconformal
symmetry group. Note the presence of the Chern-Simons term, which
implies that the corresponding action is only gauge invariant up to
boundary terms.

When coupled to the fields of the Weyl multiplet, this Lagrangian is
invariant under local superconformal transformations. However, it is
inconsistent because the field $D$ acts as a Lagrange multiplier which
requires $\sigma$ to vanish. To avoid this difficulty one must also
introduce the superconformally invariant Lagrangian of a
hypermultiplet. Introducing one hypermultiplet, the field equation for
$D$ implies that $\chi= -2\,\sigma^3$. In view of the local invariance
under scale transformations $\sigma$ can be fixed to a constant. The
phases contained in the hypermultiplet scalars can be fixed as well by
making use of the local $\mathrm{SU}(2)$ transformations of the
superconformal group, so that none of the scalar fields will
correspond to physical degrees of freedom. Furthermore one can
eliminate the auxiliary fields, $Y^{ij}$ and $T_{ab}$, by their
(algebraic) field equations, which yields $Y^{ij}=0$ and $T_{ab} =
(4\sigma)^{-1} F_{ab}$. Hence one is left with (for constant
$\sigma$),
\begin{equation}
  \label{eq:pure-sg}
  8\pi^2\mathcal{L} = - \tfrac12 \sigma^3 \mathcal{R}
  -\tfrac3{8}\sigma \,F_{\mu\nu}F^{\mu\nu} -
  \tfrac18\mathrm{i}\,e^{-1} \varepsilon^{\mu\nu\rho\sigma\tau} W_\mu 
  F_{\nu\rho} F_{\sigma\tau} \,,
\end{equation}
which, upon including the gravitino field (the other fermions are
either auxiliary or can be set to zero by a gauge choice), is equal to
the Lagrangian of pure five-dimensional supergravity. Observe that the
vector gauge field is the only field of the vector multiplet and the
hypermultiplet that describes physical degrees of freedom. The other
ones are compensating fields (associated with scale transformations,
R-symmetry and S-supersymmetry) or auxiliary fields. The field
$\sigma$ is a constant and defines the Newton constant to be equal to
$G_\mathrm{N}=\sigma^{-3}$, so that the Ricci scalar will appear in
the Lagrangian with a multiplicative factor
$(16\pi^2G_\mathrm{N})^{-1}$. This definition of Newton's constant is
different from the more conventional one, where one adopts a prefactor
$(16\pi \,G_\mathrm{N})^{-1}$, just as in four space-time dimensions.
As a result of the convention of this paper, the Bekenstein-Hawking
area law, leads to the area in Planck units, $A/G_\mathrm{N}$, with
proportionality factor $(4\pi)^{-1}$.

Let us briefly examine the relevant definition of the entropy defined
in terms of the Noether potential \cite{Wald:1993nt,Iyer:1994ys},
\begin{equation}
  \label{eq:wald-entropy}
  \mathcal{S}_\mathrm{macro}=  2\pi \int_{\Sigma_\mathrm{hor}}
  \frac{\partial\mathcal{L}}{\partial 
  \mathcal{R}_{\mu\nu\rho\sigma}}\;\varepsilon_{\mu\nu}\varepsilon_{\rho\sigma}
  \,, 
\end{equation}
where $\varepsilon_{\mu\nu}$ is the bi-normal tensor associated with
the horizon, normalized such that
$\varepsilon_{\mu\nu}\varepsilon^{\mu\nu}= -2$. For $\mathcal{L}=
-(16\pi^2 G_\mathrm{N})^{-1} \mathcal{R}$, this definition yields
$\mathcal{S}_\mathrm{macro} = \tfrac14 (A/ \pi G_\mathrm{N})$, which
is the area law with the area described in units of $\pi
G_\mathrm{N}$. This has a bearing on the various normalization factors
for the Noether potential and the entropy discussed later on.

In the next section we present a more detailed discussion of the
entropy and the angular momentum. Before doing so we briefly discuss
the electric charges, which follow from the relevant integral over the
3-cycle that encloses the black hole or the black ring,
\begin{equation}
  \label{eq:def-q}
  q_I = \frac{2}{8 \pi^2} \int \mathrm{d}\theta
  \,\mathrm{d}\varphi\, \mathrm{d} \psi \; \left[ - 3\,C_{IJK} \,
  E_{\psi\theta\varphi}^{JK} + c_I\,E_{\psi\theta\varphi}^\mathrm{W}
  \right] \,,
\end{equation}
where the relative factor 3 results from the fact that the Lagrangian
\eqref{eq:vvv} is cubic in the vector multiplets, whereas the
Lagrangian \eqref{eq:vww} is only linear. An overall factor 2 has been
included to be consistent with the usual definition of the charge in
terms of the electric displacement field. Making use of the results
\eqref{eq:E-attractor-IJ} and \eqref{eq:W-tensor-potential-attractor},
one obtains the following result,
\begin{equation}
  \label{eq:q-attractor}
  q_I = \frac3{2\,v^2} C_{IJK}\left(\sigma^J\sigma^K
  \,\mathrm{e}^g T_{01} - \sigma^J  
  [W_\psi{}^K]\, T_{23}\right)  - \frac32 c_I\,\mathrm{e}^g T_{01} \,,
\end{equation}
where we used the definition 
\begin{equation}
  \label{eq:[W]}
  [W_\psi{}^I]= \frac1{16\,\pi^2}\int \mathrm{d}\theta
  \,\mathrm{d}\varphi\, \mathrm{d} \psi\,\sin\theta \; W_\psi{}^I\,.
\end{equation}
which is gauge invariant under periodic gauge transformations. For
spinning black holes, where the gauge fields are globally defined,
\eqref{eq:q-attractor} takes the form
\begin{equation}
  \label{eq:q-attractor-bh}
  q_I = \frac{3\,\mathrm{e}^g}{2\,T_{01}}  \left[ C_{IJK}\sigma^J\sigma^K 
  - c_I T_{01}{}^2\right] \,.
\end{equation}
Observe that the above results are scale invariant. 

To derive the corresponding result for the black ring is more subtle
in view of the fact that the gauge fields are not globally defined, as
was discussed in subsection \ref{sec:gauge-fields}. This will be
discussed in subsection \ref{sec:entropy-black-rings} and the
resulting expression for the charges will be given in
\eqref{eq:attractor-ring}.

The charges can also be determined by making use of the Noether
potential associated with abelian gauge transformations. Consider, for
instance, a Lagrangian in five space-time dimensions consisting of an
invariant Lagrangian depending on the abelian field strength
$F_{\mu\nu}$, its space-time derivatives $\nabla_\rho F_{\mu\nu}$, and
matter fields denoted by $\psi$ and their derivatives
$\nabla_\mu\psi$, plus an abelian Chern-Simons term,
\begin{equation}
  \label{eq:ho-gauge-lagr}
  \mathcal{L}^\mathrm{total} = \mathcal{L}^\mathrm{inv}(F_{\mu\nu},\nabla_\rho
  F_{\mu\nu},\psi,\nabla_\mu\psi)  +
  \varepsilon^{\mu\nu\rho\sigma\tau} A_\mu 
  F_{\nu\rho}F_{\sigma\tau}\,,
\end{equation}
For this Lagrangian, the Noether potential reads as follows, 
\begin{equation}
  \label{eq:N-pot-gauge}
  \mathcal{Q}_\mathrm{gauge}^{\mu\nu}(\phi,\xi)=
  2\,\mathcal{L}_F^{\mu\nu} \xi -2\, 
  \nabla_\rho \mathcal{L}^{\rho,\mu\nu}_F \xi +
  6\,e^{-1}\varepsilon^{\mu\nu\rho\sigma\tau}\, \xi A_\rho
  F_{\sigma\tau} \,, 
\end{equation}
where $\phi$ generically denotes the various fields and $\xi$ is the
infinitesimal local parameter associated with the gauge
transformations. Here we use the notation, 
\begin{equation}
  \label{eq:var-Lagr}
  \delta\mathcal{L}^\mathrm{inv} = \mathcal{L}_F^{\mu\nu}\,\delta
  F_{\mu\nu} +  \mathcal{L}_F^{\rho,\mu\nu}\,\delta(\nabla_\rho 
  F_{\mu\nu}) +  \mathcal{L}_\psi\,\delta \psi +
  \mathcal{L}_\psi^{\mu}\,\delta (\nabla_\mu\psi)
\end{equation}
It is straightforward to verify that 
$\partial_\nu\mathcal{Q}^{\mu\nu}$ is equal to the field equation, up
to terms proportional to $\partial_\nu\xi$. The electric charge can be
written as 
\begin{equation}
  \label{eq:electric-charge}
  q= \int_{\Sigma_\mathrm{hor}} \,\varepsilon_{\mu\nu} \,
  \mathcal{Q}_\mathrm{gauge}^{\mu\nu}(\phi,\xi)   \,.  
\end{equation}
where $\varepsilon_{\mu\nu}$ is the binormal tensor associated with
the horizon and the gauge parameter $\xi$ must be taken constant so
that the underlying field configuration is invariant and the
corresponding Noether current vanishes on-shell.
%%%%%%%%%%%%%%%%%%%%%%%%%%%%%%%%%%%%%%%%%%%%%%%%%%%%%%%%%%%%%%%%%%%%
\section{Entropy and angular momentum for black holes and rings}
\label{sec:entropy-angular-momentum}
\setcounter{equation}{0}
%%%%%%%%%%%%%%%%%%%%%%%%%%%%%%%%%%%%%%%%%%%%%%%%%%%%%%%%%%%%%%%%%%%%
The evaluation of the entropy and the angular momentum proceeds from
the expression for the Noether potential associated with space-time
diffeomorphisms \cite{Wald:1993nt,Jacobson:1993vj,Iyer:1994ys}. In the
case at hand this is complicated in view of higher-derivative
interactions, but it is especially subtle because of the presence of
the Chern-Simons terms.  At the end, one must evaluate the integral of
the appropriate Noether potential over the horizon, and here one may
encounter an extra subtlety when the gauge fields are not globally
defined. This will be discussed further in section
\ref{sec:entropy-black-rings}.

In this section we start with a systematic discussion of the relevant
Noether potential based on the Lagrangian specified in section
\ref{sec:lagr-electr-charg}. This Lagrangian contains two different
Chern-Simons terms, one of the type $W\wedge F\wedge F$, which is
cubic in the abelian gauge fields, and a mixed one of the type
$W\wedge\mathrm{Tr}[\mathcal{R}\wedge\mathcal{R}]$, which is linear in
the gauge fields and quadratic in the Riemann curvature. The
derivation of the corresponding Noether potential is straightforward
but subtle. We first evaluate this potential for a Lagrangian that
depends on the Riemann tensor, on the field strengths of abelian gauge
fields, and on an anti-symmetric tensor field $T_{\mu\nu}$ with at
most first-order space-time derivatives $\nabla_\mu T_{\nu\rho}$.
This Lagrangian does {\it not} contain the two Chern-Simons terms,
which are considered separately. Its Noether potential associated with
space-time diffeomorphisms decomposes into two different terms,
\begin{equation}
  \label{eq:Q-decomp}
  \mathcal{Q}_0^{\mu\nu} = \hat{\mathcal{Q}}^{\mu\nu}(\xi^\rho) +
  \hat{\mathcal{Q}}^{\mu\nu}_\mathrm{gauge}(-\xi^\rho W_\rho{}^I)\,,
\end{equation}
corresponding to the following decomposition of the diffeomorphisms on
the gauge field,
\begin{equation}
  \label{eq:diff-W}
  \delta_\xi W_\mu{}^I = -\partial_\mu \xi^\nu W_\nu{}^I - \xi^\nu
  \partial_\nu W_\mu{}^I = \xi^\nu F_{\mu\nu}{}^I +
  \partial_\mu(-\xi^\nu W_\nu{}^I)\,. 
\end{equation}
The first term, which does not contain the effect of the last
term in \eqref{eq:diff-W}, is given by
\begin{eqnarray}
  \label{eq:gen-N-potentia}
    \hat{\mathcal{Q}}^{\mu\nu}(\xi^\rho)&=&-
  2\,\mathcal{L}_\mathrm{R}^{\mu\nu\rho\sigma}\,\nabla_\rho\xi_\sigma +
  4\,\nabla_\rho \mathcal{L}_\mathrm{R}^{\mu\nu\rho\sigma}\,\xi_\sigma
  \nonumber\\
  &&{}
  +\left[\mathcal{L}_\mathrm{T}^{\mu,\rho\sigma}
  \,T^\nu{}_{\sigma}  
    +\mathcal{L}_\mathrm{T}^{\rho,\mu\sigma}\,T^\nu{}_{\sigma}
  +\mathcal{L}_\mathrm{T}^{\nu,\mu\sigma}\,T^\rho{}_{\sigma} 
   -(\mu\leftrightarrow\nu) \right]\xi_\rho\,, 
\end{eqnarray}
where $\xi^\rho$ parametrizes the diffeomorphisms, and
$\mathcal{L}^{\mu\nu\rho\sigma}$ and $\mathcal{L}^{\mu,\nu\rho}$
denote partial derivatives of the Lagrangian according to
\begin{equation}
  \label{eq:partial-der-L}
  \delta\mathcal{L}= \mathcal{L}_\mathrm{R}^{\mu\nu\rho\sigma}
  \,\delta\mathcal{R}_{\mu\nu\rho\sigma} +
  \mathcal{L}_\mathrm{T}^{\mu,\nu\rho}\,\delta(\nabla_\mu T_{\nu\rho})\,.
\end{equation}
These derivatives are subject to the BPS attractor equations. As a
result they take the following form on the horizon, 
\begin{eqnarray}
  \label{eq:L-R+L-DT}
  8\pi^2\, \mathcal{L}_\mathrm{R}^{\mu\nu\rho\sigma} &=& (-\tfrac12 C(\sigma) 
  -\tfrac34 c_I\sigma^I\, T^2) g^{\mu[\rho}g^{\sigma]\nu}  +\tfrac12
  c_I\sigma^I 
  (T^{\mu\nu}T^{\rho\sigma} - T^{\mu[\rho}T^{\sigma]\nu})
  \,,\nonumber\\ 
  8\pi^2\,\mathcal{L}_\mathrm{T}^{\rho,\mu\nu} &=& - 3\,c_I\sigma^I(
  3\,\mathcal{D}^{[\mu} T^{\nu]\rho} + \mathcal{D}^\rho T^{\mu\nu}
  +\tfrac54 \mathrm{i}\varepsilon^{\mu\nu\sigma\lambda\tau}
  T_{\sigma\lambda}T_\tau{}^\rho) \nonumber\\
  &=&{}
  -\tfrac94 \mathrm{i}\,c_I\sigma^I\, \varepsilon^{\mu\nu\sigma\lambda\tau}
  \,T_{\sigma\lambda}T_\tau{}^\rho \,,
\end{eqnarray}
Obviously we also need the derivative $\nabla_\rho
\mathcal{L}_\mathrm{R}^{\mu\nu\rho\sigma}$, which follows form
\eqref{eq:L-R+L-DT} by means of the attractor equations. The result
reads as follows,
\begin{equation}
  \label{eq:nabla-L-R}
    8\pi^2\,\nabla_\rho\mathcal{L}_\mathrm{R}^{\mu\nu\rho\sigma}=
    \tfrac58 \mathrm{i}\,c_I\sigma^I (T^{\mu\nu}
    \varepsilon^{\sigma\rho\lambda\kappa\tau} - T^{\sigma[\mu}
    \varepsilon^{\nu]\rho\lambda\kappa\tau} )
    T_{\rho\lambda}T_{\kappa\tau} \,. 
\end{equation}

It remains to consider the second term in \eqref{eq:Q-decomp},
$\hat{\mathcal{Q}}^{\mu\nu}_\mathrm{gauge}(-\xi^\rho W_\rho{}^I)$, which denotes the
Noether potential associated with the abelian gauge transformations
with field-dependent gauge parameters $\xi^I= -\xi^\rho\,W_\rho{}^I$.
This potential was already presented in \eqref{eq:N-pot-gauge}, where
the last term corresponding to the $W\wedge F\wedge F$ Chern-Simons
term has been suppressed. We thus need the expression for
$\mathcal{L}_{\mathrm{F}\;I}^{\mu\nu}$ (c.f. \eqref{eq:var-Lagr}),
\begin{eqnarray}
  \label{eq:L-F}
  8\pi^2\,\mathcal{L}_{\mathrm{F}\;I}^{\mu\nu} &=&{} \tfrac3{2}
  C_{IJK}\sigma^JF^{\mu\nu K} - 9\, C_{IJK}\sigma^J\sigma^K\,T^{\mu\nu}
  \nonumber \\
  &&{}
  -\tfrac3{16}c_I\, R_{ab}{}^{\mu\nu}(M)\,T^{ab} +\tfrac3{4} \mathrm{i} c_I
  \varepsilon^{\mu\nu\rho\sigma\tau}
  \,T_{\rho\lambda}\left(\mathcal{D}^\lambda T_{\sigma\tau}  + 
  \tfrac3{2} \mathcal{D}_\sigma T_\tau{}^\lambda\right) \nonumber\\
  &&{}
  -c_I \left[T^{\mu\nu}(D+\tfrac38
  T^2)-\tfrac9{2}T^{\mu\rho}T_{\rho\sigma}T^{\sigma\nu} \right]
  \nonumber\\ 
  &=&{}
  -3\,C_{IJK}\sigma^J\sigma^K\, T^{\mu\nu} + \tfrac3{4}
  c_I\left(T^{\mu\nu}\,T^2 + 6\,
  T^{\mu\rho}T_{\rho\sigma}T^{\sigma\nu}\right) \,,
\end{eqnarray}
where the second equation represents the value taken at the horizon. 

By combining the above contributions we obtain an explicit expression
for \eqref{eq:Q-decomp}. In practice we need the contraction of the
Noether potential with the bi-normal tensor \eqref{eq:binormal}
associated with the horizon. Therefore we evaluate the following
expression for \eqref{eq:Q-decomp},
\begin{eqnarray}
  \label{eq:N-pot-0}
  8\pi^2\,\varepsilon_{\mu\nu}\mathcal{Q}_0^{\mu\nu} &=&{}
  -2\,\varepsilon_{01}\,C(\sigma) \,\nabla_{[0}\xi_{1]} \nonumber \\
  && {}
  - 2\,\varepsilon_{01}\, c_I\sigma^I\left[3\,T_{23}{}^2
  \,\nabla_{[0}\xi_{1]} -2\, T_{01}T_{23}\,\nabla_{[2}\xi_{3]}
  +11\,T_{01}{}^2 T_{23} \,\xi_5\right] \nonumber\\
  &&{}
  + 2\,\varepsilon_{01}\,\xi^\rho W_\rho{}^I\,T_{01}\,\left[-6\,
  C_{IJK}\sigma^J\sigma^K + 3\,c_I (T_{23}{}^2+2\,T_{01}{}^2) \right] \,.
\end{eqnarray}

Subsequently, we turn to the Chern-Simons terms contained in
\eqref{eq:vvv} and \eqref{eq:vww},
\begin{equation}
  \label{eq:combined-CS-terms}
  8\pi^2\,\mathcal{L}_\mathrm{CS}= -\tfrac1{8} \mathrm{i}\,e^{-1}
  \varepsilon^{\mu\nu\rho\sigma\tau} 
  \left[C_{IJK}\,W_\mu{}^I F_{\nu\rho}{}^JF_{\sigma\tau}{}^K
  +\tfrac1{16}  c_I\,W_\mu{}^I 
  \mathcal{R}_{\nu\rho}{}^{ab} \,\mathcal{R}_{\sigma\tau ab} \right]
  \,,
\end{equation}
which contribute to the Noether potential as follows,
\begin{eqnarray}
  \label{eq:Q-CS}
  8\pi^2\,\mathcal{Q}^{\mu\nu}_\mathrm{CS} &=& {} \tfrac1{2}
   \mathrm{i}\,e^{-1}\varepsilon^{\mu\nu\rho\sigma\tau}\,
   C_{IJK} \,\xi^\lambda W_\lambda{}^I  W_\rho{}^J  F_{\sigma\tau}{}^K
   \nonumber\\ 
   &&{}
   + \tfrac1{32} \mathrm{i}\,e^{-1}
   \varepsilon^{\mu\nu\rho\sigma\tau} c_I
   W_\rho{}^I \,
   \mathcal{R}_{\sigma\tau}{}^{\kappa\lambda}\,\nabla_\kappa\xi_\lambda
   \nonumber \\
   &&{}
   - \tfrac1{32} \mathrm{i}\,e^{-1}
   \varepsilon^{\rho\sigma\tau\lambda[\mu} \,c_I F_{\rho\sigma}{}^I \,
   \mathcal{R}_{\tau\lambda}{}^{\nu]\kappa}\, \xi_\kappa \nonumber\\
   &&{}
    + \tfrac1{64} \mathrm{i}\,e^{-1}
  \,\varepsilon^{\rho\sigma\tau\lambda\kappa} \,c_I
   F_{\rho\sigma}{}^I \, \mathcal{R}_{\tau\lambda}{}^{\mu\nu}\,
   \xi_\kappa   \,.
\end{eqnarray}
We note that the first term is similar to what one expects from the
expression for the Noether potential associated with gauge
transformations. However, it carries a different pre-factor than in
\eqref{eq:N-pot-gauge}, a feature that is well known (see e.g.
\cite{Suryanarayana:2007rk}). Evaluating this expression at the
horizon, using \eqref{eq:R-omega} and \eqref{eq:A-I-S3}, yields,
\begin{eqnarray}
  \label{eq:vareps-Q-CS}
  8\pi^2\,\varepsilon_{\mu\nu}\mathcal{Q}_\mathrm{CS}^{\mu\nu} &=&{}
  8\,\varepsilon_{01}\, T_{23}\, C_{IJK}\, \sigma^I W_5{}^J
   W_\lambda{}^K\,\xi^\lambda \nonumber\\ 
  &&{}
  - \varepsilon_{01}\, c_I W_5{}^I \left[-2\,T_{01}T_{23}
  \,\nabla_{[0}\xi_{1]} + 
  (T_{01}{}^2 + 4\,\,T_{23}{}^2) \,\nabla_{[2}\xi_{3]}\right]
  \nonumber\\
  &&{}
  - \varepsilon_{01}\,T_{01}{}^2\,c_I\left[W_3{}^I\,\nabla_{[5}\xi_{2]} -
  W_2{}^I\,\nabla_{[5}\xi_{3]} \right] \nonumber\\
  &&{} 
  + 2\,\varepsilon_{01}\, c_I \sigma^I\,T_{23} \left[6\,
  T_{01}{}^2 -T_{23}{}^2\right]  \,\xi_5 \,. 
\end{eqnarray}
Note that the above results depend explicitly on the gauge fields
$W_\mu{}^I$. For black holes, where the gauge fields are globally
defined, this is not an issue. However, for black rings the situation
is different and extra care is required. As we discuss in subsection
\ref{sec:an-alternative-form} we will employ an alternative treatment
of the mixed Chern-Simons term which will lead to expressions that
differ from \eqref{eq:combined-CS-terms}-\eqref{eq:vareps-Q-CS}.

By integrating the Noether potential over the horizon one obtains the
entropy and the angular momentum from the Noether potential associated
with the relevant Killing vector and contracted with the bi-normal
tensor \eqref{eq:binormal}. For the entropy the relevant Killing
vector is the timelike one, $\xi^\mu\partial_\mu=\partial/\partial t$,
and in the integrand one drops all terms except the ones
proportional to $\nabla_{[0}\xi_{1]}= \varepsilon_{01}$,
\begin{equation}
  \label{eq:entropy-generic}
  \mathcal{S}_\mathrm{macro}= -\pi  \int_{\Sigma_\mathrm{hor}}
  \varepsilon_{\mu\nu}\mathcal{Q}^{\mu\nu}
  (\xi)\Big\vert_{\nabla_{[\mu} \xi_{\nu]}
  =\varepsilon_{\mu\nu};\,\xi^\mu=0}\;, 
\end{equation} 
For the angular momentum the Killing vector is associated with the
corresponding periodic isometry of the space-time, and one has
\begin{equation}
  \label{eq:angular-mom-generic}
  J(\xi) = \int_{\Sigma_\mathrm{hor}}
  \varepsilon_{\mu\nu}\mathcal{Q}^{\mu\nu}
  (\xi)\;.
\end{equation} 

%%%%%%%%%%%%%%%%%%%%%%%%%%%%%%%%%%%%%%%%%%%%%%%%%%%%%%%%%
\subsection{An alternative form of the mixed Chern-Simons term}
\label{sec:an-alternative-form}
%%%%%%%%%%%%%%%%%%%%%%%%%%%%%%%%%%%%%%%%%%%%%%%%%%%%%%%%%
In the above derivation of the Noether potential, we were able to
handle the mixed Chern-Simons term by writing it in the form
$W\wedge \mathrm{Tr}[\mathcal{R}\wedge\mathcal{R}]$, so that the
Lagrangian is manifestly diffeomorphism covariant at the expense of
introducing explicit gauge fields $W_\mu{}^I$. This is acceptable in
cases where the gauge fields are globally defined, such as for
spinning black holes. In the case of a black ring, however, the
presence of magnetic charges implies that the gauge fields are only
defined in patches, making the use of the above formulae questionable.

Therefore we consider an alternative derivation, based on a
modification of the Lagrangian \eqref{eq:vww} proportional to
$\varepsilon^{\mu\nu\rho\sigma\tau} W_\mu{}^I
\mathcal{R}_{\nu\rho}{}^{ab} \mathcal{R}_{\sigma\tau ab}$, by adding a
suitable total derivative. In this way the gauge field is converted to
its field strength (which is globally defined), and the square of the
curvature tensor $\mathcal{R}$ is converted to a corresponding
Chern-Simons term. The alternative form of the mixed Chern-Simons term
is thus,\footnote{%%%%%%%%%%%%%%%%%%%%%%%%%%%%%%%%%%%%%%%
  Note that in this subsection we suppress the $W\wedge F\wedge F$
  Chern-Simons term of \eqref{eq:combined-CS-terms}, which is not
  affected by the conversion and whose effect has already been
  evaluated. }%%%%%%%%%%%%%%%%%%%%%%%%%%%%%%%%%%%%%%%%%%%%%%%%%%%%
\begin{equation}
  \label{eq:converted-CS}
  8\pi^2 \mathcal{L}_\mathrm{CS} = -\tfrac1{64} \mathrm{i} e^{-1}
  \,\varepsilon^{\mu\nu\rho\sigma\tau}\,c_I F_{\mu\nu}{}^I    
  \omega_\rho{}^{ab} \left(\partial_\sigma\omega_{\tau ab} -\tfrac23 
  \omega_{\sigma ac} \,\omega_\tau{}^c{}_b \right)\,. 
\end{equation}
From the point of view of general coordinate invariance, this change
does not seem crucial, as the Lagrangian \eqref{eq:converted-CS} still
transforms as a scalar. On the other hand, the spin-connection field
$\omega_\mu{}^{ab}$ is a composite vector field associated with local
Lorentz transformations.  As a result of the explicit spin-connection,
this form of the Lagrangian is no longer invariant under local Lorentz
transformations, but transforms into a boundary term.

In this formulation diffeomorphism invariance of the relevant field
configurations will be defined up to a local Lorentz transformation.
Therefore Lorentz transformations have to be taken into account in the
relevant Noether potential. In the previous form of the mixed
Chern-Simons term given in \eqref{eq:combined-CS-terms}, the local
Lorentz transformations were avoided because that expression can be
interpreted directly in the metric formulation without the need for
including vielbein fields. In principle, the invariance of the field
configuration could require additional components associated with
gauge transformations other than the Lorentz transformations, but the
gauge fields turn out to be invariant under the relevant
diffeomorphisms without the need for including compensating gauge
transformations. Therefore it suffices to consider only
diffeomorphisms and local Lorentz transformations in the following.

Under the combined variation of a diffeomorphism and a local Lorentz
transformation with parameters $\xi^\mu$ and $\varepsilon^{ab}$, the
Lagrangian \eqref{eq:converted-CS} corresponding to the mixed
Chern-Simons term transforms as
 \begin{eqnarray} 
  \label{eq:delta-L} 
  8\pi^2 \delta \left(\sqrt{g}\,\mathcal{L}_\mathrm{CS}\right) =
  -\partial_\mu\left(\xi^\mu \sqrt{g}\,\mathcal{L}_\mathrm{CS}  
  -\tfrac{1}{64}  
  \mathrm{i}\varepsilon^{\mu\nu\rho\sigma\tau}\,c_I
  F_{\nu\rho}{}^I\,    
   \partial_\sigma\,\varepsilon_{ab}\, \omega_\tau{}^{ab} \right)\,. 
\end{eqnarray} 
The corresponding Noether potential depending on both $\xi^\mu$ and
$\varepsilon_{ab}$, is then equal to
\begin{eqnarray} 
\label{eq:alt-mixed-Npot}
  8\pi^2 \mathcal{Q}^{\mu\nu}_\mathrm{CS} &=&
  - \tfrac1{32}\mathrm{i}e^{-1}  
  \,\varepsilon^{\mu\nu\rho\sigma\tau}\,c_I F_{\rho\sigma}{}^I
  \omega_\tau{}^{ab}   
  \left[\varepsilon_{ab} -\tfrac12 \xi^\kappa\omega_{\kappa\,ab}
  \right]\nonumber\\
  &&{}
  +\tfrac1{32}\mathrm{i}e^{-1} 
  \,\varepsilon^{\mu\nu\rho\sigma\tau}\,c_I\,\xi^\lambda W_\lambda{}^I
  \omega_\rho{}^{ab} \left(\partial_\sigma\omega_{\tau ab} -\tfrac23 
  \omega_{\sigma ac} \,\omega_\tau{}^c{}_b \right)
   \nonumber \\
   &&{}
   - \tfrac1{32} \mathrm{i}\,e^{-1}
   \varepsilon^{\rho\sigma\tau\lambda[\mu} \,c_I F_{\rho\sigma}{}^I \,
   \mathcal{R}_{\tau\lambda}{}^{\nu]\kappa}\, \xi_\kappa \nonumber\\
   &&{}
    + \tfrac1{64} \mathrm{i}\,e^{-1}
  \,\varepsilon^{\rho\sigma\tau\lambda\kappa} \,c_I
   F_{\rho\sigma}{}^I \, \mathcal{R}_{\tau\lambda}{}^{\mu\nu}\,
   \xi_\kappa   \,.
\end{eqnarray} 
We note that the last two covariant terms proportional to $F\wedge R$
are identical to the corresponding terms given in \eqref{eq:Q-CS}.
This expression should be evaluated for backgrounds that are
invariant, which implies that the transformation parameter
$\varepsilon^{ab}$ should be chosen such that the vielbein is
invariant under the diffeomorphisms.  This implies that the
diffeomorphism is again generated by a Killing vector $\xi^\mu$, and
\begin{equation} 
  \label{eq:comp-lor} 
  \varepsilon^{ab}= - \nabla^{[a}\xi^{b]}+
  \xi^\lambda\omega_\lambda{}^{ab}\,.
\end{equation} 
This result for $\varepsilon^{ab}$ should be substituted into the
expression \eqref{eq:alt-mixed-Npot} for the Noether potential. The
resulting expression is then expected to match the previous result
\eqref{eq:Q-CS} (without the contribution of the $W\wedge F\wedge F$
Chern-Simons term which has not been included above), when both the
gauge fields and the spin connection field can be globally defined.
This is not the case for the black hole and black ring solutions,
so that only one of the two expressions will be applicable in either
case. It should be of interest to compare the two formulae in more
detail by making explicit use of coordinate patches.

We should, however, briefly comment on the ambiguity in
\eqref{eq:alt-mixed-Npot} related to the fact that the extraction of
the derivative $\partial_\mu$ in \eqref{eq:delta-L} is not well
motivated for the second term, as we could have also left the
derivative on the spin connection field $\omega_\tau{}^{ab}$ and
extracted the derivative from the transformation parameter
$\varepsilon^{ab}$. The choice made above can be justified along the
lines of \cite{Tachikawa:2006sz}, which is consistent with the
original description of Wald \cite{Wald:1993nt,Iyer:1994ys}. One
considers the variation of the corresponding Noether current under a
continuous change in the space of solutions of the field equations, in
order to derive the first law of black hole mechanics.  For the
current relevant in this section, this variation equals,
\begin{eqnarray}
  \label{eq:variation-current}
  8\pi^2 \delta J^\mu(\phi,\xi,\varepsilon)&=&
  \partial_\nu\big[\xi^\mu\theta^\nu(\phi,\delta\phi)
  -\xi^\nu\theta^\mu(\phi,\delta\phi)\big]   
    + \omega(\phi;\delta_\xi\phi,\delta\phi)\nonumber\\
 &&{}
  -\tfrac18\mathrm{i}
  \partial_\nu\big[\varepsilon^{\mu\nu\rho\sigma\tau}
  c_I\,\delta W_\rho{}^I \,\partial_\sigma\varepsilon^{ab}\,\omega_{\tau
  ab}\big] \,,
\end{eqnarray}
where $\xi^\mu$ parametrizes a diffeomorphism and $\varepsilon^{ab}$ a
Lorentz transformation, while the variation $\delta_\xi$ is defined as
the combined effect of both transformations. The variations
$\delta\phi$ and $\delta W_\mu{}^I$ connect two nearby solutions. At
this point the diffeomorphism and the Lorentz transformation are
arbitrary and do not have to constitute an invariance of the field
configuration.

The first two terms on the right-hand side are generic. The first one,
proportional to the divergence of $\xi^{[\mu}\theta^{\nu]}$, should be
written as the variation of another term, which can then be included
into the Noether potential. This modification will not change the
entropy because it does not involve derivatives of $\xi^\mu$, and
furthermore it does not contribute to the variations of the angular
momenta at spatial infinity \cite{Iyer:1994ys}. Actually, the form of
this term ensures that the angular momenta can be determined from the
Noether potential and remain constant as a function of the distance
from the horizon \cite{Suryanarayana:2007rk}.

The second term is equal to $\omega(\phi,\delta_1\phi;\delta_2\phi)=
\delta_2\theta(\phi,\delta_1\phi) -
\delta_1\theta(\phi,\delta_2\phi)$, where $\delta_1\phi$ and
$\delta_2\phi$ denote independent field variations. When $\xi^\mu$ is
the time evolution field, then the integral of this quantity over a
Cauchy surface will be equal to the corresponding Hamiltonian in the
covariant phase-space approach. The variation of the ADM mass follows
from this Hamiltonian, and the modification to the Noether potential
related to $\xi^{[\mu}\theta^{\nu]}$ will thus contribute to it. It is
also relevant that $\omega(\phi;\delta_\xi\phi,\delta\phi)$ will
vanish for symmetric field configuration, because $\delta_\xi\phi=0$
in that case.

The hope is that the third term in \eqref{eq:variation-current} will
behave in the same way.  This term will also lead to modifications of
the Noether potential, and since it depends on $\xi^\mu$ as well as on
its derivatives, these modifications may affect the entropy. However,
it is easy to see that this will not be the case, because the relevant
$\varepsilon^{ab}$ at the horizon is precisely the bi-normal tensor
\eqref{eq:binormal}, whose derivatives vanish. Therefore the third
term in \eqref{eq:variation-current} will not lead to extra terms in
the entropy. For the angular momenta, the situation is similar but
more subtle. In that case the combination
$\partial_\sigma\varepsilon^{ab} \,\omega_{\tau ab}$ vanishes at the
horizon, except for $\partial_\theta\varepsilon^{ab} \,\omega_{\varphi
  ab}\propto \sin\theta\cos\theta$. Therefore this term vanishes upon
integration over the horizon for all $\delta W_\rho{}^I$ that are
allowed by the attractor equations. Hence the angular momenta at the
horizon are not modified and can be determined from the Noether
potential obtained earlier. An obvious question is, whether the
angular momentum whose variation appears in the first law, and which
is measured at spatial infinity, will coincide with the angular
momenta determined at the horizon. The answer to this question is not
known, but the results that we will present in section
\ref{sec:entropy-black-rings} indicate that the answer is affirmative.
Obviously a full derivation of the first law for the ring geometry is
subtle in the presence of higher-derivative couplings. Without the
latter, the derivation of the first law has already been pursued in
\cite{Copsey:2005se} in connection with the presence of the dipole
charges.

%%%%%%%%%%%%%%%%%%%%%%%%%%%%%%%%%%%%%%%%%%%%%%%%%%%%%%%%%
\section{Spinning BPS black holes}
\label{sec:entropy-black-holes}
\setcounter{equation}{0}
%%%%%%%%%%%%%%%%%%%%%%%%%%%%%%%%%%%%%%%%%%%%%%%%%%%%%%%%%
In this section we apply the material derived in the preceding
sections to the case of spinning black holes. Subsequently we discuss
various implications of our results and compare them to results that
have been obtained elsewhere.

We assume arbitrary non-zero values of $p^0$. Using \eqref{eq:area},
we integrate the Noether potential derived in \eqref{eq:N-pot-0} and
\eqref{eq:vareps-Q-CS} over the horizon. In this way we obtain the
following expression for the entropy,
\begin{equation}
  \label{eq:wald-bh5}
  \mathcal{S}_\mathrm{macro}^\mathrm{BH}=  \frac{\pi\,
  \mathrm{e}^g}{4\, v^2}  
  \big[ C(\sigma) + 4\,c_I\sigma^I\, T_{23}{}^2\big] \,.
\end{equation}
The moduli are expressed in terms of the angular momentum $J_\psi$ and
the charges $q_I$ and $p^0$ by the attractor equations.  The black
holes have only one component of angular momentum, associated with the
Killing vector $\xi^\mu\partial_\mu= \partial/\partial\psi$. Here we
refrain from introducing any additional normalization factor.  This leads to
$\xi_5=\mathrm{e}^g$ and
\begin{equation}
  \label{eq:nabla-xi}
  \nabla_{[0}\xi_{1]}= 2\, T_{23}\, \mathrm{e}^g\,,\qquad
  \nabla_{[2}\xi_{3]}= -2\, T_{01} \,\mathrm{e}^g\, . 
\end{equation}
Substituting these results into \eqref{eq:N-pot-0} and
\eqref{eq:vareps-Q-CS}, and setting $\varepsilon_{01}=1$, yields the
following expression for $J_\psi$,
\begin{equation}
  \label{eq:angular-spinn-bh}
  J_\psi=  \frac{T_{23} \mathrm{e}^{2g}}{T_{01}{}^2}
  \,\left[C_{IJK}\,\sigma^I\sigma^J\sigma^K - 4\,
  c_I\sigma^I\,T_{01}{}^2\right] \,.
\end{equation}
Note that there is no other non-vanishing component of angular
momentum in this case. The charges follow from \eqref{eq:q-attractor}
and \eqref{eq:magnetic-P},
\begin{equation}
  \label{eq:attractor-bh1}
  q_I = \frac{3\,\mathrm{e}^g}{2\,T_{01}}  \left[C_{IJK}\sigma^J\sigma^K 
  - c_I T_{01}{}^2\right] \,,\qquad    p^0=
  \frac{\mathrm{e}^{-g}}{4\,v^2}\,T_{01}\, .
\end{equation}

It is convenient to express these results in terms of scale invariant
variables defined by
\begin{equation}
  \label{eq:scale-inv-bh-variables}
  \phi^I=\frac{\sigma^I}{4 T_{01}}\,,\qquad
  \phi^0=\frac{e^{-g}T_{23}}{4v^2}= \frac{p^0\,T_{23}}{T_{01}}\,.
\end{equation}
In terms of these variables \eqref{eq:wald-bh5} reads
\begin{equation}
  \label{eq:wald-bh}
  \mathcal{S}_\mathrm{macro}^\mathrm{BH}=  \frac{4\pi p^0}{({\phi^0}^2
    +{p^0}^2)^2} \Big[{p^0}^2 
  C_{IJK}\phi^I\phi^J\phi^K +\frac14 \, c_I\phi^I \,{\phi^0}^2 \Big]\,,
\end{equation}
whereas the attractor equations for the electric charges $q_I$ and the
angular momentum $J_\psi$ take the form,
\begin{eqnarray}
  \label{eq:attractor-bh-q-J}
  q_I &=& \frac{6\,p^0}{{\phi^0}^2 +{p^0}^2}
  \Big[C_{IJK}\phi^J\phi^K -\frac1{16}\,c_I\Big] \,, \nonumber\\
  J_\psi &=&  \frac{4\phi^0p^0}{({\phi^0}^2 +{p^0}^2)^2}
  \Big[C_{IJK} \phi^I\phi^J\phi^K - \frac14 c_I\phi^I \Big]\,. 
\end{eqnarray}
This result shows that $\phi^0$ is proportional to the angular
momentum, as is also obvious from \eqref{eq:scale-inv-bh-variables}.
To understand the limit in which the charges become uniformly large,
we consider uniform rescalings of the charges $q_I$ and $p^0$ as well
as of the moduli $\phi^I$ and $\phi^0$. Obviously, the terms
proportional to $C_{IJK}$ in the attractor equations are consistent
with this scaling, whereas the terms proportional to $c_I$ are
suppressed inversely proportional to the square of the charges and
thus represent subleading contributions. The leading term of the
entropy then scales as the square of the charges, while the correction
terms proportional to $c_I$, which originate from the higher-order
derivative couplings, represent the subleading contributions in the
limit where all charges become large.

The above results can be compared to the corresponding results
in four space-time dimensions. The relevant holomorphic and
homogeneous function in which the supergravity action is encoded takes
the form, 
\begin{equation}
  \label{eq:holo-function}
  F(Y,\Upsilon) = \frac{D_{IJK}Y^IY^JY^K + d_IY^I\,\Upsilon}{Y^0} \,.
\end{equation}
Here $Y^I$ and $Y^0$ are holomorphic variables related to the scalar
fields of the four-dimensional vector multiplets and $\Upsilon$ is the
lowest component of the square of the Weyl multiplet, all subject to
some uniform rescaling. The quantities $D_{IJK}$ and $d_I$ should be
identified with $C_{IJK}$ and $c_I$, up to suitable proportionality
factors. In terms of these variables the attractor equations read
\cite{Lopes Cardoso:1998wt,LopesCardoso:1999ur,LopesCardoso:2000qm}, 
\begin{equation}
  \label{eq:4D-attractor}
  Y^A-\bar Y^A= \mathrm{i} p^A\,, \qquad  F_A- \bar F_A  = \mathrm{i}
  q_A\,, 
\end{equation}
where the index $A$ denotes $A=0$ or $A=I$ and where the magnetic and
electric charges are denoted by $p^A$ and $q_A$, respectively.
Furthermore the BPS condition implies $\Upsilon=-64$.

It is well-known that the theory based on \eqref{eq:holo-function} is
invariant under the following symmetry transformations, which take the
form of electric/magnetic dualities (see, e.g. \cite{de Wit:1995zg}),
\begin{eqnarray}
  \label{eq:em-dual}
  Y^0 &\to& Y^0\,,\nonumber\\
  Y^I&\to&{} Y^I + k^IY^0\,, \nonumber\\
  F_0&\to&{} F_0 - k^IF_I - 3\,D_{IJK}k^Jk^K\,Y^I -D_{IJK} k^Ik^Jk^K
  \,Y^0\,, \nonumber \\
  F_I&\to&{} F_I + 6\, D_{IJK}k^J\,Y^K +3\, D_{IJK}k^Jk^K\,Y^0 \,,
\end{eqnarray}
where the parameters $k^I$ are real. In principle there could be other
dualities as well, depending on the specific form of the coefficients
$C_{IJK}$ and $c_I$. The electric and magnetic charges will exhibit similar
transformations,
\begin{eqnarray}
  \label{eq:em-dual-charges}
  p^0 &\to& p^0\,,\nonumber\\
  p^I&\to&{} p^I + k^Ip^0\,, \nonumber\\
  q_0&\to&{} q_0 - k^Iq_I - 3\,D_{IJK}k^Jk^K\,p^I -D_{IJK} k^Ik^Jk^K
  \,p^0\,, \nonumber \\
  q_I&\to&{} q_I + 6\, D_{IJK}k^J\,p^K  + 3\,D_{IJK}k^Jk^K\, p^0 \,. 
\end{eqnarray}

Parametrizing the $Y^A$ by $Y^A=\tfrac12(\phi^A+\mathrm{i}p^A)$, so
that the magnetic attractor equations of \eqref{eq:4D-attractor} are
satisfied, we obtain the following expressions for the entropy and the
remaining attractor equations for $p^I=0$ and $p^0\not=0$,
\begin{equation}
  \label{eq:entropy4}
  \mathcal{S}^\mathrm{BH}_\mathrm{4D}=\frac{2\pi p^0}{({\phi^0}^2
  +{p^0}^2)^2} \Big[{p^0}^2 
  D_{IJK} \phi^I\phi^J\phi^K +256\,d_I\phi^I {\phi^0}^2 \Big]\,,
\end{equation}
with
\begin{eqnarray}
  \label{eq:q+q4}
  q_I{}^\mathrm{4D}&=&{}- \frac{3\, p^0}{{\phi^0}^2 +{p^0}^2}
  \Big[ D_{IJK}\phi^J\phi^K -\frac{256}{3}\,d_I \Big]\,,\nonumber\\ 
  q_0{}^\mathrm{4D}&=& \frac{2\phi^0p^0}{({\phi^0}^2 +{p^0}^2)^2}
  \Big[D_{IJK} \phi^I\phi^J\phi^K - 256\,d_I\phi^I \Big]\,. 
\end{eqnarray}
The symmetry discussed in \eqref{eq:em-dual} and
\eqref{eq:em-dual-charges} is not manifest for the above result, in
view of the fact that we have fixed the $p^I$ to zero. However, it can
be used to find the corresponding expressions for non-zero charges
$p^I$.

Without giving a precise calibration between four- and
five-dimensional quantities (which is subtle in the presence of
higher-derivative couplings) it is clear that the four- and
five-dimensional expressions are not compatible upon absorbing
suitable normalization factors in the quantities involved. The
difference seems to reside exclusively in the attractor equations for
the electric charges $q_I$, in the terms proportional to $c_I$ induced
by the higher-derivative couplings. The expressions for entropy and
angular momentum agree assuming that the charge $q_0{}^\mathrm{4D}$ is
identified with $J_\psi$. These results are different from those
obtained in \cite{Castro:2007ci}, especially in the case of non-zero
angular momentum. For details, we refer to the discussion at the end
of this subsection.

To investigate some of the consequences of this discrepancy, we again
consider the attractor equations \eqref{eq:attractor-bh-q-J}, where we
rescale the coefficients $c_I$ in the attractor equations for $q_I$ by
$c_I\to \alpha\,c_I$ to account for the two different expressions.
Hence we set the parameter $\alpha= 1$ or $\tfrac43$, depending on
whether we consider $D=5$ or 4 space-time dimensions, respectively.

Subsequently we solve the attractor equations for $\phi^I$ and
$\phi^0$ to first order in $c_I$, keeping the charges constant. To do
this we first determine the solution for the case that $c_I=0$, 
\begin{eqnarray}
  \label{eq:cI=0}
  \hat \phi^I &\equiv& \frac{\phi^I}{\sqrt{{\phi^0}^2 + {p^0}^2}}
  \approx \frac{\hat q^I}{\sqrt{p^0}} + \mathcal{O}(c_I)\,,\nonumber\\ 
  \phi^0 &\approx& \frac{J_\psi \, {p^0}^2}{2 \sqrt{p^0 Q^3 -\tfrac14
  (p^0 J_\psi)^2}} + \mathcal{O}(c_I)\,,
\end{eqnarray}
where the $\hat q^I$ are defined by the requirement that they satisfy
the attractor equations in the limit of vanishing $c_I$. Therefore we
have, 
\begin{eqnarray}
  \label{eq:def-q-Q}
  q_I &=& 6\, C_{IJK}\,\hat q^J\hat q^K\,,\nonumber\\
  Q^{3/2} &=& 2\, C_{IJ}\,\hat q^J\hat q^K \,,\nonumber\\
  C_{IJ}&=& C_{IJK}\,\hat q^K\,. 
\end{eqnarray}
To first order in $c_I$ this result changes into, 
\begin{eqnarray}
  \label{eq:cInot0}
  \hat \phi^I & \approx& \frac{1}{\sqrt{p^0}}\left\{  \hat q^I+
  \frac{\alpha\,(p^0 Q^3- \tfrac14(p^0\,J_\psi)^2)}{32\,{p^0}^2
  \,Q^{3}} \; C^{IJ}\, c_J \right\}  + 
  \mathcal{O}(c_I{}^2)\,,\nonumber\\  
  \phi^0 &\approx& \frac{J_\psi \, {p^0}^2}{2 \sqrt{p^0 Q^3 -\tfrac14
  (p^0J_\psi)^2}} \left\{1- \frac{(3\,\alpha-8)\,c_I\,\hat q^I}{16\,p^0
  \,Q^{3/2}}\right\} + \mathcal{O}(c_I{}^2)\,,
\end{eqnarray}
where the matrix $C^{IJ}$ denotes the inverse of $C_{IJ}$.
Substituting these expressions into the entropy formula
\eqref{eq:wald-bh}, one obtains,
\begin{equation}
  \label{eq:order-c-entropy}
\mathcal{S}_\mathrm{macro}^\mathrm{BH}\approx 2\pi\,\sqrt{p^0
  Q^3 -\frac{1}{4}(p^0J_\psi)^2}\,\left( 1+\frac{3\alpha}{16}
  \frac{c_I\,\hat q^I}{p^0 Q^{3/2}} \right) + \mathcal{O}(c_I{}^2)\,.   
\end{equation}
We note that the terms proportional to $c_I$ are indeed subleading in
the limit of large charges. 

The expression \eqref{eq:order-c-entropy} can be confronted with
results from the literature. For the non-rotating case, where a direct
comparison with microscopic counting is possible, the above result
with $\alpha=1$ agrees with the results reported in \cite{Vafa:1997gr,
  Huang:2007sb}. For the rotating black hole, no analytic microscopic
results are available, but our results can be compared to the
supergravity results of \cite{Castro:2007ci, Castro:2008ne}. Here
there is a clear discrepancy originating from the different form of
the attractor equations \eqref{eq:angular-spinn-bh} for the electric
charges and the angular momentum, which reflects itself in a different
dependence on $J$ in (\ref{eq:order-c-entropy}). 
%Only for zero angular momentum our results can be matched upon making
%suitable redefinitions of the charges. Note that the entropy formula
%in terms of the moduli does agree with our expression
%\eqref{eq:wald-bh5}. 
The expression \eqref{eq:order-c-entropy} can also be compared to the
results of \cite{Guica:2005ig}, where the only higher-derivative
coupling included into the action was the Euler density. For zero
angular momentum one recovers the same relative factor for the
subleading correction between the four- and five-dimensional entropies
represented by the parameter $\alpha$ in \eqref{eq:order-c-entropy}. 
For finite angular momentum the subleading corrections determined by
\cite{Castro:2007ci, Castro:2008ne} and \cite{Guica:2005ig} are
mutually different and both fail to reproduce the expression
\eqref{eq:order-c-entropy}. As already mentioned in
footnote~\ref{footnote:N=4}   there exist microscopic results for
theories with 16 supercharges \cite{Castro:2008ys,Banerjee:2008ag},
which could possibly be connected to the results of this paper in
certain asymptotic limits. 

Irrespective of the discrepancies in the rotating case, our
five-dimensional results, as well as those in \cite{Vafa:1997gr,
  Huang:2007sb, Castro:2007ci, Castro:2008ne} disagree with a naive
uplift of the four-dimensional result.  Restricting ourselves to the
static case, it appears that the $\tfrac43$ discrepancy in the
attractor equation for the charges is ubiquitous. A possible origin of
these discrepancies follows from the observation that the actions used
in four and in five dimensions are not directly related by dimensional
reduction. Upon reduction the five-dimensional Weyl multiplet
decomposes into the four-dimensional Weyl multiplet and a
four-dimensional vector multiplet, as is obvious from table
\ref{tab:countWeyl}. Therefore we expect that the resulting
four-dimensional action will also contain special higher-derivative
couplings involving vector multiplets that have, so far, not been
considered in this context. Apparently these terms do not change the
expression for the Wald entropy expressed in the moduli. The latter is
in accord with a result of \cite{deWit:2006gn}, where the construction
of higher-derivative Lagrangians for tensor multiplets leads to
additional couplings to the Riemann tensor which all vanish in the BPS
limit, so that they will not contribute to the Wald entropy.  Assuming
that this phenomenon holds in the general case, then the higher-order
couplings of the vector multiplets may still generate new
contributions to the electric charges or to the angular momentum. At
present these couplings are not completely known so it is difficult to
check whether or not this is the correct explanation for
the discrepancy.  Observe that this phenomenon will not arise for
models with 16 supercharges, because in that case the five- and
four-dimensional Weyl multiplets carry the same number of degrees of
freedom. See appendix \ref{App:conf-sg} for details.

To further explore this difference between four and five space-time
dimensions, let us also consider the case of small black holes, whose
entropy depends sensitively on the higher-derivative couplings. We
assume $C_{1ab}= \eta_{ab}$ and $c_a=0$, which represents the typical
situation for $\mathrm{K3}\times T^2$ heterotic string
compactifications. From the attractor equations (including the
parameter $\alpha$ as before) we obtain
\begin{eqnarray}
  \label{eq:het-q}
  q_1&=& \frac{6\,p^0}{{\phi^0}^2+ {p^0}^2} \Big[\eta_{ab}\phi^a\phi^b 
  -\frac\alpha{16}\,c_1 \Big] \,,\nonumber\\
  q_a &=& \frac{12\,p^0\,\phi^1}{{\phi^0}^2+ {p^0}^2} \, \eta_{ab}\phi^b \,.
\end{eqnarray}
Using the above equations one easily derives,
\begin{eqnarray}
  \label{eq:entropy-Q-K3T2BH}
    \mathcal{S}_\mathrm{macro}^\mathrm{BH}&=&
    \frac{2\pi\,p^0\,\phi^1}{({\phi^0}^2+ {p^0}^2)^2} \, \Big[
     p^0q_1({\phi^0}^2+ {p^0}^2) + \frac12
    \Big(\frac{3\,\alpha}{4}{p^0}^2 +{\phi^0}^2\Big)c_1  \Big]  \,,\\[2mm]
    \eta^{ab}q_aq_b &=& \frac{3\,p^0\,(\phi^1)^2}{{\phi^0}^2+
    {p^0}^2} \Big[8\,q_1 + \frac{3\,\alpha\,p^0}{{p^0}^2+
    {\phi^0}^2} \,c_1 \Big]  \,.
\end{eqnarray}
Let us now set $q_1=0$, so that we are describing small black holes.
In that case one finds,
\begin{eqnarray}
  \label{eq:small-bh-entropy}
    \mathcal{S}_\mathrm{macro}^\mathrm{BH}=
    \frac{\pi}{4} \,\sqrt{\vert\alpha\,c_1\,\eta^{ab}q_aq_b \vert}
    \,\left\{ 1 +  \frac{4-3\,\alpha}{3\,\alpha} 
    \,\frac{{\phi^0}^2} {{\phi^0}^2+ {p^0}^2}\right\}\,,
\end{eqnarray}
where $\phi^0$ is related to the angular momentum according to
\begin{equation}
  \label{eq:J-phi-0}
  J_\psi= - \frac{\phi^0}{{\phi^0}^2 + {p^0}^2} \,
  \frac{4-3\,\alpha}{12\,\alpha} \,\sqrt{\vert
  \alpha\,c_1\,\eta^{ab}q_aq_b \vert}   \,. 
\end{equation}
For non-zero angular momentum this last relation does not allow a
uniform rescaling of the charges in the way indicated before. For zero
angular momentum only the first term in \eqref{eq:small-bh-entropy}
contributes. In that case the entropy coincides with the
four-dimensional result for small black holes, except for an overall
relative factor equal to $\sqrt{4/3}$ induced by the
$\alpha$-dependence. This result was already discussed in
\cite{Huang:2007sb} where exact expressions for microscopic
degeneracies of small static black holes in five space-time dimensions
were derived. In this work it was found that the asymptotics of the
entropy of the small black holes in five dimensions is the same as in
four, with the same normalization. To resolve this puzzle it might
perhaps be helpful to also have microscopic results for non-zero
angular momentum, so that one has a more detailed test for
\eqref{eq:small-bh-entropy}. However, such results are quite difficult
to obtain. As is well known, in four space-time dimensions the
sub-leading contribution to the entropy of small black holes is
problematic in the supergravity description, but the leading
contribution is in perfect agreement with microstate
counting arguments. The five-dimensional result thus poses a puzzle in
this respect. 

It is also worth mentioning that, when $J_\psi\not=0$, the
resulting values of the four-dimensional charges will not correspond
to four-dimensional small black holes. Indeed, when setting
$\alpha=\tfrac43$ in \eqref{eq:J-phi-0}, we find $J_\psi=0$, so that
we are dealing with vanishing $q_0, p^1, p^a$, which do characterize a
small black hole in four dimensions. Hence the situation remains
unsatisfactory. 

%%%%%%%%%%%%%%%%%%%%%%%%%%%%%%%%%%%%%%%%%%%%%%%%%%%%%%%%%
\section{BPS black rings}
\label{sec:entropy-black-rings}
\setcounter{equation}{0}
%%%%%%%%%%%%%%%%%%%%%%%%%%%%%%%%%%%%%%%%%%%%%%%%%%%%%%%%%
In this final section we turn to the black rings, for which the
relevant Noether potential has been derived in section
\ref{sec:entropy-angular-momentum}. In particular we refer to the
treatment of the mixed Chern-Simons term in subsection
\ref{sec:an-alternative-form}, which is crucial for the black ring. In
this section we discuss the resulting expressions for the entropy, and
for the charges and angular momenta, which are then confronted with
results from the literature. As we shall see, the actual evaluation
still involves a number of non-trivial issues related to the
integration over the spacelike section $\Sigma$ of the horizon.

The relevant Noether potential consists of \eqref{eq:N-pot-0} combined
with the contributions from the Chern-Simons terms that can be
extracted from \eqref{eq:vareps-Q-CS} and \eqref{eq:alt-mixed-Npot}.
Using that $T_{01}=0$ for the black ring, it is easy to see that
\eqref{eq:N-pot-0} gives rise to the following contribution,
\begin{equation}
  \label{eq:N-pot-0-ring}
  8\pi^2\,\varepsilon_{\mu\nu}\mathcal{Q}_0^{\mu\nu} = 
  -2\,\varepsilon_{01}\left[ 
    C(\sigma)  + 3\,  c_I\sigma^I \,T_{23}{}^2
  \right]  \nabla_{[0}\xi_{1]} \,.
\end{equation} 
Subsequently we add the contributions from \eqref{eq:alt-mixed-Npot},
together with the first term in \eqref{eq:vareps-Q-CS} that originates
from the $W\wedge F\wedge F$ Chern-Simons term,
\begin{eqnarray} 
\label{eq:alt-mixed-Npot-WFF}
  8\pi^2 \mathcal{Q}^{\mu\nu}_\mathrm{CS} &=& {}\tfrac1{2}
   \mathrm{i}\,e^{-1}\varepsilon^{\mu\nu\rho\sigma\tau}\,
   C_{IJK} \,\xi^\lambda W_\lambda{}^I  W_\rho{}^J  F_{\sigma\tau}{}^K
   \nonumber\\ 
   &&{}
  - \tfrac1{32}\mathrm{i}e^{-1}  
  \,\varepsilon^{\mu\nu\rho\sigma\tau}\,c_I F_{\rho\sigma}{}^I
  \omega_\tau{}^{ab}   
  \left[\varepsilon_{ab} -\tfrac12 \xi^\kappa\omega_{\kappa\,ab}
  \right]\nonumber\\
  &&{}
  +\tfrac1{32}\mathrm{i}e^{-1} 
  \,\varepsilon^{\mu\nu\rho\sigma\tau}\,c_I\,\xi^\lambda W_\lambda{}^I
  \omega_\rho{}^{ab} \left(\partial_\sigma\omega_{\tau ab} -\tfrac23 
  \omega_{\sigma ac} \,\omega_\tau{}^c{}_b \right)
   \nonumber \\
   &&{}
   - \tfrac1{32} \mathrm{i}\,e^{-1}
   \varepsilon^{\rho\sigma\tau\lambda[\mu} \,c_I F_{\rho\sigma}{}^I \,
   \mathcal{R}_{\tau\lambda}{}^{\nu]\kappa}\, \xi_\kappa \nonumber\\
   &&{}
    + \tfrac1{64} \mathrm{i}\,e^{-1}
  \,\varepsilon^{\rho\sigma\tau\lambda\kappa} \,c_I
   F_{\rho\sigma}{}^I \, \mathcal{R}_{\tau\lambda}{}^{\mu\nu}\,
   \xi_\kappa   \,.
\end{eqnarray} 
Observe that the last two terms in \eqref{eq:alt-mixed-Npot-WFF} have
already been evaluated in \eqref{eq:vareps-Q-CS}. The third term of
\eqref{eq:alt-mixed-Npot-WFF} vanishes as can be readily deduced from
\eqref{eq:W-tensor-potential-attractor}.  Straightforwardly combining
the various contributions gives rise to the following additional
contribution to the Noether potential,
\begin{eqnarray}
  \label{eq:vareps-Q-CS-ring-comb}
  8\pi^2\,\varepsilon_{\mu\nu}\mathcal{Q}_\mathrm{CS}^{\mu\nu} &=&
  2\,\varepsilon_{01}\, T_{23}\left[4\, C_{IJK}\, \sigma^I W_5{}^J
  \,\xi^\lambda W_\lambda{}^K  - c_I \sigma^I\,T_{23}{}^2
  \,\xi_5\right] \nonumber\\
   &&{}
  -2\,\varepsilon_{01}\,c_I\sigma^I\,T_{23}{}^2\left[\nabla_{[0}\xi_{1]}
  -\tfrac12 \xi^\lambda \omega_{\lambda 01} \right] \,,
\end{eqnarray}
where we have used that $\omega_5{}^{ab}$ vanishes with the exception of
$\omega_5{}^{01}= -2\,T_{23}$. 

From \eqref{eq:vareps-Q-CS-ring-comb} we directly determine the
expression for the entropy, which coincides with the corresponding
expression \eqref{eq:wald-bh5} for the black hole, 
\begin{equation} 
  \label{eq:wald-BR-entro} 
  \mathcal{S}^\mathrm{BR}_\mathrm{macro}=  \frac{\pi\, 
  \mathrm{e}^g}{4\, v^2}  
  \big[ C(\sigma) + 4\,c_I\sigma^I\, T_{23}{}^2\big] \,. 
\end{equation} 
Observe that, in order to obtain this result, it was crucial to use
the alternative form of the Noether potential derived in subsection
\ref{sec:an-alternative-form}.  Naive application of the Noether
potential that was used earlier for the black hole, will yield a
different result. In any case, we should stress that the mixed
Chern-Simons term contributes to both the black hole and the black
string entropy. This raises some question about the derivation
presented in \cite{Guica:2005ig} where the Chern-Simons term was not
taken into account.

To obtain the expression \eqref{eq:wald-BR-entro} we had to integrate
over the horizon, which, in the case at hand, was straightforward.
However, to determine the electric charges and the angular momenta,
one is confronted with an integration of terms that depend explicitly
on gauge fields that are not globally defined. To perform the integral
one therefore has to make use of patches, as was already explained in
section \ref{sec:gauge-fields}, in such a way that the result will be
invariant under `small' gauge transformations continuously connected
to the identity. The precise procedure for doing this has already been
proposed in \cite{Hanaki:2007mb}, and we will adopt it here.

We thus define two coordinate patches on the $S^1\times S^2$ spacelike
cross section $\Sigma$ of the horizon. As we shall discuss in due
time, these patches have to be also defined away from $\Sigma$, but
for the moment we restrict our attention to $\Sigma$ itself. One patch
contains the north pole {\sf N} of the $S^2$ factor. It is
parametrized by $-1+\epsilon\leq\cos\theta\leq 1$, $0\leq\varphi<
2\pi$ and $0\leq\psi< 4\pi$. This patch has the topology of a solid
two-torus.  The second patch, which has the same topology, contains
the south pole {\sf S} of the $S^2$ factor, and is parametrized by
$-1\leq\cos\theta\leq -1+\epsilon$, $0\leq\varphi< 2\pi$ and
$0\leq\psi< 4\pi$.  The boundary of these two patches is a two-torus
defined by $\cos\theta=-1+\epsilon$, where the parameter $\epsilon$
will be taken to zero at the end of the calculation.  On these patches
we define the gauge fields, $W_\mu^{{\sf N}\,I}$ and $W_\mu^{{\sf
    S}\,I}$, respectively, which are related by gauge transformations
$\beta^I$.  These gauge transformations move the Dirac brane
singularities from the south to the north pole in a way that involves
the ring coordinate $\psi$, as was already described in subsection
\ref{sec:gauge-fields} (in particular, see
\eqref{eq:A-I-ringsections}). Hence,
\begin{eqnarray} 
  \label{eq:S-N-ringsection} 
  W_\mu^{{\sf N}\,I} \,\mathrm{d}x^\mu &=& 
  - p^I\left[\cos\theta \, \mathrm{d}\varphi 
  - \mathrm{d}(\varphi+\tfrac12\psi) \right] + 
  a^I \mathrm{d}\psi \,, \nonumber \\[2mm]
  W_\mu^{{\sf S}\,I} &=& W_\mu^{{\sf N}\,I}  + \beta_\mu{}^I\,,\qquad 
 \beta_\mu{}^I\mathrm{d}x^\mu=- 2\,p^I  \mathrm{d}(\varphi+ \tfrac12\psi)\,. 
\end{eqnarray}
Integrals over the spacelike cross section $\Sigma$ of the horizon, are
now decomposed into integrals over the sections {\sf N} and {\sf S} and
an additional integral over the boundary of the coordinate patches
that involves the gauge transformations $\beta^I$. This last term must
restore the gauge invariance of the integral under small gauge
transformations \cite{Hanaki:2007mb}. The limit $\epsilon\downarrow 0$
is taken for convenience, so that the contribution from the section
{\sf S} will vanish, and the contribution from {\sf N} will cover the
whole horizon with the exception of the singular points related to the
position of the Dirac brane.

Let us first consider the attractor equations for the electric charges
$q_I$. From the evaluation of the charges for the black hole (c.f.
\eqref{eq:q-attractor}) it is clear that the only contribution
originates from the $C_{IJK}\,W^I\wedge F^J\wedge F^K$ Chern-Simons
term, since all other contributions vanish when $T_{01}=0$.  Therefore
we focus directly on the Chern-Simons term, which requires to evaluate
the integral of $C_{IJK}\,W^J\wedge F^K$ over the spacelike cross
section $\Sigma$ of the horizon. According to the prescription
specified above, this integral is evaluated as follows,
\begin{eqnarray}
  \label{eq:1CS-charge}
  \int_\Sigma\,C_{IJK}\,W^J\wedge F^K &=&
  \int_{\sf N}\,C_{IJK}\,W^{{\sf N}\,J}\wedge F^K +
  \int_{\sf S}\,C_{IJK}\,W^{{\sf S}\,J}\wedge F^K \nonumber\\
  &&{} 
  + 2 \int_{\partial{\sf N}}\,C_{IJK}\,W^{{\sf N}\,J} \wedge \beta^K
  \,, 
\end{eqnarray}
where the factor 2 arises because $F^I= 2\,\mathrm{d}W^I$. In the
limit $\epsilon\downarrow0$, the second integral vanishes. The third
integral extends over the boundary, $\partial{\sf N}= -\partial{\sf
  S}$, of the two sections. Now, observe that $W^{{\sf N}\,J}\wedge
F^K$ is proportional to $(a^J+\ft12 p^J) p^K\, \mathrm{d}\theta
\wedge\mathrm{d}\varphi\wedge\mathrm{d}\psi$, while $W^{{\sf N}\,J}
\wedge \beta^K$ is proportional to $(a^J-\ft12(1-\epsilon) p^J) p^K\,
\mathrm{d}\varphi\wedge\mathrm{d}\psi$. As it turns out, the
contributions proportional to $p^Jp^K$ from the first and the second
integral cancel (in the limit $\epsilon\downarrow0$), whereas the
terms proportional to $a^Jp^K$ add. This confirms the conclusion below
\eqref{eq:CS-charge} that the Chern-Simons charge should be
proportional to $C_{IJK}\,a^Jp^K$. From comparison with
\eqref{eq:q-attractor}, one then easily determines the expression for
the electric charges by substituting $[W_\psi{}^K]= 2\,a^K$. The
attractor equations for the black ring charges are therefore
summarized by
\begin{equation} 
  \label{eq:attractor-ring} 
  q_I = - \frac3{T_{23}}  C_{IJK}\,\sigma^J a^K  \,,\qquad 
    p^0= 0\,, \qquad 
  p^I= \frac{\sigma^I}{4\,T_{23}} \,.
\end{equation} 

It is important to realize that the prescription of
\cite{Hanaki:2007mb} is based on the fact that $\mathrm{d} [W^J\wedge
F^K] = \tfrac12 F^J\wedge F^K$ is gauge invariant. Upon extending the
patches outside the horizon, we may calculate $F^J\wedge F^K$ over a
four-dimensional manifold by extending the radial coordinate $r$,
which can then be expressed as an integral over its three-dimensional
boundary. This is the justification for the prescription
\eqref{eq:1CS-charge}, as $\Sigma$ constitutes (part of) this
boundary.  However, we have simply ignored that the gauge fields must
in principle be extendable outside the horizon in the two patches, and
in the above calculation this feature does not seem to play a role as
we obtain a result that is invariant under small gauge
transformations.  Indeed, one can repeat the calculation without any
difficulty for a different choice of coordinate patches, such as, for
instance, defined by $\cos\theta_0\leq \cos\theta\leq 1$ for the {\sf
  N} patch and $-1\leq \cos\theta\leq \cos\theta_0$ for the {\sf S}
patch, so that the boundary is located at $\theta =\theta_0$. As it
turns out the final result will not depend on $\theta_0$ and simply
remains the same.

However, the situation is different when considering the evaluation of
the angular momenta and we shall see that the extension of the
sections away from $\Sigma$ will become an issue. The expression
for the angular momenta follows from the Noether potential
\eqref{eq:vareps-Q-CS-ring-comb}, which is again not gauge invariant so that
the integral is again subtle.  The troublesome term is the first one,
depending on $W_5{}^J$, which originates form the $W\wedge F\wedge F$
Chern-Simons term shown in the first line of
\eqref{eq:alt-mixed-Npot-WFF}. This term leads to
\begin{equation}
  \label{eq:CS-WWW}
  8\pi^2\,\varepsilon_{\mu\nu}\mathcal{Q}_\mathrm{CS}^{\mu\nu} =
  \varepsilon_{01} \, \varepsilon^{\mu\nu\rho}
  \,C_{IJK}\, \xi^\lambda W_\lambda{}^I\,W_\mu{}^JF_{\nu\rho}{}^K +
  \cdots\,,
\end{equation}
where the dots denote the remaining gauge invariant contributions in
\eqref{eq:vareps-Q-CS-ring-comb}, which can be evaluated
straightforwardly. Note that, unlike as on previous occasions, we
converted the above expression to a density over $\Sigma$, so that its
integration will require only the surface element
$\mathrm{d}\psi\wedge \mathrm{d}\varphi\wedge \mathrm{d}\theta$.

In order that the integral over $\Sigma$ of \eqref{eq:CS-WWW} is
amenable to the same prescription as used above, it is important that
$\Sigma$ and the gauge potentials are invariant under the isometries
associated with linear combinations,
$\xi^\psi\partial_\psi+\xi^\varphi\partial_\varphi$, of the two
Killing vectors associated with rotations over the angles $\psi$ and
$\varphi$. One then observes that $\mathrm{d}[(\xi\cdot W)\,W\wedge
F]$ can be written as a linear combination of two terms. One is the
contraction of the Killing vector with the five-form $W\wedge F\wedge
F$ whose integral must vanish for symmetry reasons. The second term
equals $(\xi\cdot W)\,F\wedge F$, which changes by a total derivative
under gauge transformations, again because the gauge fields are
invariant under the symmetry associated with the Killing vector.
Hence the integral over the four-dimensional manifold is invariant
under small gauge transformations, and, just as before, the integral
of \eqref{eq:CS-WWW} over its boundary $\Sigma$ can be decomposed into
integrals over the patches {\sf N} and {\sf S} and an additional
integral over the boundary $\partial{\sf N}$ of
\begin{eqnarray}
  \label{eq:boundary-patches}
  &&{}
  \varepsilon^{\mu\nu\rho}
  \,C_{IJK}[\xi^\lambda W_\lambda{}^{{\sf N}\,I}\,W_\mu{}^{{\sf
  N}\,J} -\xi^\lambda W_\lambda{}^{{\sf S}I}\,W_\mu{}^{{\sf S}\,J}]
  F_{\nu\rho}{}^K = \nonumber\\
  &&{}\qquad
  + \varepsilon^{\mu\nu\rho}\,C_{IJK}\,\partial_\mu
  \left[ \xi^\lambda W_\lambda{}^{{\sf S}\,I}\,\beta_\nu{}^{J} W_\rho{}^{{\sf
  S}\,K}-2\, \xi^\lambda\beta_\lambda{}^I\, \beta_\nu{}^J W_\rho{}^{{\sf
  S}\,K}  \right] \nonumber \\
  &&{}\qquad 
  -\tfrac32
  \varepsilon^{\mu\nu\rho}\,C_{IJK}\,\xi^\lambda\beta_\lambda{}^I
  \,W_\mu{}^{{\sf S}\,J} F_{\nu\rho}{}^K\,.
\end{eqnarray}
Here we insisted in writing the last two lines in terms of sections
$W_\mu{}^{{\sf S}\,I}$, which are well defined at the south pole.
Therefore, when writing the last term as a surface term over
$\xi^\lambda\beta_\lambda{}^I\,W_\psi{}^{{\sf S}\,J} W_\varphi{}^{{\sf
    S}\,K}$, its contribution will vanish in the limit
$\epsilon\downarrow0$ because $W_\varphi{}^{{\sf S}\,K}$ vanishes at
the south pole.

Combining the results above the integral of \eqref{eq:CS-WWW} over
$\Sigma$ can therefore be written as,
\begin{eqnarray}
  \label{eq:int-CS-WWW}
  \int_\Sigma
  \varepsilon_{\mu\nu}\mathcal{Q}_\mathrm{CS}^{\mu\nu} &=&
  \frac{\varepsilon_{01}}{4\pi^2} 
  \int \;\mathrm{d}\theta\,\mathrm{d}\varphi\,\mathrm{d}\psi 
  \,C_{IJK}\, \xi^\lambda W_\lambda{}^{{\sf N}\,I}\,W_\psi{}^{{\sf
  N}\,J} F_{\theta\varphi}{}^K\nonumber \\
  &&{}
  +\frac{\varepsilon_{01}}{4\pi^2} \, 
  \int \;\mathrm{d}\varphi\,\mathrm{d}\psi 
  \,C_{IJK}\, [\beta_\varphi^I W_\psi{}^{{\sf S}\,J} -\beta_\psi^I
  W_\varphi{}^{{\sf S}\,J} ] \, (\tfrac12
  \xi^\lambda W_\lambda{}^{{\sf S}\,K}-\xi^\lambda\beta_\lambda{}^K
  )\Big\vert_{\theta=\pi}\,.\nonumber\\  
  &&{~}
\end{eqnarray}
For both of these integrals the limit $\epsilon\downarrow0$ can
be taken without difficulty, so that the first one extends over the
whole horizon section $\Sigma$ and the second one over the boundary of
the sections on the horizon. A straightforward calculation then leads
to $12\, C_{IJK} p^Ip^J(a^K -\tfrac16 p^K)$ and $6\, C_{IJK} p^I( a^J
a^K + a^J p^K -\tfrac1{12} p^Jp^K)$, for $J_\varphi$ and $J_\psi$,
respectively.

The same calculation can be repeated for a different choice of the
patches, namely such that, in the limit $\epsilon\downarrow0$, the
{\sf S} patch will cover the whole horizon area $\Sigma$ and the
overlap of the {\sf N} patch will shrink to the north pole. This
requires to re-evaluate \eqref{eq:boundary-patches}, but up to a few
signs the calculations proceeds in the same way. However, now the
result is {\it not} the same, and one finds instead, $- 12\, C_{IJK}
p^Ip^J(a^K +\tfrac16 p^K)$ and $6\, C_{IJK} p^I( a^J a^K - a^J p^K
-\tfrac1{12} p^Jp^K)$, for $J_\varphi$ and $J_\psi$, respectively.
The reason for this discrepancy resides in the last term in
\eqref{eq:boundary-patches}, which we dropped because it does not
contribute at the south pole of the horizon.

However, one must verify whether there is no obstruction away from the
horizon. If one assumes that the south poles are directed to the
outward part of the ring, extending all the way to spatial infinity as
in \cite{Emparan:2006mm}, one expects an obstruction which will result
in an extra contribution from the integral at spatial infinity.  On
the other hand, for the inner region of the ring which contains the
north poles, there is obviously no obstruction, so that the second
result will be valid. In case the south poles are directed to the
inward part of the ring, it is the first result that would be valid.
In other words, a minimal understanding of the topological embedding
of the near-horizon region in the global solution is essential in
order to distinguish between the two prescriptions. It is possible
that only one embedding leads to a solution that is globally BPS, in
line with what was found in \cite{Goldstein:2008fq}. For a space that
is asymptotically flat, both embeddings seem possible and lead to two
inequivalent BPS solutions.

In light of the above we adopt the second result, which must be combined
with the contributions from \eqref{eq:vareps-Q-CS-ring-comb}. Then we
obtain the following result for the two independent angular momenta,
associated with the two independent rotations of the ring in
orthogonal planes,
\begin{eqnarray} 
  \label{eq:JJ-ring} 
  J_\varphi &=& -12\, C_{IJ}p^I( a^J+ \tfrac16p^J)\, \nonumber \\ 
  J_\psi-J_\varphi&=& -\frac{\mathrm{e}^{2g}}{2\,T_{23}} \left[C(\sigma)
  + 4\,c_I\sigma^I \,  T_{23}^2\right]
+ 6\,C_{IJ}(a^I + \tfrac12 p^I)(a^J +\tfrac12 p^J)\,, 
\end{eqnarray}  
where $C^{IJ}$ is the inverse of $C_{IJK} p^K$.

The above results are all invariant under scale transformations, as
they should. Note that the Wilson line moduli $a^I$ are scale
invariant. As in the case of black holes, we introduce a scale
invariant variable,
\begin{equation} 
  \label{eq:scale-inv-ring-var} 
  \phi^0 = \frac{\mathrm{e}^{-g}}{4\, T_{23}}\, , 
\end{equation} 
so that the above expressions for the entropy and the electric charges
take a manifestly scale invariant form,  
\begin{eqnarray}
\label{eq:en-q-ring} 
 \mathcal{S}^\mathrm{BR}_\mathrm{macro}&=&\frac{4\pi}{\phi^0}
  \Big[C_{IJK}\, p^Ip^Jp^K + \frac14 c_Ip^I \Big]\,, \nonumber\\[2mm]
  {}
  q_I &=& - 12\, C_{IJK}\,p^J a^K\,.
\end{eqnarray} 
The angular momenta can be expressed as follows,
\begin{eqnarray}
  \label{eq:JJ-ring2} 
  J_\psi-J_\varphi - \frac1{24}C^{IJ}(q_I - 6\, C_{IK} p^K)(q_J - 6\, C_{JL}
  p^L) 
  &=& -\frac2{{\phi^0}^2}\Big[C_{IJK}\, p^Ip^Jp^K  + \frac14\,c_Ip^I
  \Big] \,,\nonumber\\[3mm]
  J_\varphi &=& {} 
  p^I( q_I -\tfrac16 C_{IJ}p^J) \,. 
\end{eqnarray}
The choice of the linear combination of the angular momenta in the
first term is motivated by the explicit dimensional reduction of the
known two-derivative solution \cite{Elvang:2005sa}, which showed that
the rotation of the four-dimensional black hole cannot be identified
with a rotation of the $S^2$ of the black ring but necessarily
involves also a rotation along the ring. Likewise the dimensional
reduction is over a circle generated by a simultaneous rotation
around the ring and of the $S^2$. The corresponding generator equals
the linear combination of two angular momenta, $J_\psi-J_\varphi$, which
therefore corresponds to the charge associated with the Kaluza-Klein
photon. Hence we introduce a modified charge $\hat q_0$ in the usual
fashion,
\begin{equation}
  \label{eq:q-hat-0}
  \hat q_0=  J_\psi-J_\varphi - \frac1{24}C^{IJ}(q_I - 6\, C_{IK}
  p^K)(q_J - 6\, C_{JL} p^L)\,.
\end{equation}
This expression coincides precisely with the one presented in
\cite{Bena:2005ae}. 

With this definition the entropy takes its familiar form
\cite{Maldacena:1997de,Lopes Cardoso:1998wt},
\begin{equation}
  \label{eq:br-entropy}
   \mathcal{S}^\mathrm{BR}_\mathrm{macro}=2\pi\, \sqrt{\vert 2\,\hat
   q_0(C_{IJK}\, p^Ip^Jp^K + \tfrac14 c_Ip^I)\vert} \,,  
\end{equation}
This result for the corrected entropy agrees with the microscopic
counting of \cite{Bena:2004tk,Cyrier:2004hj}. However, those results
do not yet include the contribution from the higher-derivative
couplings.  As we shall briefly review below, \eqref{eq:br-entropy}
takes the same form as the entropy for a corresponding
four-dimensional black hole. Namely, it is proportional to
$\sqrt{c_L\,\hat q_0}$, where $c_L$ is the relevant central charge of
an underlying $(4,0)$ superconformal field theory that arises when
wrapping the M-theory five-brane over a cycle $S^1\times P_4$, where
$P_4$ is a holomorphic four-cycle of a Calabi-Yau manifold
\cite{Maldacena:1997de}. The modified momentum along the $S^1$ is
denoted by $\hat q_0$. The modification is due to the presence of
membrane charges. The subleading contributions are associated with the
second Chern class of the Calabi-Yau manifold, and on the field-theory
side this induces the higher-derivative couplings \cite{Lopes
  Cardoso:1998wt}. Without these subleading corrections, results for
other than Calabi-Yau compactifications have been obtained in
\cite{Bertolini:2000yaa}. The above results are generally in line with
the AdS/CFT results for the black ring attractors
\cite{Kraus:2005vz,Kraus:2005zm,Kraus:2006wn}.

Let us now confront the above results in more detail with the corresponding
results in four space-time dimensions, again based on the function
\eqref{eq:holo-function}. Hence we are dealing with a black hole with
$p^0=0$, which leads to
\begin{equation} 
  \label{eq:entropy2} 
  \mathcal{S}^\mathrm{BH}_\mathrm{4D}= -\frac{2\pi}{\phi^0} \Big[D_{IJK}\, 
  p^Ip^Jp^K  + 256\, d_Ip^I \Big]\,,  
\end{equation} 
with 
\begin{eqnarray} 
  \label{eq:q+q2} 
  q_I{}^\mathrm{4D}&=&{} \frac{6}{\phi^0}  D_{IJK}\,p^J\phi^K 
  \,,\nonumber\\  
  \hat q_0{}^{4D}\equiv q_0{}^{4D}+\tfrac1{12} D^{IJ} q_Iq_J &=& 
  \frac1{{\phi^0}^2} \Big[ D_{IJK}\, p^Ip^Jp^K  +256\,d_Ip^I \Big]\,,  
\end{eqnarray} 
where $D^{IJ}$ is the inverse of $D_{IJK}p^K$. Just as before this
gives rise to 
\begin{equation}
  \label{eq:bh-entropy-4}
   \mathcal{S}^\mathrm{BH}_\mathrm{macro}=2\pi\, \sqrt{\vert\hat
   q_0(D_{IJK}\, p^Ip^Jp^K + 256\,d_I p^I)\vert} \,.
\end{equation}
As the reader can easily verify, the expressions for $\hat q_0$ and
for the entropy are invariant under the transformations
\eqref{eq:em-dual-charges} with $p^0=0$. Also the expression for the
charges $q_I{}^\mathrm{4D}$ is consistent with this symmetry as it
acts on $\phi^I$ according to $\phi^I\to \phi^I +k^I\,\phi^0$. The
latter follows straightforwardly from \eqref{eq:em-dual}.

The same transformations can be considered in the five-dimensional
case. In five dimensions there is no electric/magnetic duality but
there is spectral flow \cite{Bena:2005ni}, giving rise to the same
transformations, upon replacing $D_{IJK}$ by $-2\,C_{IJK}$. These
transformations are precisely generated by integer shift of the Wilson
line moduli, $a^I\to a^I+k^I$. Observe that the angular momenta will
also transform under these shifts, and we find the following results,
\begin{eqnarray}
  \label{eq:q-J-spectral-flow}
   q_I&\to& q_I -12\,C_{IJK} p^Jk^K\,,\nonumber\\
  J_\varphi&\to& J_\varphi -12\,C_{IJK}p^Ip^Jk^K\,,\nonumber\\ 
  J_\psi&\to& 
  J_\psi-q_Ik^I - 6\,C_{IJK}p^Ip^Jk^K + 6\,C_{IJK}p^Ik^Jk^K \,,
\end{eqnarray}
This shows that $\hat q_0$ remains invariant.

The difference between \eqref{eq:q+q2} and \eqref{eq:q-hat-0} resides
in the shifts of the electric charges proportional to $C_{IJK}p^Jp^K$.
The presence of these shifts is consistent with many previous results,
both from field theoretic solutions and from microstate counting
\cite{Gauntlett:2004qy,Bena:2004tk,Cyrier:2004hj,Bena:2004de,Elvang:2005sa,
  Bena:2005ni,Bena:2005ae,Hanaki:2007mb}. The modified charges $q_I
-6\,C_{IJK}p^J p^K$ in \eqref{eq:JJ-ring2} are additive. This follows
from a calculation similar to the one leading to the attractor
equation for $q_I$, but now for a configuration of concentric rings.
Such a calculation has been performed in \cite{Hanaki:2007mb} and
resulted in the equations \eqref{eq:CS-charge-multi} and
\eqref{eq:CS-charge-multi-rewrite} that we discussed earlier. When
combined with the attractor equation for $q_I$ shown in
\eqref{eq:en-q-ring}, they establish the additivity of the shifted
charges. The latter is manifest in the results of
\cite{Gauntlett:2004qy,Bena:2005ni}. The modified charges should
therefore be used in the microscopic formula of
\cite{Maldacena:1997de} to match with the macroscopic result
\eqref{eq:JJ-ring2}, as was already emphasized in \cite{Bena:2004tk,
  Bena:2005ae, Bena:2005ni}.  Note, however, that in spite of the
qualitative agreement of these conclusions, we should stress that we
have adopted a different definition of the electric charges $q_I$,
which is not based on the asymptotic fall-off of the electric fields
at spatial infinity. Therefore the modified charges should be the
same, but the electric charges may still be different.

The shifts in the electric charges cannot be removed in the
four-dimensional results by a suitable duality transformation of the
form \eqref{eq:em-dual-charges}, because that transformation induces
shifts that are twice as large. The shifts are related to the terms
$\pm \tfrac{1}2 p^I\mathrm{d}\psi$ in the gauge field sections in
\eqref{eq:A-I-ringsections}. From the point of view of subsection
\ref{sec:gauge-fields}, they arise due to the non-trivial topology of
the full five-dimensional space-time.  Therefore the four-dimensional
black hole should be compared to the reduction of an infinite magnetic
string in five dimensions, which is topologically trivial. In that
case, both the terms $\pm \tfrac{1}2p^I\mathrm{d}\psi$ in
\eqref{eq:A-I-ringsections} and the shifts in the electric charges in
\eqref{eq:JJ-ring} will be absent, so that one obtains full agreement
with the four-dimensional attractor results.

%%%%%%%%%%%%%%%%%%%%%%%%%%%%%%%%%%%%%%%%%%%%%%%%%%%%%%%%%%%%%%%%%
\subsection*{Acknowledgments}
It is a pleasure to acknowledge valuable discussions with Gabriel
Cardoso, Jan de Boer, Kevin Goldstein, Kentaro Hanaki, Albrecht Klemm,
Per Kraus, Finn Larsen, Thomas Mohaupt, George Papadopoulos, Yuji
Tachikawa, Paul Townsend and Stefan Vandoren. B.d.W.  thanks the
\'Ecole Normale Sup\'erieure in Paris, where part of this work was
carried out, for hospitality and the Centre National de la Recherche
Scientifique (CNRS) for financial support. The work of S.K.  is part
of the research program of the `Stichting voor Fundamenteel Onderzoek
der Materie (FOM)', which is financially supported by the `Nederlandse
Organisatie voor Wetenschappelijk Onderzoek (NWO)'. This work has been
partly supported by EU contracts MRTN-CT-2004-005104 and
MRTN-CT-2004-512194, and by NWO grant 047017015.
%%%%%%%%%%%%%%%%%%%%%%%%%%%%%%%%%%%%%%%%%%%%%%%%%%%%%%%%%%%%%%%%%

%%%%%%%%%%%%%%%%%%%%%%%%%%%%%%%%%%%%%%%%%%%%%%%%%%%%%%%%%%%%%%%%%%%%
%%%%%%%%%%%%%%%%%%%%%%%%%%%%%%%%%%%%%%%%%%%%%%%%%%%%%%%%%%%%%%%%%%%%
\begin{appendix}
%%%%%%%%%%%%%%%%%%%%%%%%%%%%%%%%%%%%%%%%%%%%%%%%%%%%%%%%%%%%%%%%%%%%
%%
%%%%%%%%%%%%%%%%%%%%%%%%%%%%%%%%%%%%%%%%%%%%%%%%%%%%%%%%%%%%%%%%%%%%
\section{Conventions}
\label{App:5D-conv}
\setcounter{equation}{0}
%%%%%%%%%%%%%%%%%%%%%%%%%%%%%%%%%%%%%%%%%%%%%%%%%%%%%%%%%%%%%%%%%%%%XS
In the first part of this paper, especially when dealing with spinors,
we use Pauli-K\"all\'en conventions.  Five-dimensional world and
tangent-space indices are denoted by $\mu,\nu,\ldots$ and
$a,b,\ldots$, respectively, and take the values $1,2,\ldots,5$. We
employ hermitean 4-by-4 gamma matrices $\gamma_a$, which satisfy
\begin{eqnarray} 
  \label{eq:spinor-conv} 
  C\gamma_a C^{-1} &=& \gamma_a{}^{\rm T}\,,\qquad C^{\rm T} =
  -C\nonumber\,, \qquad 
  C^\dagger = C^{-1} \,, \\ 
  \gamma_{abcde} &=& {\bf 1}\,\varepsilon_{abcde} \,.
\end{eqnarray}  
Here $C$ denotes the charge-conjugation matrix and gamma matrices with
$k$ multiple indices denote the fully antisymmetrized product of $k$
gamma matrices in the usual fashion, so that we have, for instance,
$\gamma_a\,\gamma_b={\bf 1} \,\delta_{ab} + \gamma_{ab}$. In view of
the last equation of \eqref{eq:spinor-conv}, gamma matrices with more
than two multiple indices are not independent, and can be decomposed
into the unit matrix, $\gamma_a$ and $\gamma_{ab}$. Note that $C$,
$C\gamma_a$ and $C\gamma_{ab}$ constitute a complete basis of 6
antisymmetric and 10 symmetric (unitary) matrices in spinor space. The
gamma matrices commute with the automorphism group of the Clifford
algebra, ${\rm USp}(2N)$, where $N$ denotes the number of independent
spinors.  Spinors can be described either as Dirac spinors, or as
symplectic Majorana spinors. The latter description has the advantage
that it makes the action of the ${\rm USp}(2N)$ R-symmetry group
manifest. We will thus employ symplectic Majorana spinors $\psi^i$
with $i=1,2,\ldots,2N$, subject to the reality constraint,
\begin{equation}
  \label{eq:Majorana}
  C^{-1} \,\bar\chi_i {}^{\rm T}= \Omega_{ij}\,\chi^j\,,
\end{equation}
where $\Omega$ is the symplectic ${\rm USp}(2N)$ invariant tensor.
The Dirac conjugate is defined by $\bar\psi= \psi^\dagger\gamma_5$.
Observe that we adhere to the convention according to which raising or
lowering of $\mathrm{USp}(2N)$ indices is effected by complex
conjugation.

The gravitini $\psi_\mu{}^i$ and associated supersymmetry parameters
$\epsilon^i$ transform in the ${\bf 2N}$ representation of ${\rm
  USp}(2N)$. In principle we may also consider spinors transforming
under more complicated representations of $\mathrm{USp}(2N)$. For such
a spinor $\chi^{ij\cdots}{}_{mn\cdots}$ the symplectic Majorana
constraint would read
\begin{equation}
  \label{eq:chi-Majorana}
  C^{-1} \,(\bar\chi_{ij\cdots}{}^{mn\cdots})^{\rm T}= \Omega_{ik}\,
  \Omega_{jl}\;\Omega^{mp}\,\Omega^{nq}\cdots
  \,\chi^{kl\cdots}{}_{pq\cdots} \,,  
\end{equation}
Of course, the symplectic Majorana condition is defined up to a phase,
and we made a specific choice for that in \eqref{eq:Majorana} and
\eqref{eq:chi-Majorana}. For fermionic bilinears, with spinor fields
$\psi^i$ and $\varphi^i$ and a spinor matrix $\Gamma$ constructed from
products of gamma matrices, we note the following result, 
\begin{equation}
  \label{eq:bilinear}
  \bar \psi_i\Gamma \varphi^j= -
  \Omega_{ik}\,\Omega^{jl}\,\bar\varphi_l\, 
  C^{-1}\, \Gamma^{\rm T}\,C\, \psi^k = (\bar\varphi_j\,
  \gamma_5\,\Gamma^{\dagger}\,\gamma_5 \,\psi^i)^\dagger\,. 
\end{equation}
Hence $\mathrm{i}\,\bar \psi_i\, \varphi^j$, $\bar \psi_i\gamma_a \varphi^j$
and $\mathrm{i}\,\bar \psi_i\gamma_{ab} \varphi^j$ are pseudo-hermitean
(provided $a,b,\ldots= 1,\ldots,4$; in Pauli-K\"all\'en convention the
time component associated with $a=5$ acquires an extra minus sign) .
Generalization of this result to spinors transforming according to more
complicated $\mathrm{USp}(2N)$ representations is straightforward.

Multiplication of symplectic Majorana spinors with spinor matrices
$\Gamma$ consisting of products of gamma matrices are not automatically
symplectic Majorana spinors. This follows from
\begin{equation}
  \label{eq:1sM-Gamma-chi}
  \overline{\Gamma\chi^i}^{\;\mathrm{T}} = \Omega_{ij}\,C\; \gamma_5(C^{-1}
  \Gamma^{\mathrm{T}} C)^\dagger\gamma_5 \;\chi^j \,.  
\end{equation}
This means that $\mathrm{i}\gamma_a\chi^i$, $\gamma_{ab}\chi^i$,
$\mathrm{i}\gamma_{abc}\chi^i$, $\gamma_{abcd}\chi^i$ are also
symplectic Majorana spinors with the same reality phase as
\eqref{eq:Majorana}.

%%%%%%%%%%%%%%%%%%%%%%%%%%%%%%%%%%%%%%%%%%%%%%%%%%%%%%%%%%%%%%%%%%%%
\section{Conformal supergravity in 5 space-time dimensions}
\label{App:conf-sg}
\setcounter{equation}{0}
%%%%%%%%%%%%%%%%%%%%%%%%%%%%%%%%%%%%%%%%%%%%%%%%%%%%%%%%%%%%%%%%%%%%XS
The independent bosonic fields in $N=1$ conformal supergravity
multiplet in five space-time dimensions consist of the f\"unfbein
field $e_\mu{}^a$, the $\mathrm{SU}(2)$ gauge fields $V_\mu{}^i{}_j$,
the gauge field $b_\mu$ associated with scale transformations, and an
anti-symmetric tensor field $T_{ab}$ and a scalar field $D$.
Furthermore there are composite gauge fields $\omega_\mu{}^{ab}$ and
$f_\mu{}^a$ associated with the local Lorentz transformations and the
conformal boosts. The independent fermionic fields are the gravitino
fields $\psi_\mu{}^i$ and an ordinary spinor $\chi^i$. The
composite gauge field $\phi_\mu{}^i$ is associated with the so-called
{\it special} supersymmetry transformations. Together these fields
constitute the Weyl supermultiplet
\cite{de Wit:1979pq,Bergshoeff:1980is,Howe:1981nz}. 

The field content of four-dimensional $N=2$ and five-dimensional $N=1$
supergravity is rather similar, in view of the fact that spinors carry
four components in both case, Furthermore the R-symmetry groups are
almost the same, and equal to $\mathrm{SU}(2)\times\mathrm{U}(1)$ and
$\mathrm{USp}(2)$, respectively. However, the number of degrees of
freedom are different, as is shown in table \ref{tab:countWeyl}. The
reason can be understood from the fact that the Weyl multiplet is
conjugate to the smallest massive supersymmetry representation
containing spin-2 and spin-3/2 as the highest spin states. For
comparison we also display the situation for the $N=4$ Weyl multiplet
in four dimensions, and the $N=2$ Weyl multiplet in five dimensions,
with corresponding R-symmetry groups $\mathrm{U}(4)$ and
$\mathrm{USp}(4)$, respectively. These two multiplets comprise the
same number of degrees of freedom. 
\pagebreak
%%%%%%%%%%%%%%%%%%%%%%%%%%%%%%%%%%%%%%%%%%%%%%%%%%%%%%%%%%%%%%
\begin{table}[ht]
\centering
\begin{tabular}{lccccc} \hline
& \multicolumn{2}{c} {\mbox{8 supercharges}} 
& &\multicolumn{2}{c} {\mbox{16 supercharges}} \\ \hline
 \mbox{field}  & d=4  & d=5 & & d=4  & d=5 \\
\hline
$e_\mu {}^a$    & 5 &  9 & & 5 &  9 \\
$V_{\mu i}{}^j$   & 9 & 12  & & 45  & 40 \\
$A_\mu$ &3 &  - & &-  & 4 \\
$T_{ab}{}^{[ij]}$  & 6 &  10  & & 36  & 50 \\
$D_{[kl]}{}^{[ij]}$    & 1   & 1 & & 20   & 14\\
$E_{(ij)}$  & - &  - & & 20 & 10 \\
$\phi$    & - &   - & & 2   & 1\\
\hline &&&\\[-5.7mm]
$\psi_\mu{}^i$  & 16 &  24 & & 32  & 48 \\[1mm] 
$\chi^i{}_{[kl]}$  & 8  & 8  & & 80 & 64 \\
$\Lambda_i$    & - & - & & 16 &  16 \\[1mm]
\hline
\mbox{bosons+fermions}  & 24+24  & 32+32 & & 128+128  & 128+128  \\
\hline
\end{tabular}
  \caption{\small Bosonic and fermionic degrees of freedom of the Weyl
    multiplets in four and  five dimensions for the case of four and
    sixteen supercharges. All degrees of freedom can be assigned to a
    product representation of the group of spatial rotations 
    and the R-symmetry group. Decomposing the states in the second
    column under the group of 3-dimensional rotations yields a
    reducible multiplet comprising the states of the four-dimensional Weyl
    multiplet (given in the first column) and of an extra vector (or
    tensor) multiplet.  } 
  \label{tab:countWeyl}
\end{table}
%%%%%%%%%%%%%%%%%%%%%%%%%%%%%%%%%%%%%%%%%%%%%%%%%%%%%%%%%%%%%%%%%%%%

%%%%%%%%%%%%%%%%%%%%%%%%%%%%%%%%%%%%%%%%%%%%%%%%%%%%%%%%%%%%%%%%%
\end{appendix}
%%%%%%%%%%%%%%%%%%%%%%%%%%%%%%%%%%%%%%%%%%%%%%%%%%%%%%%%%%%%%%%%%
%%%%%%%%%%%%%%%%%%%%%%%%%%%%%%%%%%%%%%%%%%%%%%%%%%%%%%%%%%%%%%%%%
%%%%%%%%%%%%%%%%%%%%%%%%%%%%%%%%%%%%%%%%%%%%%%%%%%%%%%%%%%%%%%%%%%

\providecommand{\href}[2]{#2}
\begingroup\raggedright
\endgroup
\end{document}